\documentclass[%
 reprint,
preprintnumbers,
 amsmath,amssymb,
 aps,
prb,
]{revtex4-2}

\usepackage{graphicx}
\usepackage{dcolumn}
\usepackage{bm}

\usepackage{color}
\usepackage{braket} 
\usepackage{longtable}
\usepackage{txfonts}
\usepackage{multirow}

\begin{document}

\preprint{APS/123-QED}

\title{Multipole classification in 122 magnetic point groups 
\\ for unified understanding of multiferroic responses and transport phenomena
}

\author{Megumi Yatsushiro$^{1,3}$, Hiroaki Kusunose$^{2}$, Satoru Hayami$^{3}$}
 \affiliation{$^1$Department of Physics, Hokkaido University, Sapporo 060-0810, Japan \\
 $^2$Department of Physics, Meiji University, Kawasaki 214-8571, Japan \\
 $^3$ Department of Applied Physics, The University of Tokyo, Tokyo 113-8656, Japan}
 
\begin{abstract}
Mutual interplay between the electronic degrees of freedom in solids, such as charge, spin, orbital, sublattice, and bond degrees of freedom, is a source of cross-correlated phenomena with unconventional electronic ordered states. 
Such degrees of freedom can be described by four types of multipoles (electric, magnetic, magnetic toroidal, and electric toroidal) in a unified way, which enable us to tightly connect the microscopic degrees of freedom with macroscopic physical responses in a transparent manner.
We complete a classification of the multipoles in all 122 magnetic point groups based on the group theory.
The classification is useful to identify potentially active multipoles not only in ordinary ferromagnetic and antiferromagnetic orderings but also in exotic orderings breaking time-reversal symmetry, e.g., a loop-current state.
Moreover, the classification gives an insight into the microscopic origin of the cross-correlated responses and quantum transports.
By analyzing response functions up to the second order, we summarize the indispensable multipole moments for various responses, such as the linear magnetoelectric, piezoelectric, and elastic responses, and the nonlinear conductivity and Nernst coefficient.
Our results highly promote a further discovery of functional multiferroic materials,  
guided by the bottom-up material design based on the symmetry-adapted multipoles. 
\end{abstract}

\maketitle

\section{Introduction  \label{sec:intro}}
Magnetic orderings have long been studied, in which various fascinating phenomena emerge
such as the anomalous Hall effect~\cite{Nagaosa_review}, multiferroicity like 
the magnetoelectric effect~\cite{khomskii2009trend,  
doi:10.1080/00018730902920554, 
doi:10.1080/00018732.2015.1114338, 
spaldin2019advances, 
 fiebig2005revival, eerenstein2006multiferroic}, 
nonreciprocal transports~\cite{tokura2018nonreciprocal}, and so on.
In ferromagnetic orderings, for example, the uniform magnetization affects the electron kinetics through the Berry-phase-related mechanism, which results in the anomalous Hall effect, Kerr effect, and Nernst effect~\cite{Nagaosa_review, xiao2010berry, gradhand2012first}. 
Meanwhile, antiferromagnetic (AFM) orderings can also give rise to the large anomalous Hall effect even 
with a negligibly small magnetization~\cite{PhysRevLett.112.017205,K_bler_2014,PhysRevB.103.L180407}, which have been discussed in the noncoplanar AFM ordering~\cite{shindou2001orbital}, the noncollinear AFM ordering in Mn$_3$Sn~\cite{nakatsuji2015large,suzuki2017cluster,Yang_2017} and the almost collinear-type AFM ordering in RuO$_2$~\cite{eaaz8809, feng2020observation}, $\kappa$-type organic compound~\cite{PhysRevB.102.075112}, 
$\gamma$-FeMn~\cite{wang2021giant}, and $\alpha$-Mn~\cite{PhysRevResearch.2.043090} and the magneto-optical effect in La$M$O$_3$($M=\mbox{Cr, Mn, and Fe}$)~\cite{PhysRevB.55.8060}.
These studies suggest that the relation between the magnetic orderings and the electromagnetic responses is nontrivial, and there are potential candidates of functional materials that can be utilized for future spintronics devices.

The symmetry is a powerful tool to connect the magnetic orderings and their physical phenomena~\cite{birss1964symmetry}. 
In the case of the magnetic materials, 
the physical properties are classified by the group theory, such as the
magnetic group (Shubnikov group)~\cite{shubnikov1951symmetry, tavger1956magnetic, belov1957kristallografiyd, opechowski1965magnetism, birss1964symmetry, litvin2013magnetic} or the magnetic representation theory~\cite{bertaut1968representation}.
It helps us to investigate the macroscopic physical responses including the unconventional anomalous Hall effect in the AFM orderings as mentioned above.
However, it is insufficient to understand the microscopic key parameters for the magnetic responses within the symmetry argument.

One of the promising quantities to connect the microscopic electronic degrees of freedom and the macroscopic physical phenomena is the electronic multipoles~\cite{kuramoto2009multipole, Santini_RevModPhys.81.807,hayami2018classification, watanabe2018group, suzuki2018first}, since the multipole degree of freedom constitutes a complete set for an arbitrary Hilbert space with the charge, spin, and orbital degrees of freedom in electrons~\cite{hayami2018microscopic, kusunose2020complete}. 
The multipoles have been mainly used to describe peculiar atomic electronic degrees of freedom in $d$- or $f$-electron systems~\cite{kuramoto2009multipole, Santini_RevModPhys.81.807, suzuki2018first}, e.g., electric quadrupole in CeB$_6$~\cite{takigawa1983nmr, luthi1984elastic, EFFANTIN1985145, ERKELENS198761, nakamura1994quadrupole, sakai1997new, shiina1997magnetic} and magnetic octupole in NpO$_2$~\cite{PhysRevLett.85.2188, Paixao_PhysRevLett.89.187202, sakai2005invariant, tokunaga2006nmr, PhysRevB.78.104425, PhysRevB.82.241103}. 
Meanwhile, the concept of multipole covers not only the atomic electronic degrees of freedom but also ones over multi sites and orbitals, such as the hybrid multipole for the interorbital degrees of freedom between different orbitals~\cite{hayami2018microscopic,yatsushiro2019atomic,kusunose2020complete}, the cluster multipole for the on-site degrees of freedom in a cluster~\cite{suzuki2017cluster,suzuki2019multipole}, the bond multipole for the off-site bond degrees of freedom in a cluster~\cite{hayami2019electric,hayami2020bottom}, and the $k$ multipole for the band modulations and spin splittings in the electronic band structures~\cite{hayami2018classification, watanabe2018group}. 
The microscopic description in terms of the electronic multipole degrees of freedom is useful to understand
the macroscopic responses in accordance with the crystallographic symmetry~\cite{PhysRevB.76.214404, spaldin2008toroidal, doi:10.7566/JPSJ.86.034704, hayami2018classification, watanabe2018group, watanabe2018symmetry, thole2018magnetoelectric,jpsj_yatsushiro2020odd}, such as the magnetoelectric effect in the magnetic toroidal dipole orderings, e.g., Cr$_2$O$_3$~\cite{popov1999magnetic}, LiCoPO$_4$~\cite{van2007observation, zimmermann2014ferroic}, and UNi$_4$B~\cite{hayami2014toroidal,saito2018evidence} and in the magnetic quadrupole orderings, e.g., Co$_4$Nb$_2$O$_9$~\cite{PhysRevB.93.075117, PhysRevB.96.094434, yanagi2018theory, yanagi2018manipulating}, 
anomalous Hall effect in the magnetic octupole orderings, e.g., Mn$_3$Sn~\cite{nakatsuji2015large,suzuki2017cluster}, 
magnetopiezoelectric effect in the magnetic quadrupole/hexadecapole orderings, e.g., Ba$_{1-x}$K$_x$Mn$_2$As$_2$~\cite{watanabe2017magnetic} and EuMn$_2$Bi$_2$~\cite{shiomi2019observation}, 
and magnetostriction effect in the magnetic octupole orderings~\cite{Arima_doi:10.7566/JPSJ.82.013705,patri2019unveiling}.
Recent studies also clarified that the multipole description gives the systematic microscopic understanding of the band modulation in the AFM orderings~\cite{naka2019spin,hayami2019momentum, hayami2020spontaneous, hayami2020bottom}, e.g., 
$\kappa$-(BETD-TTF)$_2$Cu[N(CN)$_2$]Cl~\cite{naka2019spin, hayami2020multipole} and  
Ba$_3$MnNb$_2$O$_9$~\cite{hayami2020spontaneous, PhysRevB.90.224402}. 
Furthermore, the multipole description has been also applied to analyze the nature of the multipole phase transitions based on the Laudau free energy expansion~\cite{kusunose2008description,kusunose2020complete} and the behavior of the effective hyperfine fields for NMR/NQR spectra~\cite{sakai1997new,sakai1999antiferro,sakai2005invariant,yatsushiro2020nqr}.

In this way, the multipole description is essential not only to understand the physical phenomena but also explore further intriguing physics on the basis of microscopic electronic degrees freedom.
As the multipoles are closely related to the spherical harmonics as shown below, they can be systematically classified into the crystal symmetry. 
Nevertheless, the classification of multipoles has been performed only for the 32 crystallographic point groups~\cite{hayami2018classification}, and it is still lacking for the 122 magnetic point groups with the anti-unitary time-reversal operation. 
In the present paper, we complete a multipole classification under the 122 magnetic point groups by using four types of multipoles, electric (E), magnetic (M), electric toroidal (ET), and magnetic toroidal (MT) multipoles. 
The multipole classification enables us to rearrange any AFM orderings or more exotic orderings, such as nematic, excitonic, and loop-current orderings into the ferroic multipole orderings. 
The systematic classification by multipoles has an advantage of understanding the nature of potential order parameters and the relevant 
cross-correlated physical phenomena on the basis of microscopic electronic degrees of freedom. 
We also discuss the relation between the multipoles and the linear/nonlinear response functions based on the Kubo formula beyond the symmetry analysis. 

This paper is organized as follows.
In Sec.~\ref{sec:multipole}, we review the expressions of the four types of multipoles.
In Sec.~\ref{sec:M_MT_representation}, the multipole classification in a magnetic point group is presented by exemplifying a cubic system.
We show the active E, ET, M, and MT multipoles in all magnetic point groups in Sec.~\ref{sec:active_MP}.
In Sec.~\ref{sec:res}, the relation between the response tensors and the active multipole moments is discussed based on the group theoretical analysis and the Kubo formula. 
We summarize the present paper in Sec.~\ref{sec:summary}.
In Appendix~\ref{sec:ap_real}, we give the cubic and tesseral harmonics used for the multipole expressions.
Appendix~\ref{sec:ap_MPG_subgroup} represents the unitary subgroup for each black and white point group.
Appendix~\ref{sec:ap_classification} gives the classification tables of multipoles, which is not presented in Sec.~\ref{sec:M_MT_representation}.
In Appendix~\ref{sec:ap_MPG_corepresentation}, we review the irreducible corepresentation of the magnetic point group.
In Appendix~\ref{sec:ap_Laue}, we summarize the relation between the magnetic point groups and (magnetic) Laue groups.
The derivation of the multipole expressions of the response tensors in Sec.~\ref{sec:res} is given in Appendix~\ref{sec:ap_res_mp}.
In Appendix~\ref{sec:ap_hexagonal}, the multipole expressions for the response tensors in the hexagonal and trigonal systems are shown.

\section{Four types of Multipoles \label{sec:multipole}}

The multipole moments are introduced by a multipole expansion of the electric scalar potential $\phi({\bm r})$ and magnetic vector potential ${\bm A}({\bm r})$ representing a spatial distribution of a source electric charge $\rho_{\rm e}({\bm r})$ and current ${\bm j}_{\rm e}({\bm r})$ in classical electromagnetism~\cite{LandauLifshitz198001,schwartz1955theory,BlattWeisskopf201111,dubovik1974multipole, Jackson3rd1999,nanz2016toroidal}. 
The spatial distribution of $\rho_{\rm e}({\bm r})$ in $\phi({\bm r})$ leads to E multipole (time-reversal even polar tensor), while that of ${\bm j}_{\rm e}({\bm r})$ in ${\bm A}({\bm r})$ leads to M multipole (time-reversal odd axial tensor) and MT multipole (time-reversal odd polar tensor). 
Furthermore, ET multipole (time-reversal even axial tensor) is introduced by considering the spatial distribution of a magnetic current ${\bm j}_{\rm m}({\bm r})$~\cite{dubovik1986axial, dubovik1990toroid,hayami2018microscopic}. 
These four types of multipoles show the different spatial inversion and time-reversal parities, as summarized in Table~\ref{table:multipole_PT}.

In condensed matter physics, four types of multipoles can describe a spatial distribution of the electronic charge and spin in the atomic orbitals by using their quantum-mechanical operator expressions as~\cite{kusunose2008description, hayami2018microscopic, kusunose2020complete} 
\begin{align}
\label{eq:E_def}
\hat{Q}_{lm} &= -e\sum_jO_{lm}({\bm r}_j), \\
\label{eq:M_def}
\hat{M}_{lm} &= -\mu_{\rm B} \sum_j {\bm m}_l({\bm r}_j) \cdot {\bm \nabla} O_{lm}({\bm r}_j), \\ 
\label{eq:MT_def}
\hat{T}_{lm} &= -\mu_{\rm B} \sum_j {\bm t}_l({\bm r}_j) \cdot {\bm \nabla} O_{lm}({\bm r}_j), \\ 
\label{eq:ET_def}
\hat{G}_{lm} &= -e\sum_j \sum_{\alpha\beta}^{x,y,z} g_l^{\alpha\beta}({\bm r}_j) \nabla_\alpha \nabla_\beta O_{lm}({\bm r}_j),
\end{align}
where $\hat{Q}_{lm}$, $\hat{M}_{lm}$, $\hat{T}_{lm}$, and $\hat{G}_{lm}$ denote the E, M, MT, and ET multipoles with the azimuthal quantum number $l$ (rank of multipoles) and magnetic quantum number $m$, respectively.
In Eqs.~\eqref{eq:E_def}--\eqref{eq:ET_def}, $-e$ and $-\mu_{\rm B}$ are the electron charge and Bohr magneton, respectively, which are taken to be unity hereafter, i.e., $-e, -\mu_{\rm B} \to1$. 
$O_{lm}({\bm r})$ is related with the spherical harmonics as a function of angle $\hat{\bm r}={\bm r}/|{\bm r}|$, $Y_{lm}(\hat{\bm r})$, which is given by
\begin{align}
O_{lm}({\bm r}) = \sqrt{\frac{4\pi}{2l+1}}r^l Y_{lm}^* (\hat{\bm r}). 
\end{align}
In the following discussion for the crystallographic system, we
adopt the real expressions of $O_{lm}$ following Refs.~\cite{hutchings1964point,kusunose2008description} and use the multipole notation as $X_0$ for monopole ($l=0$), $(X_{x},X_{y}, X_{z})$ for dipole ($l=1$), $(X_{u},X_{\varv}, X_{yz}, X_{zx}, X_{xy})$ for quadrupole ($l=2$), $(X_{xyz}, X^{\alpha}_{x}, X^{\alpha}_{y}, X^{\alpha}_z, X^{\beta}_{x}, X^{\beta}_{y}, X^{\beta}_z)$ for octupole ($l=3$), and $(X_{4}, X_{4u}, X_{4\varv}, X_{4x}^\alpha. X_{4y}^\alpha, X_{4z}^\alpha, X_{4x}^\beta, X_{4y}^\beta, X_{4z}^\beta)$ for hexadecapole ($l=4$) following Ref.~\cite{hayami2018classification} (see also Appendix~\ref{sec:ap_real} for their expressions).
${\bm m}_l({\bm r}_j)$, ${\bm t}_l({\bm r}_j)$, and $g_l^{\alpha\beta} ({\bm r}_j)$ represent the magnetic moment, magnetic toroidal moment, and electric toroidal tensor, respectively, which are expressed as
\begin{align}
{\bm m}_l({\bm r}_j) &= \frac{2{\bm l}_j}{l+1}+{\bm \sigma}_j,\\
{\bm t}_l({\bm r}_j) &= \frac{{\bm r}_j}{l+1}\times \left( \frac{2{\bm l}_j}{l+2}+{\bm \sigma}_j \right),  \\
g_l^{\alpha\beta} ({\bm r}_j) &= m_l^{\alpha} ({\bm r}_j)t_l^\beta({\bm r}_j), 
\end{align}
where ${\bm l}_j$ and ${\bm \sigma}_j/2$ are the dimensionless orbital and spin angular-momentum operators of an electron at ${\bm r}_j$, respectively.
The parities with respect to the spatial inversion and time-reversal operations, $\mathcal{P}$ and $\mathcal{T}$, for each multipole are given as follows: $(\mathcal{P}, \mathcal{T}) = [(-1)^l,+1]$  for $Q_{lm}$, $[(-1)^{l+1},-1]$ for $M_{lm}$, $[(-1)^l,-1]$ for $T_{lm}$, and $[(-1)^{l+1},+1]$ for $G_{lm}$, which are summarized in Table~\ref{table:multipole_PT}.
The four types of multipoles constitute a complete set for an arbitrary electronic degree of freedom in the atomic orbitals~\cite{kusunose2008description, hayami2018microscopic, kusunose2020complete}.

\begin{table}[h!]
\centering
\caption{
Four types of multipoles and their spatial inversion ($\mathcal{P}$), time-reversal ($\mathcal{T}$), and  $\mathcal{PT}$ parities.
The relevant source fields are also presented
\label{table:multipole_PT}}
\begin{tabular}{cccccc}
\hline\hline
type & notation & $\mathcal{P}$ & $\mathcal{T}$ & $\mathcal{PT}$ & source \\
\hline
E & $Q_{lm}$ & $(-1)^l$ & $+1$ & $(-1)^l$ & $\rho_{\rm e}$ 
\,(${\bm j}_{\rm m}$) \\
M & $M_{lm}$ & $(-1)^{l+1}$ & $-1$ & $(-1)^l$ & ${\bm j}_{\rm e}$ \\
MT & $T_{lm}$ & $(-1)^{l}$ & $-1$ & $(-1)^{l+1}$ & ${\bm j}_{\rm e}$ \\
ET & $G_{lm}$ & $(-1)^{l+1}$ & $+1$ & $(-1)^{l+1}$ &${\bm j}_{\rm m}$  \\
\hline\hline
\end{tabular}
\end{table}

The atomic-scale multipole operators in Eqs.~\eqref{eq:E_def}--\eqref{eq:ET_def} can also be applied to describe anisotropic distributions on a cluster system with the sublattice degree of freedom, which is termed as a cluster multipole~\cite{spaldin2008toroidal, hayami2016emergent, suzuki2017cluster, suzuki2018first, suzuki2019multipole}. 
For example, the expressions of M and MT multipoles are obtained by reading ${\bm r}_{j}$
 in Eqs.~\eqref{eq:M_def} and \eqref{eq:MT_def}
 with the position of the $j$-th atom $\bm{R}_{j}$ in a magnetic unit cell.
By adopting the virtual atomic cluster method in Ref.~\onlinecite{suzuki2019multipole}, one can systematically describe any $N$-site magnetic structures in terms of cluster M and MT multipoles as follows~\cite{suzuki2019multipole}:
\begin{align}
\label{eq:Mc_def}
\hat{M}^{\rm (c)}_{lm} &= \sum_{j=1}^{N} {\bm \sigma}_{j} \cdot {\bm \nabla}_{j}
 O_{lm} ({\bm R}_{j}), \\
\label{eq:MTc_def}
\hat{T}^{\rm (c)}_{lm} &= \frac{1}{l+1}\sum_{j=1}^{N} (\hat{\bm R}_{j} \times {\bm \sigma}_{j}) \cdot {\bm \nabla}_{j}
 O_{lm} ({\bm R}_{j}),
\end{align}
where the superscript (c) is introduced to represent cluster multipoles for clarity. 
Equations~(\ref{eq:Mc_def}) and (\ref{eq:MTc_def}) give the multipole order parameters under the AFM orderings, which provide an understanding of physical phenomena.
For example, the magneto-striction effect under the all-in-all-out magnetic ordering in the pyrochlore structure is understood by the emergence of the cluster M octupole~\cite{Arima_doi:10.7566/JPSJ.82.013705} and the linear magneto-electric effect in the zigzag chain is owing to the active cluster MT dipole~\cite{Yanase_JPSJ.83.014703,hayami2015spontaneous, shinozaki_doi:10.7566/JPSJ.89.033703}.

\section{Irreducible representations of multipoles under magnetic point groups \label{sec:M_MT_representation}}
The four types of multipoles in Eqs.~\eqref{eq:E_def}--\eqref{eq:ET_def} express arbitrary electronic degrees of freedom in the Hilbert space spanned by the electron wave functions in the 32 point-group irreducible representations~\cite{hayami2018classification}.  
In this section, we generalize the multipole classification from the 32 crystallographic point groups to the 122 {\it magnetic} point groups by taking into account the anti-unitary time-reversal operation, $\theta$.

The 122 magnetic point groups are categorized into three groups~\cite{bradley2009mathematical}:
\begin{itemize}
\item[(I)] ordinary crystallographic
point groups without $\theta$ (32),
\item[(II)] gray point groups (32),
\item[(III)] black and white point groups (58),
\end{itemize}
where the numbers in parentheses represent the number of magnetic point groups.
Supposing ${\bm G}$ represents any point group belonging to the type-(I) crystallographic
point groups, the type-(II) gray point group, ${\bm M}^{{\rm (II)}}$, is defined so as to contain the double elements of $\bm{G}$, which is represented by 
\begin{align}
\label{eq:GPG}
{\bm M}^{{\rm (II)}}={\bm G}+\theta{\bm G}.
\end{align}
Meanwhile, the type-(III) black and white point group, ${\bm M}^{{\rm (III)}}$, consists of half of the elements of ${\bm M}^{{\rm (II)}}$, which is represented as 
\begin{align}
\label{eq:BWPG}
{\bm M}^{{\rm (III)}}={\bm H}+\theta ({\bm G}-{\bm H}),
\end{align}
where ${\bm H}$ is a halving unitary subgroup of ${\bm G}$. 
The correspondence between ${\bm M}^{{\rm (III)}}$ and ${\bm H}$ in each black and white point group is shown in Table~\ref{table:subgroup} in Appendix~\ref{sec:ap_MPG_subgroup}.

On the basis of the group theory, we classify four types of multipoles under 122 magnetic point groups. 
We focus on the multipole classification under the magnetic point groups with time-reversal operations, i.e., type-(II) and type-(III) groups. 
First, let us classify the multipoles under the 32 type-(II) gray point groups, which is summarized in Table~\ref{table:gray_m3m1} in this section and Tables~\ref{table:gray_cubic}--\ref{table:gray_trigonal_2} in Appendix~\ref{sec:ap_classification_GPG}. 
In the following, we show the examples by taking the cubic gray point group $m\bar{3}m1'$ in Table~\ref{table:gray_m3m1}.

Table~\ref{table:gray_m3m1} presents the M and MT multipoles up to rank 4 classified into the irreducible corepresentation (IRREP) under the cubic gray point group $m\bar{3}m1'$.
In the table, the superscript ``$-$" in the IRREP stands for the odd time-inversion property, as discussed in detail in Appendix~\ref{sec:ap_MPG_corepresentation}. 
The IRREPs of the E and ET multipoles are obtained by replacing $(T,M) \to (Q,G)$ and $\Gamma^-\to \Gamma^+$ where $\Gamma={\rm A}_{\rm 1g/u}, {\rm A}_{\rm 2g/u}, {\rm E}_{\rm g/u}, {\rm T}_{\rm 1g/u}, {\rm T}_{\rm 2g/u}$.
Table~\ref{table:gray_m3m1} also shows the reduction to the subgroups in each IRREP. 
In the table,
``P. axis'' stands for the primary axis of the point group operations.
For example, the symmetry operations of $4/mm'm'$ with P. axis $[001]$ are
\begin{align}
E, C_{4z}, C_{4z}^3, C_{4z}^2, \theta C'_{2x}, \theta C'_{2y}, \theta C''_{2[110]}, \theta C''_{2[\bar{1}10]}, \notag\\
I, IC_{4z}, IC_{4z}^3, \sigma_{\perp z}, \theta \sigma_{\perp x}, \theta \sigma_{\perp y}, \theta \sigma_{\perp [110]}, \theta \sigma_{\perp [\bar{1}10]}, 
\end{align}
where we specify the operation axis or plane in the subscript.
On the other hand, for $4/mm'm'$ with P. axis $[100]$, the operation axes and planes are transformed in cyclic in accordance with the change of the P. axis, which are represented as
\begin{align}
E, C_{4x}, C_{4x}^3, C_{4x}^2, \theta C'_{2y}, \theta C'_{2z}, \theta C''_{2[011]}, \theta C''_{2[0\bar{1}1]}, \notag\\
I, IC_{4x}, IC_{4x}^3, \sigma_{\perp x}, \theta \sigma_{\perp y}, \theta \sigma_{\perp z}, \theta \sigma_{\perp [011]}, \theta \sigma_{\perp [0\bar{1}1]}.
\end{align}

\begin{table}[h!]
\centering
\caption{Irreducible corepresentations (IRREP) of magnetic (M) and magnetic toroidal (MT) multipoles ($l\leq 4$) in the cubic gray point group $m\bar{3}m1'$.
The superscript ``$-$'' of the IRREP means the odd parity with respect to the time-reversal operation.
The corresponding magnetic point group (MPG) with its primary axis (P. axis) is also shown. 
\label{table:gray_m3m1}}
\begin{tabular}{ccccc}
\hline\hline
IRREP & MT & M & MPG & P. axis \\
\hline 
A$_{\rm 1g}^-$ & $T_0, T_4$ &  & $m\bar{3}m$ & $\left<100 \right>$ \\
A$_{\rm 2g}^-$ && $M_{xyz}$ & $m\bar{3}m'$ &  $\left<100 \right>$ \\
E$_{\rm g}^-$ & $T_u, T_{4u}$ && $4/mmm$ & $[001]$ \\
 & $T_\varv, T_{4\varv}$ & &$4'/mmm'$ & $[001]$ \\
T$_{\rm 1g}^-$ & $T_{4x}^\alpha$ & $M_x, M_x^\alpha$ & $4/mm'm'$ & $[100]$ \\
  & $T_{4y}^\alpha$ & $M_y, M_y^\alpha$ & $4/mm'm'$ & $[010]$  \\
  & $T_{4z}^\alpha$ & $M_z, M_z^\alpha$ & $4/mm'm'$ & $[001]$  \\
T$_{\rm 2g}^-$ & $T_{yz}, T_{4x}^\beta$ & $M_x^\beta$ & $4'/mm'm$ & $[100]$ \\
  & $T_{zx}, T_{4y}^\beta$ & $M_y^\beta$ & $4'/mm'm$ & $[010]$ \\
  & $T_{xy}, T_{4z}^\beta$ & $M_z^\beta$ & $4'/mm'm$ & $[001]$ \\
\hline
A$_{\rm 1u}^-$ && $M_0, M_4$ & $m'\bar{3}'m'$ & $\left<100 \right>$\\
A$_{\rm 2u}^-$ & $T_{xyz}$ && $m'\bar{3}'m$ & $\left<100 \right>$ \\
E$_{\rm u}^-$ && $M_u, M_{4u}$ & $4/m'm'm'$ & $[001]$ \\
 && $M_\varv, M_{4\varv}$ & $4'/m'm'm$ & $[001]$ \\
T$_{\rm 1u}^-$ & $T_x, T_x^\alpha$ & $M_{4x}^\alpha$ & $4/m'mm$ & $[100]$ \\
  & $T_y, T_y^\alpha$ & $M_{4y}^\alpha$ & $4/m'mm$ & $[010]$  \\
  & $T_z, T_z^\alpha$ & $M_{4z}^\alpha$ & $4/m'mm$ & $[001]$ \\
T$_{\rm 2u}^-$ & $T_x^\beta$ & $M_{yz}, M_{4x}^\beta$ & $4'/m'mm'$ & $[100]$ \\
  & $T_y^\beta$ & $M_{zx}, M_{4y}^\beta$ & $4'/m'mm'$ & $[010]$ \\
  & $T_z^\beta$ & $M_{xy}, M_{4z}^\beta$ & $4'/m'mm'$ & $[001]$ \\
\hline\hline
\end{tabular}
\end{table}

\begin{figure}[htb!]
\centering
\includegraphics[width=85mm]{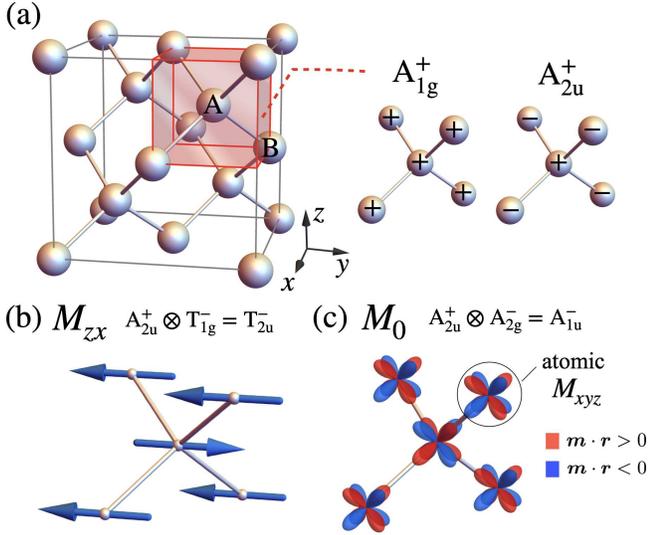}
\caption{(a) Diamond structure with two sublattices A and B (left panel). 
The IRREPs and the corresponding potential distributions in the two sublattices are shown in the right panel.
(b) The staggered magnetic dipole along $y$ axis and (c) the staggered $xyz$-type magnetic octupole, which are regarded as the cluster M quadrupole $M_{zx}$ and the M monopole $M_0$, respectively. 
The arrows in (b) represent the spin direction and the color in (c) represents the distribution of the magnetic monopole charge defined by ${\bm m}\cdot {\bm r}$. 
\label{fig:diamond}
}
\end{figure}

The classification in Table~\ref{table:gray_m3m1} provides the identification of hidden/unknown multipole order parameters in a systematic way.
Let us take an example of the diamond structure in Fig.~\ref{fig:diamond}(a) with the two sublattices A and B under the space group $Fd\bar{3}m$ and the magnetic point group $m\bar{3}m1'$~\footnote{
In general, the multipole description can also be applied to any types of orderings discussed in the main text, although more sophisticated consideration based on the magnetic superspace group is required in incommensurate orderings~\cite{Perez_Mato_2012}, since information regarding the ordering vector is also necessary to identify the symmetry in the ordered state.
}.
Supposing the two-sublattice ordering, the IRREPs for the sublattice degree of freedom,
$\Gamma_{\rm sub}$, are given by
\begin{align}
\Gamma_{\rm sub}={\rm A}_{\rm 1g}^+ \oplus {\rm A}_{\rm 2u}^+, 
\end{align}
where ${\rm A}_{\rm 1g}^+$ corresponds to the uniform potential distribution and ${\rm A}_{\rm 2u}^+$ corresponds to the staggered one, as shown in the right panel of Fig.~\ref{fig:diamond}(a).

When the two-sublattice magnetic order occurs, one can obtain the IRREPs with totally six components by taking the product of $\Gamma_{\rm sub}$ and the IRREPs for the M dipole $(M_x, M_y, M_z)$, $\Gamma_{\rm M1}={\rm T}_{\rm 1g}^-$, which are represented by~\footnote{The reduction of the inner Kronecker product of two IRREPs is shown in Refs.~\onlinecite{bradley1968magnetic, bradley2009mathematical}.}
\begin{align}
\Gamma_{\rm sub} \otimes \Gamma_{\rm M1} 
={\rm T}_{\rm 1g}^- \oplus {\rm T}_{\rm 2u}^-.
\end{align}
The IRREP ${\rm T}_{\rm 1g}^-$ in the right-hand side represents the uniform alignment of the M dipole, i.e., the ferromagnetic order. 
On the other hand, the IRREP ${\rm T}_{\rm 2u}^-$, which corresponds to the staggered magnetic structure shown in Fig.~\ref{fig:diamond}(b), is regarded as the M quadrupole ordering from the Table~\ref{table:gray_m3m1}~\footnote{Here and hereafter, we refer the name of the multipole orderings by adopting the lowest-rank multipoles belonging to the same irreducible (co)representation.}.
Thus, it is easy to predict the emergent physical phenomena related with the M quadrupole, such as the transverse magnetoelectric effect, once the staggered magnetic ordering occurs in the diamond structure~\cite{PhysRevB.97.024414} 
(see also Sec.~\ref{sec:res}).

Similarly, the classification in Table~\ref{table:gray_m3m1} is used not only for the AFM ordering but also for the unconventional electronic orderings, such as the spin nematics~\cite{andreev1984spin, papanicolaou1988unusual, Chen_PhysRevLett.27.1383, lacroix2011introduction},
excitonic states~\cite{jerome1967excitonic,Halperin_RevModPhys.40.755,Kune__2015, Kunes_PhysRevB.90.235112,Kaneko_PhysRevB.94.125127,yamaguchi2017multipole}, staggered flux states~\cite{affleck1988large,Nayak_PhysRevB.62.4880,Chakravarty_PhysRevB.63.094503,
Allais_PhysRevB.90.155114}, loop-current states~\cite{hayami2019electric, hayami2020bottom,
PhysRevB.73.155113, PhysRevB.80.214501, zhao2016evidence, PhysRevX.11.011021}, 
and other higher-rank multipole orderings~\cite{kuramoto2009multipole, Santini_RevModPhys.81.807, suzuki2019multipole}.
For instance, when considering the orderings of the atomic M octupoles $(M_{xyz}, M_x^\alpha, M_y^\alpha, M_z^\alpha, M_x^\beta, M_y^\beta, M_z^\beta)$ with the IRREPs $\Gamma_{\rm M3} = {\rm A}_{\rm 2g}^- \oplus {\rm T}_{\rm 1g}^- \oplus {\rm T}_{\rm 2g}^-$, 
the possible IRREPs within the two sublattices are given by
\begin{align}
\Gamma_{\rm sub} \otimes \Gamma_{\rm M3} =( {\rm A}_{\rm 2g}^- \oplus {\rm T}_{\rm 1g}^- \oplus {\rm T}_{\rm 2g}^-) \oplus ( {\rm A}_{\rm 1u}^- \oplus {\rm T}_{\rm 1u}^- \oplus {\rm T}_{\rm 2u}^-).
\end{align}
The former (latter) parenthesis represents the uniform (staggered) alignment of M octupoles. 
From Table~\ref{table:gray_m3m1}, one can find the multipole order parameters, e.g., the staggered $M_{xyz}$ ordering with ${\rm A}_{\rm 1u}^-$ is regarded as the M monopole $M_0$, as schematically shown in Fig.~\ref{fig:diamond}(c).
Then, the staggered $M_{xyz}$ ordering is expected to exhibit physical phenomena in the presence of $M_0$, such as the longitudinal magnetoelectric effect.

Next, we classify four types of multipoles under the type-(III) black and white point groups.
The results are summarized in Table~\ref{table:m3m1_maxim} in this section and Tables~\ref{table:BW_cubic}--\ref{table:BW_trigonal} in Appendix~\ref{sec:ap_classification_BWPG}. 
We here show the multipole classification under the maximum subgroups of $m\bar{3}m1'$, among which do not have the time-reversal symmetry, in Table~\ref{table:m3m1_maxim}. 
The IRREPs in Table~\ref{table:m3m1_maxim} present the transformation property with respect to their unitary subgroup and the $\pm$ symbol stands for the parity with respect to the anti-unitary operation except for the ${\rm B}_{\rm 2g/u}$ representation of $4'/mmm'$ (see also the discussion in Appendix~\ref{sec:ap_MPG_corepresentation}).

Table~\ref{table:m3m1_maxim} provides a complete correspondence of the IRREPs for the group-subgroup relation.
For example, in the staggered $M_{xyz}$-type octupole ordering as exemplified above in Fig.~\ref{fig:diamond}(c),
the symmetry reduces as $m\bar{3}m1' \to m'\bar{3}'m'$. In this case,
 the IRREPs of the parent point group $m\bar{3}m1'$ are read by those of the subgroup $m'\bar{3}'m'$ as follows: $({\rm A}_{\rm 1g}^\pm, {\rm A}_{\rm 1u}^\mp) \to {\rm A}_1^\pm$, $({\rm A}_{\rm 2g}^\pm, {\rm A}_{\rm 2u}^\mp) \to {\rm A}_2^\pm$, $({\rm E}_{\rm g}^\pm, {\rm E}_{\rm u}^\mp) \to {\rm E}^\pm$, $({\rm T}_{\rm 1g}^\pm, {\rm T}_{\rm 1u}^\mp) \to {\rm T}_1^\pm$, $({\rm T}_{\rm 2g}^\pm, {\rm T}_{\rm 2u}^\mp) \to {\rm T}_2^\pm$. 
Since some of the multipoles are turned to belong to the same IRREP by the lowering of the symmetry, additional cross couplings between multipoles that are not present in the parent point group are expected, e.g., the coupling between the MT dipole $(T_x, T_y, T_z)$ and the ET dipole $(G_x, G_y, G_z)$ appears because of the same IRREP $T_1^+$.

\begin{table*}[h!]
\centering
\caption{
IRREPs of the multipoles in the time-reversal breaking maximum subgroups of the gray point group $m\bar{3}m1'$.
$\{ X_{u}+ iX_{\varv}, X_{u}- iX_{\varv}\}$ and $\{ X_{4u}+ iX_{4\varv}, X_{4u}- iX_{4\varv}\} $ ($X=Q$ or $G$) are the basis of ${\rm E}_{\rm g/u}^{\rm (1,2)+}$ in $m\bar{3}m'$, while $\{ X_{u}+ iX_{\varv}, -X_{u}+ iX_{\varv}\}$ and $\{ X_{4u}+ iX_{4\varv}, -X_{4u}+ iX_{4\varv}\} $ ($X=T$ or $M$) for ${\rm E}_{\rm g/u}^{\rm (1,2)-}$.
$\{X_x+iX_y, X_x-iX_y\}$, $\{X_{yz}-iX_{zx}, X_{yz}+iX_{zx}\}$, $\{X_x^\alpha+iX_y^\alpha, X_x^\alpha-iX_y^\alpha\}$, $\{X_x^\beta-iX_y^\beta, X_x^\beta+iX_y^\beta\}$, $\{X_{4x}^\alpha+iX_{4y}^\alpha, X_{4x}^\alpha-iX_{4y}^\alpha\}$, and $\{X_{4x}^\beta-iX_{4y}^\beta, X_{4x}^\beta+iX_{4y}^\beta\}$ ($X=Q,G, T, M$) are the basis of ${\rm E}_{\rm g/u}^{\rm (1,2)\pm}$ in $4/mm'm'$.
The IRREPs stands for the transformation property in the unitary subgroups shown in the second row.
The superscript $\pm$ of the IRREPs represents the parity with respect to the anti-unitary operation in the third low, whose
operation axis is shown in the fourth row. 
\label{table:m3m1_maxim}}
\begin{tabular}{ccccccccccccccc}
\hline \hline
\multicolumn{2}{c}{magnetic point group} & $m\bar{3}m1'$ & $m\bar{3}m$ & $m\bar{3}m'$ & $m'\bar{3}'m'$ & $m'\bar{3}'m$ & $4/mmm$ & $4'/mmm'$ & $4/mm'm'$ & $4/m'm'm'$ & $4'/m'm'm$ & $4/m'mm$ \\ \cline{3-13}
\multicolumn{2}{c}{
unitary subgroup} & $m\bar{3}m$ &  & $m\bar{3}$ & $432$ & $\bar{4}3m$ &  & $mmm$ & $4/m$ & $422$ & $\bar{4}2m$ & $4mm$  \rule[0pt]{0pt}{8pt}\\
\multicolumn{2}{c}{anti-unitary operation} & $\theta$ & & $\theta C_{2}^{\prime}$ & $\theta I$ & $\theta I$ & & $\theta C_{2}^{\prime\prime}$ & $\theta C_{2}^\prime$ & $\theta I$ & $\theta I$ & $\theta I$  \\
& & & & $[110]$ & & & & $[110]$  & $[100]$ \\
\hline

E & ET  & & \\ \cline{1-2}
$Q_0, Q_4$ & &
A$_{\rm 1g}^+$ & A$_{\rm 1g}$ & A$_{\rm g}^+$ & A$_{\rm 1}^+$ & A$_{\rm 1}^+$ & A$_{\rm 1g}$ & A$_{\rm g}^+$ & A$_{\rm g}^+$ &  A$_1^+$ &  A$_1^+$ & A$_1^+$\\

 & $G_{xyz}$ & 
 A$_{\rm 2g}^+$ &  A$_{\rm 2g}$ & A$_{\rm g}^-$ &  A$_{\rm 2}^+$ & A$_{\rm 2}^+$ & B$_{\rm 1g}$ & A$_{\rm g}^-$ & B$_{\rm g}^+$ & B$_1^+$ & B$_1^+$ & B$_1^+$  \\
 
 $Q_u, Q_{4u}$ & & 
 E$_{\rm g}^+$ &  E$_{\rm g}$ & E$_{\rm g}^{(1,2)+}$ & E$^+$ & E$^+$ & A$_{\rm 1g}$ & A$_{\rm g}^+$ & A$_{\rm g}^+$ & A$_1^+$ & A$_1^+$ & A$_1^+$  \\
\hline\hline
\end{tabular}
\end{table*}

\section{Active multipoles under magnetic point groups \label{sec:active_MP}}
\begin{table*}[h!]
\centering
\caption{
Active multipoles in the crystallographic point group (CPG), gray point group (GPG), and black and white point group (BWPG) according to $\mathcal{P}$, $\mathcal{T}$, and $\mathcal{PT}$ symmetries.
``even/odd-parity'' represents the spatial-inversion parity of multipoles.
\label{table:type_of_active_MP}
}
\begin{tabular}{ccccccccc}
\hline \hline
type & magnetic point group & $\mathcal{P}$ & $\mathcal{T}$ & $\mathcal{PT}$ & even-parity &  odd-parity  & even-parity & odd-parity \\ \cline{6-9}
 & & & & & E, ET & E, ET &  MT, M & MT, M \\ 
 \hline 
(I) CPG & $m\bar{3}m$, $m\bar{3}$, $4/mmm$,  $4/m$, $mmm$, $2/m$, $\bar{1}$,
& $\bigcirc$& $\times$& $\times$ & $\checkmark$ &  & $\checkmark$ \rule[0pt]{0pt}{8pt}\\
& $6/mmm$, $6/m$, $\bar{3}m$, $\bar{3}$\\
\cline{2-9}
  & $432$, $\bar{4}3m$, $23$,  $422$, $\bar{4}2m$, $4mm$, $4$, $\bar{4}$, $222$, $mm2$, $2$, $m$, $1$,
 & $\times$& $\times$& $\times$ & $\checkmark$ & $\checkmark$ & $\checkmark$ & $\checkmark$ \rule[0pt]{0pt}{8pt}\\
 & $622$, $\bar{6}m2$, $6mm$, $6$, $\bar{6}$, $32$, $3m$, $3$  \\
 \hline
(II) GPG & $m\bar{3}m1'$, $m\bar{3}1'$, $4/mmm1'$,  $4/m1'$, $mmm1'$, $2/m1'$, $\bar{1}1'$,
 & $\bigcirc$& $\bigcirc$& $\bigcirc$ & $\checkmark$ \rule[0pt]{0pt}{8pt} \\ 
 & $6/mmm1'$, $6/m1'$, $\bar{3}m1'$, $\bar{3}1'$  \\
\hline\hline
\end{tabular}
\end{table*}

\begin{table*}[h!]
\centering
\caption{
Active even-parity E, ET, MT, and M multipoles in the centrosymmetric crystallographic point groups.
The triclinic point group $\bar{1}$ ($C_i$), in which all the even-parity multipoles are active, is omitted.
 \label{table:active_MP_CPG_1}}
\vspace{2mm}
\begin{tabular}{cccccccccccccc}\hline \hline
 &  \multicolumn{6}{c}{even-parity} &&  \multicolumn{6}{c}{even-parity} \\ 
 \cline{2-7} \cline{9-14}

 &  \multicolumn{3}{c}{E} &&  \multicolumn{2}{c}{ET} &&  \multicolumn{3}{c}{MT} &&  \multicolumn{2}{c}{M} \\ 
 \cline{2-4} \cline{6-7} \cline{9-11} \cline{13-14}
 
& $l=0$ & $2$ & $4$ &&
$1$ & $3$ &&
$0$ & $2$ & $4$ &&
$1$ & $3$ \\\hline
 
$m\bar{3}m$ ($O_{\rm h}$) & $Q_0$ & & $Q_4$ &&
& &&
$T_0$ & & $T_4$ \\
  
$m\bar{3}$ ($T_{\rm h}$) & $Q_0$ & & $Q_4$ &&
& $G_{xyz}$ &&
$T_0$ & & $T_4$ &&
& $M_{xyz}$\\
  
$4/mmm$ ($D_{\rm 4h}$) & $Q_0$ & $Q_u$ & $Q_4, Q_{4u}$ &&
& &&
$T_0$ & $T_u$ & $T_4, T_{4u}$ \\
 
$4/m$ ($C_{\rm 4h}$) & $Q_0$ & $Q_u$ & $Q_4, Q_{4u}, Q_{4z}^\alpha$ &&
$G_z$ & $G_z^\alpha$ &&
$T_0$ & $T_u$ & $T_4, T_{4u}, T_{4z}^\alpha$ &&
$M_z$ & $M_z^\alpha$ \\
 
$mmm$ ($D_{\rm 2h}$) & $Q_0$ & $Q_u$, $Q_\varv$ & $Q_4$, $Q_{4u}$, $Q_{4\varv}$ &&
& $G_{xyz}$ &&
$T_0$ & $T_u$, $T_\varv$ & $T_4$, $T_{4u}$, $T_{4\varv}$ &&
& $M_{xyz}$\\
  
$2/m$ ($C_{\rm 2h}$) & $Q_0$ & $Q_u$, $Q_\varv$, $Q_{zx}$ & $Q_4$, $Q_{4u}$, $Q_{4\varv}$, $Q_{4y}^\alpha$, $Q_{4y}^\beta$ &&
$G_y$ & $G_{xyz}$, $G_y^\alpha$, $G_{y}^\beta$ &&
$T_0$ & $T_u$, $T_\varv$, $T_{zx}$ & $T_4$, $T_{4u}$, $T_{4\varv}$, $T_{4y}^\alpha$, $T_{4y}^\beta$ &&
$M_y$ & $M_{xyz}$, $M_y^\alpha$, $M_{y}^\beta$\\
  
$6/mmm$ ($D_{\rm 6h}$) & $Q_0$ & $Q_u$ & $Q_{40}$ &&
& &&
$T_0$ & $T_u$ & $T_{40}$ &&
 & \\
  
$6/m$ ($C_{\rm 6h}$) & $Q_0$ & $Q_u$ & $Q_{40}$ &&
$G_z$ & $G_z^\alpha$ &&
$T_0$ & $T_u$ & $T_{40}$ &&
$M_z$ & $M_z^\alpha$ \\
 
$\bar{3}m$ ($D_{\rm 3d}$) & $Q_0$ & $Q_u$ & $Q_{40}$, $Q_{4b}$ &&
& $G_{3b}$ &&
$T_0$ & $T_u$ & $T_{40}$, $T_{4b}$ &&
& $M_{3b}$ \\
 
$\bar{3}$ ($C_{\rm 3i}$) & $Q_0$ & $Q_u$ & $Q_{40}$, $Q_{4a}$, $Q_{4b}$ &&
$G_z$ & $G_{3a}$, $G_{3b}$, $G_z^\alpha$ &&
$T_0$ & $T_u$ & $T_{40}$, $T_{4a}$, $T_{4b}$ &&
$M_z$ & $M_{3a}$, $M_{3b}$, $M_z^\alpha$ \\

 \hline\hline
 \end{tabular}
\end{table*}

\begin{table*}[h!]
\centering
\caption{
Active E, ET, MT, and M multipoles in the noncentrosymmetric crystallographic point groups.
The triclinic point group $1$ ($C_1$), in which all the multipoles are active, is omitted.
 \label{table:active_MP_CPG_2}}
\vspace{2mm}
\begin{tabular}{cccccccccccccccccccccccccccc}\hline \hline
& \multicolumn{6}{c}{even-parity} && \multicolumn{6}{c}{odd-parity} && \multicolumn{6}{c}{even-parity} && \multicolumn{6}{c}{odd-parity} 
\\
\cline{2-7}
\cline{9-14}
\cline{16-21}
\cline{23-28}

& \multicolumn{3}{c}{E} &&
   \multicolumn{2}{c}{ET} &&
   \multicolumn{2}{c}{E} &&
   \multicolumn{3}{c}{ET} &&
   \multicolumn{3}{c}{MT} &&
   \multicolumn{2}{c}{M} &&
   \multicolumn{2}{c}{MT} &&
   \multicolumn{3}{c}{M} \\
   \cline{2-4} \cline{6-7}
   \cline{9-10} \cline{12-14}
   \cline{16-18} \cline{20-21}
   \cline{23-24} \cline{26-28}
 
 & $l=0$ & $2$ & $4$ &&
  $1$ & $3$ &&
  $1$ & $3$ &&
  $0$ & $2$ & $4$ &&
  $0$ & $2$ & $4$ &&
  $1$ & $3$ &&
  $1$ & $3$ &&
  $0$ & $2$ & $4$ \\\hline 
 
$432$ ($O$) & $Q_0$ & & $Q_4$ &&
& &&
& &&
$G_0$ & & $G_4$ &&
$T_0$ & & $T_4$ &&
& &&
& &&
$M_0$ & & $M_4$ \\
\hline
 
$\bar{4}3m$ ($T_{\rm d}$) & $Q_0$ & & $Q_4$ &&
& &&
& $Q_{xyz}$ &&
& & && 
$T_0$ & & $T_4$ &&
& &&
&
$T_{xyz}$ & & & \rule[0pt]{0pt}{8pt} \\
\hline

$23$ ($T$) & $Q_0$ & & $Q_4$ &&
& $G_{xyz}$ &&
& $Q_{xyz}$ &&
$G_0$ & & $G_4$ &&
$T_0$ & & $T_4$ &&
& $M_{xyz}$ &&
& $T_{xyz}$ &&
$M_0$ & & $M_4$\\
\hline

$422$ ($D_{\rm 4}$) & $Q_0$ & $Q_u$ & $Q_4$ &&
& &&
& &&
$G_0$ & $G_u$ & $G_4$ &&
$T_0$ & $T_u$ & $T_4$ &&
& &&
& &&
$M_0$ & $M_u$ & $M_4$ \\

& & &  $Q_{4u}$ &&
& &&
& &&
& & $G_{4u}$ &&
& & $T_{4u}$ &&
& &&
& &&
& & $M_{4u}$\\
\hline\hline
\end{tabular}
\end{table*}

\begin{table*}[h!]
\centering
\caption{
Active even-parity E and ET multipoles in the centrosymmetric gray point groups.
The triclinic point group $\bar{1}1'$, in which all the even-parity E and ET multipoles are active, is omitted.
 \label{table:active_MP_GPG_1}}
\vspace{2mm}
\begin{tabular}{ccccccc}\hline \hline
 & \multicolumn{6}{c}{even-parity} \\ \cline{2-7}
 & \multicolumn{3}{c}{E} & & \multicolumn{2}{c}{ET} \\ \cline{2-4} \cline{6-7}
 
 & $l=0$ & $2$ & $4$ & & $1$ & $3$   \\\hline
 $m\bar{3}m1'$ & $Q_0$ & & $Q_4$ \rule[0pt]{0pt}{8pt}\\
 $m\bar{3}1'$ & $Q_0$ & & $Q_4$ & & & $G_{xyz}$ \\
 $4/mmm1'$ & $Q_0$ & $Q_u$ & $Q_4, Q_{4u}$ \\
 $4/m1'$ & $Q_0$ & $Q_u$ & $Q_4, Q_{4u}, Q_{4z}^\alpha$ & & $G_z$ & $G_z^\alpha$ \\
 $mmm1'$ & $Q_0$ & $Q_u$, $Q_\varv$ & $Q_4$, $Q_{4u}$, $Q_{4\varv}$ & & & $G_{xyz}$\\
 $2/m1'$ & $Q_0$ & $Q_u$, $Q_\varv$, $Q_{zx}$ & $Q_4$, $Q_{4u}$, $Q_{4\varv}$, $Q_{4y}^\alpha$, $Q_{4y}^\beta$ & & $G_y$ & $G_{xyz}$, $G_y^\alpha$, $G_{y}^\beta$\\
 $6/mmm1'$  & $Q_0$ & $Q_u$ & $Q_{40}$ & & \\
 $6/m1'$ & $Q_0$ & $Q_u$ & $Q_{40}$ & & $G_z$ & $G_z^\alpha$ \\
 $\bar{3}m1'$ & $Q_0$ & $Q_u$ & $Q_{40}$, $Q_{4b}$ & & & $G_{3b}$ \\
 $\bar{3}1'$ & $Q_0$ & $Q_u$ & $Q_{40}$, $Q_{4a}$, $Q_{4b}$ & & $G_z$ & $G_{3a}$, $G_{3b}$, $G_z^\alpha$ \\
 \hline\hline
 \end{tabular}
\end{table*}

\begin{table*}[h!]
\centering
\caption{
Active E and ET multipoles in the noncentrosymmetric gray point groups.
The triclinic point group $11'$, in which all the E and ET multipoles are active, is omitted.
 \label{table:active_MP_GPG_2}}
\vspace{2mm}
\begin{tabular}{cccccccccccccc}\hline \hline
& \multicolumn{6}{c}{even-parity} && \multicolumn{6}{c}{odd-parity} \\
\cline{2-7} \cline{9-14}

& \multicolumn{3}{c}{E} & & \multicolumn{2}{c}{ET} & & \multicolumn{2}{c}{E} & & \multicolumn{3}{c}{ET} \\
\cline{2-4} \cline{6-7} \cline{9-10} \cline{12-14}

& $l=0$ & $2$ & $4$ && $1$ & $3$ && $1$ & $3$ && $0$ & $2$ & $4$ \\\hline

$4321'$ & $Q_0$ & & $Q_4$ &&
& &&
& &&
$G_0$ & & $G_4$ \\
   
$\bar{4}3m1'$ & $Q_0$ & & $Q_4$ &&
& &&
& $Q_{xyz}$ &&
& &  \\

$231'$ & $Q_0$ & & $Q_4$ &&
& $G_{xyz}$ &&
& $Q_{xyz}$ &&
$G_0$ & & $G_4$ \\

$4221'$ & $Q_0$ & $Q_u$ & $Q_4, Q_{4u}$ &&
& &&
& &&
$G_0$ & $G_u$ & $G_4$, $G_{4u}$\\

$\bar{4}2m1'$ & $Q_0$ & $Q_u$ & $Q_4, Q_{4u}$ &&
& &&
& $Q_{xyz}$ &&
& $G_\varv$ & $G_{4\varv}$\\
\hline\hline
 \end{tabular}
\end{table*}

\begin{table*}[h!]
\centering
\caption{
Active even-parity E, ET, MT, and M multipoles in the centrosymmetric black and white point groups.
 \label{table:active_MP_BWPG_1}}
\vspace{2mm}
\begin{tabular}{cccccccccccccc}\hline \hline
& \multicolumn{6}{c}{even-parity} & & \multicolumn{6}{c}{even-parity} \\ \cline{2-7} \cline{9-14}
 &  \multicolumn{3}{c}{E} && \multicolumn{2}{c}{ET} &&  \multicolumn{3}{c}{MT} &&  \multicolumn{2}{c}{M} \\
 \cline{2-4}
 \cline{6-7}
 \cline{9-11}
 \cline{13-14}
 
 & $l=0$ & $2$ & $4$ && 
 $1$ & $3$ && 
 $0$ & $2$ & $4$ && 
 $1$ & $3$  \\\hline
 
$m\bar{3}m'$ & $Q_0$ & & $Q_4$ &&
& &&
& & && 
& $M_{xyz}$ \rule[0pt]{0pt}{8pt}\\

$4/mm'm'$ & $Q_0$ & $Q_u$ & $Q_4$, $Q_{4u}$ && 
& &&
& & $T_{4z}^\alpha$ && 
$M_z$ & $M_z^\alpha$ \\

$4'/mmm'$ & $Q_0$ & $Q_u$ & $Q_4$, $Q_{4u}$ &&
& &&
& $T_\varv$ & $T_{4\varv}$ && 
& $M_{xyz}$ \\

$4'/m$ & $Q_0$ & $Q_u$ & $Q_4$, $Q_{4u}$, $Q_{4z}^\alpha$ && 
$G_z$ & $G_z^\alpha$ &&
& $T_\varv$, $T_{xy}$ & $T_{4\varv}$, $T_{4z}^\beta$ && 
& $M_{xyz}$, $M_z^\beta$ \\

$m'm'm$ & $Q_0$ & $Q_u$. $Q_\varv$ & $Q_4$, $Q_{4u}$, $Q_{4\varv}$ && 
& $G_{xyz}$ && 
& $T_{xy}$ & $T_{4z}^\alpha$, $T_{4z}^\beta$ && 
$M_z$ & $M_z^\alpha$, $M_z^\beta$ \\

$2'/m'$ & $Q_0$ & $Q_u$, $Q_{\varv}$, $Q_{zx}$ & $Q_4$, $Q_{4u}$, $Q_{4\varv}$, $Q_{4y}^\alpha$, $Q_{4y}^\beta$ && 
$G_y$ & $G_{xyz}$, $G_y^\alpha$, $G_y^\beta$ && 
& $T_{xy}$, $T_{yz}$ & $T_{4z}^\alpha$, $T_{4x}^\alpha$, $T_{4z}^\beta$, $T_{4x}^\beta$ && 
$M_z$, $M_x$ & $M_z^\alpha$, $M_x^\alpha$, $M_z^\beta$, $M_x^\beta$ \\

$6/mm'm'$ & $Q_0$ & $Q_u$ & $Q_{40}$ && 
& && 
& & && 
$M_z$ & $M_z^\alpha$ \\

$6'/m'mm'$ & $Q_0$ & $Q_u$ & $Q_{40}$ && 
& &&
& & $T_{4a}$ && 
& $M_{3a}$ \\

$6'/m'$ & $Q_0$ & $Q_u$ & $Q_{40}$ && 
$G_z$ & $G_z^\alpha$ && 
& & $T_{4a}$, $T_{4b}$ && 
& $M_{3a}$, $M_{3b}$ \\

$\bar{3}m'$ & $Q_0$ & $Q_u$ & $Q_{40}$, $Q_{4b}$ && 
& $G_{3b}$ &&
& & $T_{4a}$ && 
$M_z$ & $M_{3a}$, $M_z^\alpha$ \\
\hline\hline
\end{tabular}
\end{table*}

\begin{table*}[h!]
\centering
\caption{
Active even-parity E and ET multipoles and odd-parity MT and M multipoles in the noncentrosymmetric black and white point groups with the $\mathcal{PT}$ symmetry.
The triclinic point group $\bar{1}'$ is omitted, where all the even-parity E and ET multipoles and odd-parity MT and M multipoles are active.
\label{table:active_MP_BWPG_2}}
\vspace{2mm}
\begin{tabular}{cccccccccccccc}\hline \hline
& \multicolumn{6}{c}{even-parity} && \multicolumn{6}{c}{odd-parity} \\
\cline{2-7} 
\cline{9-14}

& \multicolumn{3}{c}{E} &&  \multicolumn{2}{c}{ET} && \multicolumn{2}{c}{MT} && \multicolumn{3}{c}{M} \\
\cline{2-4} 
\cline{6-7}
\cline{9-10}
\cline{12-14}

 & $l=0$ & $2$ & $4$ && $1$ & $3$ && $1$ & $3$ && $0$ & $2$ & $4$  \\\hline
 
$m'\bar{3}'m'$ & $Q_0$ & & $Q_4$ && 
& &&
& && 
$M_0$ & & $M_4$ \rule[0pt]{0pt}{8pt} \\
  
$m'\bar{3}'m$ & $Q_0$ & & $Q_4$ && 
& &&
& $T_{xyz}$ && 
& & \\
  
$m'\bar{3}'$ & $Q_0$ & & $Q_4$ && 
& $G_{zyz}$ &&
& $T_{xyz}$ && 
$M_0$ & & $M_4$ \\

$4/m'm'm'$ & $Q_0$ & $Q_u$ & $Q_4$, $Q_{4u}$ && 
& &&
& && 
$M_0$ & $M_u$ & $M_{4}$, $M_{4u}$ \\

$4'/m'm'm$ & $Q_0$ & $Q_u$ & $Q_4$, $Q_{4u}$ && 
& &&
& $T_{xyz}$ && 
& $M_{\varv}$ & $M_{4\varv}$  \\
\hline\hline
\end{tabular}
\end{table*}

\begin{table*}[h!]
\centering
\caption{Active multipoles in the noncentrosymmetric black and white point groups without the $\mathcal{PT}$ symmetry.
 \label{table:active_MP_BWPG_3}}
\vspace{2mm}
\begin{tabular}{cccccccccccccccccccccccccccc}\hline \hline
& \multicolumn{6}{c}{even-parity} && \multicolumn{6}{c}{odd-parity} && \multicolumn{6}{c}{even-parity} && \multicolumn{6}{c}{odd-parity} \\
\cline{2-7}
\cline{9-14}
\cline{16-21}
\cline{23-28}

 & \multicolumn{3}{c}{E} && \multicolumn{2}{c}{ET} && \multicolumn{2}{c}{E} && \multicolumn{3}{c}{ET} 
 && \multicolumn{3}{c}{MT} && \multicolumn{2}{c}{M} && \multicolumn{2}{c}{MT} && \multicolumn{3}{c}{M}\\
 \cline{2-4} \cline{6-7}
 \cline{9-10} \cline{12-14}
 \cline{16-18} \cline{20-21}
 \cline{23-24} \cline{26-28}
 
 & $l=0$ & $2$ & $4$ && 
 $1$ & $3$ && 
 $1$ & $3$ && 
 $0$ & $2$ & $4$ && 
 $0$ & $2$ & $4$ && 
 $1$ & $3$ && 
 $1$ & $3$ && 
 $0$ & $2$ & $4$ \\\hline
 
$\bar{4}'3m'$ & $Q_0$ & & $Q_4$ && 
& &&
& $Q_{xyz}$ && 
& & &&
& & && 
& $M_{xyz}$ &&
& && 
$M_0$ & & $M_4$ \rule[0pt]{0pt}{8pt}\\
\hline
  
$4'32'$ & $Q_0$ & & $Q_4$ && 
& &&
& && 
$G_0$ & & $G_4$ && 
& & && 
& $M_{xyz}$ &&
& $T_{xyz}$ && 
& & \\
\hline

$42'2'$ & $Q_0$ & $Q_u$  & $Q_4$ &&
& &&
& && 
$G_0$ & $G_u$ & $G_{4}$ && 
& & $T_{4z}^\alpha$ && 
$M_z$ & $M_z^\alpha$ &&
$T_z$ & $T_z^\alpha$ && 
& & $M_{4z}^\alpha$ \\

& & & $Q_{4u}$ && 
& && 
& && 
& & $G_{4u}$ \\
\hline
  
$4'22'$ & $Q_0$ & $Q_u$ & $Q_4$ && 
& &&
& && 
$G_0$ & $G_u$ & $G_{4}$ &&
&  $T_{\varv}$ & $T_{4\varv}$ && 
& $M_{xyz}$ &&
& $T_{xyz}$ && 
& $M_{\varv}$ & $M_{4\varv}$ \\
\hline\hline
\end{tabular}
\end{table*}

In this section, we focus on the active multipoles, which belong to the totally symmetric 
IRREP in each magnetic point group. 
The active multipoles are roughly classified according to the spatial and time 
inversion properties in each magnetic point group, as shown in Table~\ref{table:type_of_active_MP}. 
The E and ET multipoles can be active for all the magnetic point groups, while the M and MT multipoles can become active only for the type-(I) crystallographic point groups and type-(III) black and white point groups without the pure time-reversal symmetry. 

Let us start from discussing the active multipoles under type-(I) crystallographic point groups, where not only E and ET multipoles but also M and MT multipoles become active because of the  time-reversal symmetry breaking.
The type of active multipoles depends on the spatial parity, as shown in Table~\ref{table:type_of_active_MP}. 
In the 11 crystallographic point groups with the spatial inversion symmetry $m\bar{3}m$, $m\bar{3}$, $4/mmm$,  $4/m$, $mmm$, $2/m$, $\bar{1}$, $6/mmm$, $6/m$, $\bar{3}m$, and $\bar{3}$, even-parity E, ET, M, and MT multipoles are active.  
The active multipoles in each crystallographic point group is shown in Table~\ref{table:active_MP_CPG_1}.
On the other hand,  in the 21 noncentrosymmetric crystallographic point groups $432$, $\bar{4}3m$, $23$,  $422$, $\bar{4}2m$, $4mm$, $4$, $\bar{4}$, $222$, $mm2$, $2$, $m$, $1$, $622$, $\bar{6}m2$, $6mm$, $6$, $\bar{6}$, $32$, $3m$, and $3$, odd-parity E, ET, M, and MT multipoles become active in addition to the even-parity ones, which are shown in Table~\ref{table:active_MP_CPG_2}. 
It is noted that the active E and MT (ET and M) multipoles have the same angle dependence; the same components of $Q_{lm}$ and $T_{lm}$ ($G_{lm}$ and $M_{lm}$) are activated simultaneously.
This is because the type-(I) crystallographic point group has no point group operations accompanied by the time-reversal operation. 

In the case of the type-(II) gray point groups, no M and MT multipoles are activated because of the presence of the time-reversal symmetry~\cite{hayami2018classification}. 
The even-parity E and ET multipoles become active in all the type-(II) gray point groups, while the odd-parity E and ET ones become active in the noncentrosymmetric 21 point groups; $4321'$, $\bar{4}3m1'$, $231'$, $4221'$, $\bar{4}2m1'$, $4mm1'$, $41'$, $\bar{4}1'$, $2221'$, $mm21'$, $21'$, $m1'$, $11'$, $6221'$, $\bar{6}m21'$, $6mm1'$, $61'$, $\bar{6}1'$, $321'$, $3m1'$, and $31'$. 
The active even-parity E and ET multipoles with rank $0$--$4$ in the centrosymmetric 11 gray point groups are summarized in Table~\ref{table:active_MP_GPG_1}.
Meanwhile, the E and ET multipoles in the noncentrosymmetric 21 gray point groups are shown in Table~\ref{table:active_MP_GPG_2}.

In the type-(III) black and white point groups, 
four types of multipoles can become active, similar to the type-(I) crystallographic point groups.
However, there is a major difference in the symmetry operations; there are proper product operations of the point-group and time-reversal operations in the type-(III) black and white point groups, resulting in no relations between the active E and MT (ET and M) multipoles. 
Thus, the type-(III) black and white point groups are classified into three types according to the presence/absence of the $\mathcal{P}$ and $\mathcal{PT}$ symmetries.
The first one is the 10 black and white point groups with $(\mathcal{P}, \mathcal{PT})=(\bigcirc, \times)$, $m\bar{3}m'$, $4/mm'm'$, $4'/mm'm$, $4'/m$, $m'm'm$, $2'/m'$, $6/mm'm'$, $6'/m'mm'$, $6'/m'$, and $\bar{3}m'$, where the even-parity E, ET, M, and MT multipoles are active, as listed in Table~\ref{table:active_MP_BWPG_1}. 
The second one is the 21 black and white point groups with $(\mathcal{P}, \mathcal{PT})=(\times, \bigcirc)$, $m'\bar{3}'m'$, $m'\bar{3}'m$, $m'\bar{3}'$, $4/m'm'm'$, $4'/m'm'm$, $4/m'mm$, $4'/m'$, $4/m'$, 
$m'm'm'$, $m'mm$, $2'/m$, $2/m'$, $\bar{1}'$, $6/m'm'm'$, $6'/mmm'$, $6/m'mm$, $6'/m$, $6/m'$, $\bar{3}'m'$, $\bar{3}'m$, and $\bar{3}'$, where the even-parity E and ET multipoles and the odd-parity M and MT multipoles become active, as shown in Table~\ref{table:active_MP_BWPG_2}. 
The last one is the 27 black and white point groups with $(\mathcal{P}, \mathcal{PT})=(\times, \times)$, $4'32'$, $\bar{4}'3m'$, $42'2'$, $4'22'$, $\bar{4}2'm'$, $\bar{4}'2m'$, $\bar{4}'2'm$, $4m'm'$, $4'm'm$, $4'$, $\bar{4}'$, $2'2'2$, $m'm'2$, $m'm2'$, $2'$, $m'$, $62'2'$, $6'22'$, $\bar{6}m'2'$, $\bar{6}'m'2$, $\bar{6}'m2'$,  $6m'm'$, $6'mm'$, $6'$, $\bar{6}'$, $32'$, and $3m'$, where all types of the multipoles become active irrespective of the spatial-inversion and time-reversal parities, as summarized in Table~\ref{table:active_MP_BWPG_3}.

Let us remark on the active multipoles from the standpoint of the (magnetic) Laue group, which is often used for the diffraction measurement.
The even-parity E and ET multipoles are well classified by 11 Laue groups, $m\bar{3}m$, $m\bar{3}$, $4/mmm$, $4/m$, $mmm$, $6/mmm$, $6/m$, $\bar{3}m$, $\bar{3}$, $2/m$, and $\bar{1}$, whose correspondence to the magnetic point groups is summarized in Table~\ref{table:Laue} in Appendix~\ref{sec:ap_Laue}.
Meanwhile, the even-parity M and MT multipoles are well classified by 32 magnetic Laue groups, $m\bar{3}m1'$, $m\bar{3}1'$, $4/mmm1'$, $4/m1'$, $mmm1'$, $6/mmm1'$, $6/m1'$, $\bar{3}m1'$, $\bar{3}1'$, $2/m1'$, $\bar{1}1'$, 
$m\bar{3}m$, $m\bar{3}$, $4/mmm$, $4/m$, $mmm$, $6/mmm$, $6/m$, $\bar{3}m$, $\bar{3}$, $2/m$, $\bar{1}$,
$m\bar{3}m'$, $4/mm'm'$, $4'/mm'm$, $4'/m$, $m'm'm$, $6/mm'm'$, $6'/m'mm'$, $6'/m'$, $\bar{3}m'$, and $2'/m'$. 
The correspondence between the magnetic Laue groups and magnetic point groups is shown in Tables~\ref{table:MLG_GPG} and \ref{table:MLG_CPG} in Appendix~\ref{sec:ap_Laue}.

Tables~\ref{table:active_MP_CPG_1}--\ref{table:active_MP_BWPG_3} are useful to identify what type of ferroic state~\cite{wadhawan2000introduction} realizes in a magnetic material, e.g., ferroelectric, ferromagnetic~\cite{aizu1966possible, aizu1969possible, aizu1970possible}, ferrotoroidal~\cite{litvin2008ferroic}, and ferroelectric toroidal (ferroaxial) states~\cite{hlinka2016symmetry},
since active E dipole $Q_i$ ($i=x,y,z$), M dipole $M_i$, MT dipole $T_i$, and ET dipole $G_i$ correspond to the ferroelectric, ferromagnetic, ferrotoroidal, and ferroaxial order parameters, respectively~\cite{cheong2018broken}. 
One can find all the candidate magnetic point groups with these active dipole moments from Tables~\ref{table:active_MP_CPG_1}--\ref{table:active_MP_BWPG_3} as follows:
\begin{itemize}
\item E dipole: \\
$4mm1'$, $41'$, $mm21'$, $6mm1'$, $61'$, $3m1'$, $31'$, $21'$, $m1'$, $11'$, 
$4mm$, $4$, $mm2$, $6mm$, $6$, $3m$, $3$, $2$, $m$, $1$, 
$4m'm'$, $4'm'm$, $4'$, $m'm'2$, $m'm2'$, $6m'm'$, $6'mm'$, $6'$, $3m'$, $2'$, $m'$, 
\item M dipole: \\
$4/m$, $\bar{4}$, $4$, $6/m$, $\bar{6}$, $6$, $\bar{3}$, $3$, $\bar{1}$, $1$, $2/m$, $2$, $m$, 
$4/mm'm'$, $42'2'$, $4m'm'$, $\bar{4}2'm'$, 
$m'm'm$, $2'2'2$, $m'm'2$, $m'm2'$, $6/mm'm'$, $62'2'$, $6m'm'$, $\bar{6}m'2'$, $\bar{3}m'$, $3m'$, $32'$, $2'/m'$, $2'$, $m'$, 
\item MT dipole: \\
$4mm$, $4$, $mm2$, $6mm$, $6$, $3m$, $3$, $2$, $m$, $1$, 
$4/m'mm$, $4/m'$, $mmm'$, $6/m'mm$, $6/m'$, $\bar{3}'m$, $\bar{3}'$, $2'/m$, $2/m'$, $\bar{1}'$, 
$42'2'$, $\bar{4}'2'm$, $\bar{4}'$, $2'2'2$, $m'm2'$, $62'2'$, $\bar{6}'m2'$, $\bar{6}'$, $32'$, $m'$, $2'$, 
\item ET dipole: \\
$4/m1'$, $41'$, $\bar{4}1'$, $6/m1'$, $61'$, $\bar{6}1'$, $\bar{3}1'$, $31'$, $2/m1'$, $21'$, $m1'$, $\bar{1}1'$, $11'$, 
$4/m$, $4$, $\bar{4}$, $6/m$, $6$, $\bar{6}$, $\bar{3}$, $3$, $2/m$, $2$, $m$, $\bar{1}$, $1$, 
$4'/m'$, $4/m'$, $4'/m$, $4'$, $\bar{4}'$, $6'/m'$, $6/m'$, $6'/m$, $6'$, $\bar{6}'$, $\bar{3}'$, $2'/m'$, $2/m'$, $2'/m$, $2'$, $m'$, $\bar{1}'$.
\end{itemize}
Thus, our results contain the previous classification for the ferroelectric, ferromagnetic, ferrotoroidal, and ferroaxial states based on the symmetry analyses~\cite{gallego2019automatic, litvin2008ferroic, doi:10.1080/00150199408245120, schmid1999possibility, schmid2008some}.
Furthermore, Tables~\ref{table:active_MP_CPG_1}--\ref{table:active_MP_BWPG_3} provide unconventional order parameters more than the dipole moments,
e.g., E/M/MT/ET quadrupole and octupole moments. 
This classification gives a complete guide to identify the electronic order parameters in a systematic way.

Tables~\ref{table:active_MP_CPG_1}--\ref{table:active_MP_BWPG_3} also enable us to understand the cross-correlated phenomena through the couplings between the different multipoles.
Let us take an example of the cubic crystal with $m\bar{3}m1'$ symmetry, where the E monopole $Q_0$ and E hexadecapole $Q_4$ are active up to rank 4 as shown in Table~\ref{table:active_MP_GPG_1}.
Once the spontaneous symmetry breaking to $m'\bar{3}'m'$ occurs, magnetic monopole $M_0$ and magnetic hexadecapole $M_4$ are additionally activated as shown in Table~\ref{table:active_MP_BWPG_2}.

The identification of active multipoles under the target magnetic point groups enables us to construct the free energy and the Hamiltonian in terms of the multipole degrees of freedom.
For example, the Landau free energy expansion can be performed in terms of multipole degrees of freedom for any types of order parameters, which is useful to analyze mutual coupling between them~\cite{kusunose2008description, kusunose2020complete}.
For the above example of the symmetry breaking from $m\bar{3}m1'$ to $m'\bar{3}'m'$, it is easily to notice the additional multipole couplings between $(Q_0, Q_4)$ and $(M_0, M_4)$, which contribute to the free energy, from the IRREP.

Moreover, one can find the additional active multipoles induced by external fields, such as electric and magnetic fields.
For example, when the symmetry is lowered from $m'\bar{3}'m'$ to $4m'm'$ under the magnetic field along the $z$ axis, $H_z$,
the additional active multipoles up to rank 4 are $Q_z, Q_u, Q_z^\alpha, Q_{4u}, G_{4z}^\alpha,
M_z, M_u, M_z^\alpha, M_{4u}$, and  $T_{4z}^\alpha$, as shown in Table~\ref{table:active_MP_BWPG_3}.
Then, the additional multipole couplings by the magnetic field arise, which become source of the field-induced cross-correlated phenomena.
For example, since $Q_z$ and $Q_u$ correspond to the electric polarization $P_z$ and the ($3z^2-r^2$)-type symmetric strain $\varepsilon_u$, respectively, 
one can expect that the magnetoelectric coupling $H_zP_z$ and magnetoelastic coupling $H_z\varepsilon_u$ appear in the free energy expansion. 

In addition, our result can be used when constructing the so-called hyperfine coupling to investigate the field dependence of NQR/NMR spectra~\cite{yatsushiro2020nqr}. 
Besides, such multipole couplings in each magnetic point group are also related with the band deformations~\cite{hayami2019momentum, hayami2020spontaneous, hayami2020bottom, hayami2021spin} and field responses~\cite{hayami2018classification, watanabe2018group, oiwa2021systematic}.  
The latter will be discussed in the next section.

\clearpage
\section{Linear and nonlinear responses under active multipole \label{sec:res}} 

According to Neumann's principle, macroscopic physical responses are determined not by the space-group symmetry but by the crystallographic point-group symmetry~\cite{neumann1885vorlesungen, curie1894symetrie}. 
This can be generalized to magnetic point groups: macroscopic responses in magnets, such as the Hall conductivity, the linear magnetoelectric effect, and nonlinear conductivity, are determined by the magnetic point-group symmetry~\cite{birss1964symmetry, kleiner1966space, kleiner1967space, kleiner1969space, grimmer1993general, seemann2015symmetry, mook2020origin, gallego2019automatic}. 
In this section, we show physical responses under the active multipole in any magnetic point groups. 
The correspondence between the response tensor components and multipoles will provide useful information to systematically discuss the essential microscopic model parameters for the responses in the magnetic materials~\cite{oiwa2021systematic}.
First, we discuss the relation between the response tensors and multipoles based on the point group symmetry in Sec.~\ref{sec:res_symmetry}.
Then, we discuss the role of the anti-unitary time-reversal operation on the response  function by using the linear and second-order nonlinear response theory in Secs.~\ref{sec:res_linear} and \ref{sec:res_nonlinear}, respectively.

\subsection{Correspondence between tensor component and multipole \label{sec:res_symmetry}}
First, we discuss the relation between the response tensor components and multipoles on the basis of group theory.
The response tensor $\chi^{[n_{B} \times n_{F}]}$ is defined as
\begin{align}
B^{[n_B]}=\chi^{[n_{B} \times n_{F}]} F^{[n_F]}, 
\end{align}
where $B^{[n_B]}$ and $F^{[n_F]}$ are the rank-$n_{B}$ (output) response and the rank-$n_{F}$ external (input) field, respectively, which are typically represented by the electric, magnetic, elastic, and their product degrees of freedom. 
For example, $F^{[n_F]}$ is an electric field $\bm{E}$, a magnetic field $\bm{H}$, a (symmetric) stress ${\bm \tau}$ and their combination, while $B^{[n_B]}$ is the electric polarization ${\bm P}$, magnetization ${\bm M}$, symmetric strain $\varepsilon_{ij}=(\partial_i u_j + \partial_j u_i)/2$, and rotation ${\bm \omega}=({\bm \nabla}\times {\bm u})/2$ where $\bm{u}$ is the displacement vector~\footnote{It is noted that ${\bm \omega}$ in the long-wavelength limit does not contribute to the free energy, since it represents a uniform rotation.}. 
$B^{[n_B]}$ also represents quantities for the transport phenomena, such as the electric (thermal) current ${\bm J}$ (${\bm J}^{\rm Q}$) and the spin current $J^{\rm s}_{ij}= \sigma_i J_j $. 
Each external field and response have the correspondence to the multipoles, e.g., electric field $\bm{E}$ $\leftrightarrow$ E dipole and symmetric strain ${\bm \varepsilon} \leftrightarrow$ E monopole and E quadrupole.
The representative relation is 
summarized in Table~\ref{table:field_response_MP}, where the correspondence between the external field (response) and multipole is shown in the upper (lower) panel.

\begin{table}[htb!]
\centering
\caption{Correspondence of the external fields and the responses to the multipoles. 
The spatial-inversion parity of the external field or the response is shown in the column of $\mathcal{P}$.
In the column of multipole, $X_{lm}$ ($l=0,1,2$) means the rank-$l$ multipole ($X=Q,G,M,T$).
 \label{table:field_response_MP}}
\begin{tabular}{ccccc}
\hline \hline
$n_F$ & $\mathcal{P}$ & external field &  multipole \\ \hline
1 & $+$ & magnetic field ${\bm H}$ &  M dipole ($M_{1m}$)\\
& $-$ & electric field ${\bm E}$ & E dipole ($Q_{1m}$)\\
  2
& $+$ &(symmetric) stress ${\bm \tau}$ & E monopole ($Q_{0}$)\\
&&& E quadrupole ($Q_{2m}$)\\
\hline
$n_B$  & & response &  \\ \hline
 1 & $+$ & magnetization ${\bm M}$ & M dipole ($M_{1m}$)\\
 & & rotation ${\bm \omega}$& ET dipole ($G_{1m}$)\\
& $-$ & electric polarization ${\bm P}$ & E dipole ($Q_{1m}$)\\
 & & electric (thermal) current ${\bm J}$(${\bm J}^{\rm Q}$) & MT dipole ($T_{1m}$)\\
 
2 & $+$ & symmetric strain ${\bm \varepsilon}$ & E monopole ($Q_{0}$)\\
& & & E quadrupole ($Q_{2m}$)\\
& $-$ & spin current ${\bm J}^{\rm s}$ & ET monopole ($G_0$)\\
& & & E dipole ($Q_{1m}$)\\
& & & ET quadrupole ($G_{2m}$)\\
\hline\hline
\end{tabular}
\end{table}

According to the spatial parities of $B^{[n_B]}$ and $F^{[n_F]}
$, $\chi^{[n_{B} \times n_{F}]}$ becomes a polar or axial tensor; 
$\chi^{[n_{B} \times n_{F}]}$ is the polar (axial) tensor for the parity $\mathcal{P}=(-1)^{n_{B} + n_{F}}$ [$\mathcal{P}=(-1)^{n_{B} + n_{F}+1}$].
In the following, we show the correspondence between multipoles and rank-1, 2, 3, and 4 tensors are shown in Secs.~\ref{sec:res_symmetry_rank1}--\ref{sec:res_symmetry_rank4}, respectively. 
See also Appendix~\ref{sec:ap_res_mp} for details of the derivation.
In the following, we mainly focus on the response tensors in cubic, tetragonal, orthorhombic, monoclinic, and triclinic systems, and show those in
hexagonal and trigonal systems in Appendix~\ref{sec:ap_hexagonal}.

\subsubsection{Rank-1 tensor \label{sec:res_symmetry_rank1}}
The rank-1 response tensor $\chi^{[0\times 1]}$ for the scalar response $B^{[0]}=(B)$ with $n_{B}=0$ and vector field $F^{[1]}=(F_x, F_y, F_z)$ with $n_{F}=1$ is related with the dipole $(X_x,X_y,X_z)$ as
\begin{align}
\label{eq:chi01}
\chi^{[0\times 1]} =
\begin{pmatrix}
X_x \
X_y \
X_z\
\end{pmatrix}, 
\end{align}
where $X$ stands for the polar multipoles ($Q$ or $T$) [axial multipoles ($G$ or $M$)] when $\chi^{[0\times 1]}$ is the polar (axial) tensor.
The dipoles $X_x$, $X_y$, and $X_z$ in Eq.~\eqref{eq:chi01} is $X_i= \chi^{[0\times 1]}_{0;i}$ ($i=x,y,z$).
The response tensor $\chi^{[1\times 0]}$ is obtained by transposing $\chi^{[0\times 1]}$, which is expressed by the same type of multipole as $\chi^{[0\times 1]}$.

The electrocaloric (magnetocaloric) effect where the entropy variation $\Delta S$ is induced by the electric field (the magnetic field) as $\Delta S = \sum_ip_i E_i$ ($\Delta S= \sum_iq_i H_i$), is described by one of the rank-1 response tensor. 
As $\Delta S$ corresponds to E monopole ($Q_0$), the tensor component of $p_i$ ($q_i$) is described by the E dipole $(Q_x, Q_y, Q_z)$ or MT dipole $(T_x, T_y, T_z)$ [the ET dipole $(G_x, G_y, G_z)$ or M dipole $(M_x, M_y, M_z)$]. 
Here and hereafter in Sec.~\ref{sec:res_symmetry}, we do not distinguish the multipoles with the opposite time-reversal parity for simplicity, which depends on the microscopic process in the presence/absence of the dissipation, as discussed in Secs.~\ref{sec:res_linear} and \ref{sec:res_nonlinear}.

\subsubsection{Rank-2 tensor \label{sec:res_symmetry_rank2}}
We consider two types of rank-2 tensors, $\chi^{[1\times1]}$ and $\chi^{[0\times2]}$.
$\chi^{[1\times1]}$ is the response tensor for $B^{[1]}=(B_x, B_y, B_z)$ and 
$F^{[1]}=(F_x, F_y, F_z)$, which is related with the rank-0 to 2 multipoles as
monopole $X_0$, dipole $(Y_x,Y_y,Y_z)$, and quadrupole $(X_u, X_\varv, X_{yz}, X_{zx}, X_{xy})$.
The tensor component of $\chi^{[1\times1]}$ is given by
\begin{align}
\label{eq:rank11}
\chi^{[1\times1]} &= 
\begin{pmatrix}
X_0-X_u+X_\varv & X_{xy}+Y_z & X_{zx} - Y_y \\
X_{xy} - Y_z & X_0 - X_u-X_\varv & X_{yz}+Y_x \\
X_{zx} +Y_y & X_{yz}-Y_x & X_0+2X_u \\
\end{pmatrix},
\end{align}
where $X=Q$ or $T$ ($G$ or $M$) and $Y=G$ or $M$ ($Q$ or $T$) for the polar (axial) tensor.
See Appendix~\ref{sec:ap_res_mp_11} for details.
When $\chi^{[1\times1]}$ is a polar tensor, such as the magnetic susceptibility tensor for $F^{[1]}={\bm H}$ and  $B^{[1]}={\bm M}$, the dielectric susceptibility tensor for $F^{[1]}={\bm E}$ and  $B^{[1]}={\bm P}$, and the electric conductivity tensor for $F^{[1]}={\bm E}$ and  $B^{[1]}={\bm J}$, the corresponding multipoles are the E (MT) monopole and E (MT) quadrupoles for $X$ and ET (M) dipoles for $Y$. 
Meanwhile, when $\chi^{[1\times1]}$ is an axial tensor, such as the magnetoelectric tensor for $F^{[1]}={\bm E}$ and $B^{[1]}={\bm M}$ or $F^{[1]}={\bm H}$ and $B^{[1]}={\bm P}$, the ET (M) monopole and ET (M) quadrupoles for $X$ and E (MT) dipoles for $Y$ are relevant.

$\chi^{[0\times2]}$ is another rank-2 tensor for $B^{[0]}=(B)$ and $F^{[2]}=(F_{xx}, F_{yy}, F_{zz}, F_{yz}, F_{zx}, F_{xy})$ where $F_{ij}=F_{ji}$.  
As $F^{[2]}$ is decomposed into the monopole and quadrupole components, the tensor component of $\chi^{[0\times2]}$ is given by
\begin{align}
\label{eq:rank20}
\chi^{[0\times2]}=
\begin{pmatrix}
X_0-X_{u}+X_{\varv} \\ 
X_0-X_{u}-X_{\varv} \\ 
X_0+2X_{u} \\ 
X_{yz} \\
X_{zx} \\ 
X_{xy}\\
\end{pmatrix}^{\rm T}. 
\end{align}
Thus, the active monopole and quadrupole contribute to $\chi^{[0\times2]}$.
See Appendix~\ref{sec:ap_res_mp_20} for details.
For example, the piezocaloric tensor for $F^{[2]}=\bm{\tau}$ and $B^{[0]}=\Delta S$ corresponds to $\chi^{[0\times2]}$, where the E (MT) monopole and quadrupole are relevant.
The multipole expression of $\chi^{[2\times0]}$ is obtained by transposing $\chi^{[0\times2]}$.

\subsubsection{Rank-3 tensor \label{sec:res_symmetry_rank3}}
We consider two types of rank-3 tensors, $\chi^{[1\times2]}$ and $\chi^{[0\times3]}$.
$\chi^{[1\times2]}$ is the rank-3 tensor for $B^{[1]}=(B_x, B_y, B_z)$ and $F^{[2]}=(F_{xx}, F_{yy}, F_{zz}, F_{yz}, F_{zx}, F_{xy})$, which is expressed by dipole $(X_x, X_y, X_z)$, quadrupole $(Y_{u}, Y_{\varv}, Y_{yz}, Y_{zx}, Y_{xy})$, and octupole $(X_{xyz}, X_x^\alpha, X_y^\alpha, X_z^\alpha, X_x^\beta, X_y^\beta, X_z^\beta)$ as
\begin{widetext}
\begin{align}
\label{eq:rank12}
\chi^{[1\times2]}&=
\begin{pmatrix}
\tilde{X}_x+4 X_{x}^\alpha 
 & \tilde{X}'_y
 -2Y_{zx}-2 X_{y}^\alpha-2X_{y}^\beta
 & \tilde{X}'_z
 +2Y_{xy}-2  X_{z}^\alpha+2X_{z}^\beta\\
\tilde{X}'_x
+2 Y_{yz} -2X_{x}^\alpha+2X_{x}^\beta
 &\tilde{X}_y+4 X_{y}^\alpha 
 &\tilde{X}'_z
 -2Y_{xy}-2  X_{z}^\alpha-2X_{z}^\beta\\
\tilde{X}'_x
-2Y_{yz}-2X_{x}^\alpha-2X_{x}^\beta
 & \tilde{X}'_y
 +2Y_{zx}-2X_{y}^\alpha+2X_{y}^\beta 
 &\tilde{X}_z+4 X_{z}^\alpha \\
 %
Y_{u}
+Y_{\varv}+ X_{xyz}
 & -3 X_{z}
 +Y_{xy}-2 X_{z}^\alpha-2 X_{z}^\beta
 & -3 X_{y}
 -Y_{zx}-2 X_{y}^\alpha+2 X_{y}^\beta \\
  -3 X_{z}
  -Y_{xy}-2 X_{z}^\alpha+2 X_{z}^\beta
 & -Y_{u}
 +Y_{\varv}+X_{xyz}
 & -3 X_{x}
 +Y_{yz}-2 X_{x}^\alpha-2 X_{x}^\beta \\
 -3 X_{y}
 +Y_{zx}-2 X_{y}^\alpha-2 X_{y}^\beta
 & -3 X_{x}
 -Y_{yz}-2 X_{x}^\alpha+2 X_{x}^\beta
  & -2 Y_{\varv} +X_{xyz}\\
\end{pmatrix}^{\rm T}.
\end{align}
\end{widetext}
It is noted that both $\tilde{X}_i$ and $\tilde{X}_i^\prime$ ($i=x,y,z$) stand for the dipole but they are independent with each other.
See Appendix~\ref{sec:ap_res_mp_12} for details. 
$\chi^{[1\times2]}$ is polar for the piezoelectric tensor $(F^{[2]}={\bm \tau}, B^{[1]}={\bm P})$ and second-order nonlinear conductivity $(F^{[2]}_{ij}=E_i  E_j, B^{[1]}={\bm J})$ where $X=Q$ or $T$ and $Y=G$ or $M$, while it is axial for the piezomagnetic tensor $(F^{[2]}={\bm \tau}, B^{[1]}={\bm M})$ where $X=G$ or $M$ and $Y=Q$ or $T$.
The multipole expression of the tensor $\chi^{[2\times1]}$, e.g., the spin conductivity tensor $(F^{[1]}={\bm E}, B^{[2]}={\bm J}^{\rm s})$, are obtained by transposing the piezomagnetic tensor.

$\chi^{[0\times3]}$ is the rank-3 response tensor for $B^{[0]}=(B)$ and $F^{[3]}=(F_{xxx}, F_{yyy}, F_{zzz}, F_{yyz}, F_{zzx}, F_{xxy}, F_{yzz}, F_{zxx}, F_{xyy}, F_{xyz})$ where $F_{ijk}=F_{jik}=F_{ikj}$. 
As $F^{[3]}$ is decomposed into the dipole and octupole components, $\chi^{[0\times3]}$ is also related  to them, which is shown as
\begin{align}
\label{eq:rank30}
\chi^{[0\times3]}=
\begin{pmatrix}
3X_x+2X_x^\alpha \\
3X_y+2X_y^\alpha \\
3X_z+2X_z^\alpha \\ 
X_z-X_z^\alpha -X_z^\beta \\
X_x-X_x^\alpha -X_x^\beta \\ 
X_y-X_y^\alpha -X_y^\beta \\
X_y-X_y^\alpha +X_y^\beta \\ 
X_z-X_z^\alpha +X_z^\beta \\
X_x-X_x^\alpha +X_x^\beta \\ 
X_{xyz}\\
\end{pmatrix}^{\rm T}.
\end{align}
See Appendix~\ref{sec:ap_res_mp_30} for details.
$\chi^{[0\times3]}$, such as the third-order electrocaloric effect, is relevant with $X=Q$ or $T $($G$ or $M$) for the polar (axial) tensor.
The multipole expression of $\chi^{[3\times0]}$ is obtained by transposing $\chi^{[0\times3]}$.

\subsubsection{Rank-4 tensor \label{sec:res_symmetry_rank4}}
We consider two types of rank-4 tensors, $\chi^{[1\times3]}$ and $\chi^{[2\times2]}$.
$\chi^{[1\times3]}$ is the rank-4 response tensor for  
$B^{[1]}=(B_x, B_y, B_z)$ and 
$F^{[3]}=(F_{xxx}, F_{yyy}, F_{zzz}, F_{yyz}, F_{zzx}, F_{xxy}, F_{yzz}, F_{zxx}, F_{xyy}, F_{xyz})$ where $F_{ijk}=F_{jik}=F_{ikj}$.
The relevant multipoles are from rank 0 to 4; 
monopole $X_0$, dipole $(Y_x, Y_y, Y_z)$, quadrupole $(X_{u}, X_{\varv}, X_{yz}, X_{zx}, X_{xy})$, octupole $(Y_{xyz}, Y_x^\alpha, Y_y^\alpha, Y_z^\alpha, Y_x^\beta, Y_y^\beta, Y_z^\beta)$, and hexadecapole $(X_4, X_{4u}, X_{4\varv}, X_{4x}^\alpha, X_{4y}^\alpha, X_{4z}^\alpha, X_{4x}^\beta, X_{4y}^\beta, X_{4z}^\beta)$.
The tensor component of $\chi^{[1\times3]}$ is given by as
\begin{widetext}
\begin{align}
\label{eq:rank13}
&
\chi^{[1\times3]}=
\begin{pmatrix}
 3( {X}_{0}-\tilde{X}_u+\tilde{X}_\varv)+2 {X}_{4}-{X}_{4u}+{X}_{4\varv}
 &3 (-{Y}_{z}-\tilde{X}_{xy}
 + {Y}_{z}^\alpha
 - {Y}_{z}^\beta) +{X}_{4z}^\alpha-{X}_{4z}^\beta
 & 3(
 {Y}_{y}-\tilde{X}_{zx} 
 -{Y}_{y}^\alpha 
 -{Y}_{y}^\beta) -{X}_{4y}^\alpha-{X}_{4y}^\beta \\
 3( 
 {Y}_{z}-\tilde{X}_{xy} 
 -{Y}_{z}^\alpha 
 -{Y}_{z}^\beta)-{X}_{4z}^\alpha-{X}_{4z}^\beta 
 & 3( {X}_{0}-\tilde{X}_u-\tilde{X}_\varv)+2 {X}_{4}-{X}_{4u}-{X}_{4\varv} 
 & 3 (-{Y}_{x}-\tilde{X}_{yz} 
 +{Y}_{x}^\alpha 
 -{Y}_{x}^\beta) +{X}_{4x}^\alpha-{X}_{4x}^\beta \\
 3 (-{Y}_{y}-\tilde{X}_{zx} 
 +{Y}_{y}^\alpha 
 -{Y}_{y}^\beta)+{X}_{4y}^\alpha-{X}_{4y}^\beta 
 & 3(
 {Y}_{x}-\tilde{X}_{yz} 
 -{Y}_{x}^\alpha 
 -{Y}_{x}^\beta) -{X}_{4x}^\alpha-{X}_{4x}^\beta 
 & 3 ({X}_{0}+2\tilde{X}_u)+2 {X}_{4}+2 {X}_{4u} \\
-{Y}_{y}- \tilde{X}_{zx}
-4 {Y}_{y}^\alpha
+2 {Y}_{y}^\beta+2 {X}_{4y}^\beta 
&
{Y}_{x}+\tilde{X}'_{yz} 
-{Y}_{x}^\alpha 
+{Y}_{x}^\beta +{X}_{4x}^\alpha-{X}_{4x}^\beta
& {X}_{0}+\tilde{X}''_u-5 {X}_{\varv}
-{Y}_{xyz} -{X}_{4}-{X}_{4u}+{X}_{4\varv}\\
 {X}_{0}+\tilde{X}'_u-\tilde{X}'_\varv
 -{Y}_{xyz} -{X}_{4}-{X}_{4u}-{X}_{4\varv}
 &-{Y}_{z}-\tilde{X}_{xy}
 -4 {Y}_{z}^\alpha
 +2 {Y}_{z}^\beta+2 {X}_{4z}^\beta 
 & 
 {Y}_{y}+\tilde{X}'_{zx} 
 -{Y}_{y}^\alpha 
 +{Y}_{y}^\beta+{X}_{4y}^\alpha-{X}_{4y}^\beta \\
 {Y}_{z}+\tilde{X}'_{xy} 
 -{Y}_{z}^\alpha 
 +{Y}_{z}^\beta+{X}_{4z}^\alpha-{X}_{4z}^\beta 
 & {X}_{0}-\tilde{X}_u+\tilde{X}''_\varv
 -{Y}_{xyz}-{X}_{4}+2 {X}_{4u} 
 & -{Y}_{x}-\tilde{X}_{yz}
 -4 {Y}_{x}^\alpha
 +2 {Y}_{x}^\beta+2 {X}_{4x}^\beta \\
{Y}_{z}-\tilde{X}_{xy}
+4 {Y}_{z}^\alpha
+2 {Y}_{z}^\beta +2 {X}_{4z}^\beta
 & {X}_{0}+\tilde{X}'_u+\tilde{X}'_\varv
 +{Y}_{xyz} -{X}_{4}-{X}_{4u}+{X}_{4\varv}
 & -{Y}_{x}+\tilde{X}'_{yz}
 +{Y}_{x}^\alpha
 +{Y}_{x}^\beta-{X}_{4x}^\alpha-{X}_{4x}^\beta \\
-{Y}_{y}+\tilde{X}'_{zx} 
+{Y}_{y}^\alpha 
+{Y}_{y}^\beta -{X}_{4y}^\alpha-{X}_{4y}^\beta
 & 
 {Y}_{x}-\tilde{X}_{yz}
 +4 {Y}_{x}^\alpha
 +2{Y}_{x}^\beta +2 {X}_{4x}^\beta
 & {X}_{0}+\tilde{X}''_u+5 {X}_{\varv}
 +{Y}_{xyz}-{X}_{4}-{X}_{4u}-{X}_{4\varv} \\
 {X}_{0}-\tilde{X}_u-\tilde{X}''_\varv
 +{Y}_{xyz} -{X}_{4}+2 {X}_{4u}
 &-{Y}_{z}+\tilde{X}'_{xy} 
 +{Y}_{z}^\alpha 
 +{Y}_{z}^\beta-{X}_{4z}^\alpha-{X}_{4z}^\beta 
 &
 {Y}_{y}- \tilde{X}_{zx}
 +4 {Y}_{y}^\alpha
 +2 {Y}_{y}^\beta+2 {X}_{4y}^\beta \\
5 {X}_{yz}
-2{Y}_{x}^\beta+ 2 {X}_{4x}^\beta
  & 5 {X}_{zx}
  -2 {Y}_{y}^\beta +2 {X}_{4y}^\beta
  & 5 {X}_{xy}
  -2 {Y}_{z}^\beta+2 {X}_{4z}^\beta \\
\end{pmatrix}^{\rm T}.
\end{align}
Note that $(\tilde{X}_u, \tilde{X}'_u, \tilde{X}''_u)$, $(\tilde{X}_\varv, \tilde{X}'_\varv, \tilde{X}''_\varv)$, and $(\tilde{X}_{yz}, \tilde{X}'_{yz})$ (cyclic) are introduced to express the two independent quadrupoles. 
See Appendix~\ref{sec:ap_res_mp_13} for details.
$\chi^{[1\times3]}$ corresponds to the response tensors, such as the third-order nonlinear electric conductivity. 
The relevant multipoles are $X=Q$ or $T$ and $Y=G$ or $M$ for the polar tensor,
while those are $X=G$ or $M$ and $Y=Q$ or $T$ for the axial tensor.
The multipole expression of $\chi^{[3\times1]}$ is obtained by transposing $\chi^{[1\times3]}$.

$\chi^{[2\times2]}$ is another rank-4 tensor for 
$B^{[2]}=(B_{xx}, B_{yy}, B_{zz}, B_{yz}, B_{zx}, B_{xy})$ where $B_{ij}=B_{ji}$ and 
$F^{[2]}=(F_{xx}, F_{yy}, F_{zz}, F_{yz}, F_{zx}, F_{xy})$ where $F_{ij}=F_{ji}$. 
The tensor component of $\chi^{[2\times2]}$ is related to the rank 0--4 multipoles, which is given by 
\begin{align}
\chi^{[2\times 2]}=
\begin{pmatrix}
\chi_{ll} & \chi_{lt}\\
\chi_{tl} & \chi_{tt}\\
\end{pmatrix},
\end{align}
\begin{align}
\label{eq:rank22ll}
\chi_{ll}&=
\begin{pmatrix}
\tilde{X}_0+\tilde{X}_u+\tilde{X}_\varv+2 {X}_{4}-{X}_{4u}+{X}_{4\varv} 
 & \tilde{X}_0^{\prime} +\tilde{X}_u^{\prime}-2 {X}_{\varv}^{(-)}+{X}_{xyz}-{X}_{4}+2 {X}_{4u} 
 & \tilde{X}_0^{\prime}+\tilde{X}_u^{(+)}+\tilde{X}_\varv^{(-)}-{X}_{xyz}- X_{4u\varv+}
 \\
 \tilde{X}_0^{\prime}+\tilde{X}_u^{\prime}+2 {X}_{\varv}^{(-)}-{X}_{xyz}-{X}_{4}+2 {X}_{4u} 
 & \tilde{X}_0 +\tilde{X}_u-\tilde{X}_\varv+2 {X}_{4}-{X}_{4u}-{X}_{4\varv} 
 & \tilde{X}_0^{\prime}+\tilde{X}_u^{(+)}-\tilde{X}_\varv^{(-)}+{X}_{xyz}-X_{4u\varv-} 
 \\
\tilde{X}_0^{\prime}+\tilde{X}_u^{(-)}+\tilde{X}_\varv^{(+)}+{X}_{xyz}-X_{4u\varv+}
 & \tilde{X}_0^{\prime}+\tilde{X}_u^{(-)}-\tilde{X}_\varv^{(+)}-{X}_{xyz}-X_{4u\varv-}
 & \tilde{X}_0 -2\tilde{X}_u +2 {X}_{4}+2 {X}_{4u}  \\
 \end{pmatrix}, \\
\label{eq:rank22lt}
\chi_{lt}&=
\begin{pmatrix}
\tilde{X}_{yz}^{(+)}-2 {Y}_{x}^\beta+2 {X}_{4x}^{\beta} 
 &-2 {Y}_{y}+\tilde{X}_{zx}^{\prime(+)}+{Y}_{y}^\alpha+{Y}_{y}^\beta-{X}_{4y}^\alpha-{X}_{4y}^{\beta} 
 & 2 {Y}_{z}+\tilde{X}_{xy}^{\prime(+)}-{Y}_{z}^\alpha+{Y}_{z}^\beta+{X}_{4z}^\alpha-{X}_{4z}^{\beta} \\
2 {Y}_{x}+\tilde{X}_{yz}^{\prime(+)}-{Y}_{x}^\alpha+{Y}_{x}^\beta+{X}_{4x}^\alpha-{X}_{4x}^{\beta} 
 & \tilde{X}_{zx}^{(+)}-2 {Y}_{y}^\beta+2 {X}_{4y}^{\beta} 
 & -2 {Y}_{z}+\tilde{X}_{xy}^{\prime(+)}+{Y}_{z}^\alpha+{Y}_{z}^\beta-{X}_{4z}^\alpha-{X}_{4z}^{\beta} \\
  -2 {Y}_{x}+\tilde{X}_{yz}^{\prime(+)}+{Y}_{x}^\alpha+{Y}_{x}^\beta-{X}_{4x}^\alpha-{X}_{4x}^{\beta} 
 & 2 {Y}_{y}+\tilde{X}_{zx}^{\prime(+)}-{Y}_{y}^\alpha+{Y}_{y}^\beta+{X}_{4y}^\alpha-{X}_{4y}^{\beta} 
 & \tilde{X}_{xy}^{(+)}-2 {Y}_{z}^\beta+2 {X}_{4z}^{\beta} \\
 \end{pmatrix}, \\
\label{eq:rank22tl}
\chi_{tl}&=
\begin{pmatrix}
\tilde{X}_{yz}^{(-)}+2 {Y}_{x}^\beta+2 {X}_{4x}^{\beta} 
 & -2 {Y}_{x}+\tilde{X}_{yz}^{\prime(-)}+{Y}_{x}^\alpha-{Y}_{x}^\beta+{X}_{4x}^\alpha-{X}_{4x}^{\beta} 
 &2 {Y}_{x}+\tilde{X}_{yz}^{\prime(-)}- {Y}_{x}^\alpha-{Y}_{x}^\beta-{X}_{4x}^\alpha-{X}_{4x}^{\beta} \\
2 {Y}_{y} +\tilde{X}_{zx}^{\prime(-)} -{Y}_{y}^\alpha-{Y}_{y}^\beta-{X}_{4y}^\alpha-{X}_{4y}^{\beta} 
 & \tilde{X}_{zx}^{(-)}+2 {Y}_{y}^\beta+2 {X}_{4y}^{\beta} 
 & -2 {Y}_{y}+\tilde{X}_{zx}^{\prime(-)}+{Y}_{y}^\alpha-{Y}_{y}^\beta+{X}_{4y}^\alpha-{X}_{4y}^{\beta} \\
 -2 {Y}_{z}+\tilde{X}_{xy}^{\prime(-)}+{Y}_{z}^\alpha-{Y}_{z}^\beta+{X}_{4z}^\alpha-{X}_{4z}^{\beta} 
 & 2 {Y}_{z}+\tilde{X}_{xy}^{\prime(-)}-{Y}_{z}^\alpha-{Y}_{z}^\beta-{X}_{4z}^\alpha-{X}_{4z}^{\beta} 
 & \tilde{X}_{xy}^{(-)}+2 {Y}_{z}^\beta+2 {X}_{4z}^{\beta} \\
\end{pmatrix},\\
\label{eq:rank22tt}
\chi_{tt}&=
\begin{pmatrix}
 3 {X}_{0}+3 {X}_{u}-3 {X}_{\varv}- X_{4u\varv-}
 & -{Y}_{z}+3 {X}_{xy}-2 {Y}_{z}^\alpha+2 {X}_{4z}^{\beta} 
 & {Y}_{y}+3 {X}_{zx}+2 {Y}_{y}^\alpha+2 {X}_{4y}^{\beta} \\
 {Y}_{z}+3 {X}_{xy}+2 {Y}_{z}^\alpha+2 {X}_{4z}^{\beta} 
 & 3 {X}_{0}+3 {X}_{u}+3 {X}_{\varv}-X_{4u\varv+}
 & -{Y}_{x}+3 {X}_{yz}-2 {Y}_{x}^\alpha+2 {X}_{4x}^{\beta} \\
  -{Y}_{y}+3 {X}_{zx}-2 {Y}_{y}^\alpha+2 {X}_{4y}^{\beta} 
 & {Y}_{x}+3 {X}_{yz}+2 {Y}_{x}^\alpha+2 {X}_{4x}^{\beta} 
 & 3 {X}_{0}-6 {X}_{u}-{X}_{4}+2 {X}_{4u} \\
 \end{pmatrix},
\end{align}
\end{widetext}
where $X_{4u\varv\pm}=X_{4}+X_{4u}\pm X_{4\varv}$.
We also introduce $(\tilde{X}_0, \tilde{X}_0^{\prime})$, $(\tilde{X}_u, \tilde{X}_u^{(\pm)}, \tilde{X}_u^{\prime})$, $(\tilde{X}_\varv, \tilde{X}_\varv^{(\pm)})$, and $(\tilde{X}_{yz}^{(\pm)}, \tilde{X}_{yz}^{\prime(\pm)})$ (cyclic) 
for notational simplicity.
See Appendix~\ref{sec:ap_res_mp_22} for details.
$\chi^{[2\times 2]}$ represents the rank-4 tensor, such as elastic stiffness tensor and magneto-Seebeck tensor, which is related to the multipoles $X=Q$ or $T$ and $Y =G$ or $M$ for the polar tensor and to $X=G$ or $M$ and $Y =Q$ or $T$ for the axial tensor.

\subsection{Linear response theory \label{sec:res_linear}}
The multipoles with the opposite time-reversal parities, E and MT or ET and M, are not distinguished by the above symmetry analyses. 
They are distinguished from the microscopic process by considering the effect of the dissipation. 
To demonstrate that, we adopt the Kubo formula in the following discussion~\cite{watanabe2017magnetic,hayami2018classification}.

When we consider the external perturbation Hamiltonian 
$\mathcal{H}_{\rm ext}=-\sum_j\hat{A}_j F_j(t)$, where $F_j(t) = \int_{-\infty}^{\infty} \frac{d\omega}{2\pi} F_{j,\omega} e^{-i\omega t+\delta t}$ is the $j$-th component of an external field for $\delta>0$, the linear complex susceptibility $\chi_{i; j} (\omega)$ satisfies the relation
\begin{align}
\braket{\hat{B}_{i,\omega}}
 &=\int_{-\infty}^\infty \frac{d\omega'}{2\pi}\delta(\omega-\omega') \chi_{i;j}(\omega') F _{j,\omega'}.
 \label{eq:suscep}
\end{align} 
Considering the uniform external field with the wave vector ${\bm q}\to {\bm 0}$, 
and then taking the static limit $\omega\to 0$, the linear response function for the periodic system is represented as
\begin{align}
\chi_{i; j} &\equiv \chi_{i; j} (\omega \to 0) \notag\\
&= - \frac{i\hbar}{V} \sum_{{\bm k} nm} \frac{f[\varepsilon_n({\bm k})] - f[\varepsilon_m({\bm k})]}{\varepsilon_n({\bm k}) - \varepsilon_m({\bm k})}
\frac{B_{i{\bm k}}^{nm} \dot{A}_{j{\bm k}}^{mn}}{i\hbar\delta + \varepsilon_n({\bm k}) - \varepsilon_m({\bm k})},
\end{align}
where $\dot{A}=dA/dt$.
$X_{{\bm k}}^{nm}\equiv\braket{n{\bm k}|\hat{X}|m{\bm k}}$ is the matrix element between the Bloch states $\ket{n{\bm k}}$ and $\ket{m{\bm k}}$ with the band indices $n$ and $m$, respectively, and the wave vector ${\bm k}$.
$f[\varepsilon_n({\bm k})]$ is the Fermi distribution function with the eigenenergy $\varepsilon_n({\bm k})$ of the eigenstate $\ket{n{\bm k}}$.
$V$, $\hbar$, and $\delta$ is the system volume, the reduced Planck 
constant, and the broadening factor, respectively.
We here assume the relaxation-time approximation and mimic the constant $1/\delta$ as the relaxation time.
$\chi_{i;j}$ can be decomposed as 
\begin{align}
\chi_{i;j} 
&= \chi_{i;j}^{\rm (J)} + \chi_{i;j}^{\rm (E)}, \\ 
\label{eq:chiJ}
\chi_{i;j}^{\rm (J)}
&= 
-\frac{\hbar^2 \delta}{V} \sum_{{\bm k} nm} 
\frac{f[\varepsilon_n({\bm k})] - f[\varepsilon_m({\bm k})]}{\varepsilon_n({\bm k}) - \varepsilon_m({\bm k})} \frac{B_{i{\bm k}}^{nm} \dot{A}_{j{\bm k}}^{mn}}
{(\hbar\delta)^2 + [\varepsilon_n({\bm k}) - \varepsilon_m({\bm k})]^2},
 \\
\label{eq:chiE}
\chi_{i;j}^{\rm (E)}
&
= -\frac{i\hbar}{V}\sum_{{\bm k} nm}^{\neq} \frac{f[\varepsilon_n({\bm k})] - f[\varepsilon_m({\bm k})]}{(\hbar\delta)^2 + [\varepsilon_n({\bm k}) - \varepsilon_m({\bm k})]^2} B_{i{\bm k}}^{nm}\dot{A}_{j{\bm k}}^{mn}, 
\end{align}
where $\chi_{i;j}^{\rm (J)}$ includes the intraband (dissipative) 
contribution proportional to $1/\delta$, while $\chi_{i;j}^{\rm (E)}$ is the interband (nondissipative) 
one, which remains finite in the clean limit of $\delta \to 0$.

$\chi^{\rm (J)}$ and $\chi^{\rm (E)}$ have the opposite time-reversal property~\cite{watanabe2017magnetic, hayami2018classification}.
When the time-reversal symmetry is preserved, they are transformed as 
\begin{align}
\label{eq:TR_linear}
\chi_{i;j}^{\rm (J)} = -t_{B_i}t_{A_j}\chi_{i;j}^{\rm (J)}, \ \chi_{i;j}^{\rm (E)} = t_{B_i}t_{A_j}\chi_{i;j}^{\rm (E)},
\end{align}
where $X_{\bm k}^{nm}=t_X X_{-{\bm k}}^{\bar{m}\bar{n}}$ for $t_X=\pm 1$ ($X=A_{j}, B_{i}$). 
The $\bar{n}$-th band stands for the time-reversal partner of the $n$-th band.
Equation~\eqref{eq:TR_linear} means that $\chi_{i;j}^{\rm (J)}$ [$\chi_{i;j}^{\rm (E)}$] can be finite when $t_{B_i}t_{A_j}=-1(+1)$. 
In other words,
$\chi_{i;j}^{\rm (J)}$ [$\chi_{i;j}^{\rm (E)}$] becomes nonzero when the M and MT (E and ET) 
multipoles are active for $t_{B_i}t_{A_j}=+1$, while $\chi_{i;j}^{\rm (J)}$ [$\chi_{i;j}^{\rm (E)}$] becomes nonzero when the E and ET (M and MT) multipoles are active for $t_{B_i}t_{A_j}=-1$. 
The multipoles contributing to $\chi_{i;j}^{\rm (J)}$ and $\chi_{i;j}^{\rm (E)}$ are summarized in
Table~\ref{table:tensor_parity_MP_linear}.
A similar argument holds for the static isothermal susceptibility such as magnetic susceptibility, which is obtained by $\omega\to 0$ and then ${\bm q}\to {\bm 0}$ for $\chi_{i; j} (\omega)$ in Eq.~(\ref{eq:suscep}) in the non-degenerate system~\footnote{
The general form of the static isothermal susceptibility $\chi_{BA}^{T}$ is given by $\chi_{BA}^{T} =  \beta \sum_{nm}^= \varw_n B_{nm}A_{mn} + \sum_{nm}^{\neq} \frac{\varw_m-\varw_n}{E_n-E_m}B_{nm}A_{mn}$, where $\beta=1/k_{\rm B}T$, $\varw_n$ is the Boltzmann weight of eigenstate $n$, and the notation $=$ ($\neq$) stands for the summation taken over $E_n=E_m$ ($E_n\neq E_m$).
The first term corresponds to the Curie term in the degenerate system.}.

Let us discuss $\chi_{i;j}^{\rm (J)}$ and $\chi_{i;j}^{\rm (E)}$ by taking an example.
We consider the uniform and staggered magnetic orderings with magnetic moments along the $z$ axis in the diamond structure in Secs.~\ref{sec:res_linear_uniform} and \ref{sec:res_linear_staggered}, respectively.

\tabcolsep = 5pt
\begin{table*}[htb!]
\centering
\caption{
Correspondence between the linear response functions $\chi^{\rm (J, E)}$ and multipoles $X_{lm}$ ($l=1$--$4$, $X=Q,G,T,M$). 
Nonzero $\chi^{\rm (J, E)}$ is indicated by the checkmark($\checkmark$).
In the rightmost column, the parenthesis represents the corresponding multipoles to the response $B$ and the external field $F$ (see also Table~\ref{table:field_response_MP}). 
\label{table:tensor_parity_MP_linear}}
\begin{tabular}{cccccccccccc}
\hline \hline 
& & & & & \multicolumn{5}{l}{$(\mathcal{P}, \mathcal{T}, \mathcal{PT})=$} & 
examples \\
 rank & & $t_{B_i}t_{A_j}$ & & multipole & $(\bigcirc, \bigcirc, \bigcirc)$ & $(\times, \bigcirc, \times)$& $(\bigcirc, \times, \times)$  & $(\times, \times, \bigcirc)$ & $(\times, \times, \times)$ 
 & ($B \leftrightarrow  F$) \\ \hline 

1 & polar & $+1$ & $\chi^{\rm (J)}$ & $T_{1m}$ & & & & $\checkmark$ &  $\checkmark$ & electrocaloric tensor \\
 & & & $\chi^{\rm (E)}$ & $Q_{1m}$ &  & $\checkmark$ & & & $\checkmark$ & $(Q_0 \leftrightarrow Q_{1m})$ \\ \cline{3-11}
 & & $-1$ & $\chi^{\rm (J)}$ & $Q_{1m}$ &  & $\checkmark$ & & & $\checkmark$ & toroidalcaloric tensor \\
 & & & $\chi^{\rm (E)}$ & $T_{1m}$ & & & & $\checkmark$ & $\checkmark$ & $(Q_0 \leftrightarrow T_{1m})$ \\ \cline{2-11}
 
& axial & $+1$ & $\chi^{\rm (J)}$ & $M_{1m}$ & & & $\checkmark$ & & $\checkmark$ &  \\
& & & $\chi^{\rm (E)}$ & $G_{1m}$ & $\checkmark$ & $\checkmark$ & $\checkmark$ & $\checkmark$ & $\checkmark$ & e.g. $(Q_0 \leftrightarrow G_{1m})$ \\
\cline{3-11}

& & $-1$ & $\chi^{\rm (J)}$ & $G_{1m}$ & $\checkmark$ & $\checkmark$ & $\checkmark$ & $\checkmark$ & $\checkmark$ & magnetocaloric tensor \\
& & & $\chi^{\rm (E)}$ & $M_{1m}$ & & & $\checkmark$ & & $\checkmark$ & $(Q_0 \leftrightarrow M_{1m})$ \\
\hline

2 & polar & $+1$ & $\chi^{\rm (J)}$ & $T_{0}, M_{1m}, T_{2m}$ & & & $\checkmark$ & & $\checkmark$ & magnetic susceptibility tensor \\
& & & $\chi^{\rm (E)}$ & $Q_{0}, G_{1m}, Q_{2m}$ & $\checkmark$ & $\checkmark$ & $\checkmark$ & $\checkmark$ &  $\checkmark$ & $(M_{1m} \leftrightarrow M_{1m})$  \\
\cline{3-11}

& & $-1$ & $\chi^{\rm (J)}$  & $Q_{0}, G_{1m}, Q_{2m}$ & $\checkmark$ & $\checkmark$ & $\checkmark$ & $\checkmark$ &  $\checkmark$ & electric conductivity tensor \\ 
& & & $\chi^{\rm (E)}$ & $T_{0}, M_{1m}, T_{2m}$ & & & $\checkmark$ & & $\checkmark$ & $(T_{1m} \leftrightarrow E_{1m})$ \\ 
\cline{2-11}

 & axial & $+1$ & $\chi^{\rm (J)}$ & $M_{0}, T_{1m}, M_{2m}$ & & & & $\checkmark$ & $\checkmark$ &  \\
 & & & $\chi^{\rm (E)}$ & $G_{0}, Q_{1m}, G_{2m}$ & & $\checkmark$ & & &  $\checkmark$ & e.g. $(T_{1m} \leftrightarrow M_{1m})$ \\
 \cline{3-11}
 
 & & $-1$ & $\chi^{\rm (J)}$ & $G_{0}, Q_{1m}, G_{2m}$ & & $\checkmark$ & & &  $\checkmark$ & magnetoelectric tensor \\
  & & & $\chi^{\rm (E)}$ & $M_{0}, T_{1m}, M_{2m}$ & & & & $\checkmark$ & $\checkmark$ & $(M_{1m} \leftrightarrow E_{1m})$ \\
 \hline
 
 3 & polar & $+1$ & $\chi^{\rm (J)}$ & $T_{1m}, M_{2m}, T_{3m}$ & & & & $\checkmark$ &  $\checkmark$ & piezoelectric tensor \\
 & & & $\chi^{\rm (E)}$ & $Q_{1m}, G_{2m}, Q_{3m}$ &  & $\checkmark$ & & & $\checkmark$ & $(Q_{1m} \leftrightarrow Q_0, Q_{2m})$ \\
 \cline{3-11}
 
 & & $-1$ & $\chi^{\rm (J)}$ & $Q_{1m}, G_{2m}, Q_{3m}$ &  & $\checkmark$ & & & $\checkmark$ &  \\
 & & & $\chi^{\rm (E)}$ & $T_{1m}, M_{2m}, T_{3m}$ & & & & $\checkmark$ & $\checkmark$ &e.g. $(T_{1m} \leftrightarrow Q_0, Q_{2m})$ \\ \cline{2-11}

& axial & $+1$ & $\chi^{\rm (J)}$ & $M_{1m}, T_{2m}, M_{3m}$ & & & $\checkmark$ & & $\checkmark$ &  spin conductivity tensor \\
& & & $\chi^{\rm (E)}$ & $G_{1m}, Q_{2m}, G_{3m}$ & $\checkmark$ & $\checkmark$ & $\checkmark$ & $\checkmark$ & $\checkmark$ & $(G_0, Q_{1m}, G_{2m} \leftrightarrow Q_{1m})$ \\
\cline{3-11}

& & $-1$ & $\chi^{\rm (J)}$ & $G_{1m}, Q_{2m}, G_{3m}$ & $\checkmark$ & $\checkmark$ & $\checkmark$ & $\checkmark$ & $\checkmark$ & piezomagnetic tensor \\
& & & $\chi^{\rm (E)}$ & $M_{1m}, T_{2m}, M_{3m}$ & & & $\checkmark$ & & $\checkmark$ & $(M_{1m} \leftrightarrow Q_0, Q_{2m})$ \\
\hline

4 & polar & $+1$ & $\chi^{\rm (J)}$ & $T_{0}, M_{1m}, T_{2m},$ & & & $\checkmark$ & & $\checkmark$ & elastic stiffness tensor  \\
& & & & $M_{3m}, T_{4m}$ & & & & & & $(Q_0, Q_{2m} \leftrightarrow Q_0, Q_{2m})$\\

& & & $\chi^{\rm (E)}$  & $Q_{0}, G_{1m}, Q_{2m},$ & $\checkmark$ & $\checkmark$ & $\checkmark$ & $\checkmark$ &  $\checkmark$ & \\
& & & & $G_{3m}, Q_{4m}$ \\
\cline{3-11}

& & $-1$ & $\chi^{\rm (J)}$  &  $Q_{0}, G_{1m}, Q_{2m},$ & $\checkmark$ & $\checkmark$ & $\checkmark$ & $\checkmark$ &  $\checkmark$ & \\ 
& & & & $G_{3m}, Q_{4m}$  & & & & & & e.g. $(T_0, T_{2m} \leftrightarrow Q_0, Q_{2m})$\\

& & & $\chi^{\rm (E)}$& $T_{0}, M_{1m}, T_{2m},$ & & & $\checkmark$ & & $\checkmark$&  \\ 
& & & & $M_{3m}, T_{4m}$ \\
\cline{2-11}

& axial & $+1$ & $\chi^{\rm (J)}$ & $M_{0}, T_{1m}, M_{2m},$ & & & & $\checkmark$ & $\checkmark$ &  \\
& & & & $T_{3m}, M_{4m}$ & & & & & & e.g. $(G_0, G_{2m} \leftrightarrow Q_0, Q_{2m})$ \\

& & & $\chi^{\rm (E)}$ & $G_{0}, Q_{1m}, G_{2m},$ & & $\checkmark$ & & &  $\checkmark$  &\\
& & & & $Q_{3m}, G_{4m}$  \\
\cline{3-11}
 
& & $-1$ & $\chi^{\rm (J)}$ & $G_{0}, Q_{1m}, G_{2m},$ & & $\checkmark$ & & &  $\checkmark$ & \\
& & & & $Q_{3m}, G_{4m}$ & & & & & & e.g. $(M_0, M_{2m} \leftrightarrow Q_0, Q_{2m})$\\
 
& & & $\chi^{\rm (E)}$ & $M_{0}, T_{1m}, M_{2m},$ & & & & $\checkmark$ & $\checkmark$ & \\
& & & & $T_{3m}, M_{4m}$ \\
 
\hline\hline
\end{tabular}
\end{table*}

\begin{figure}[htb!]
\centering
\includegraphics[width=85mm]{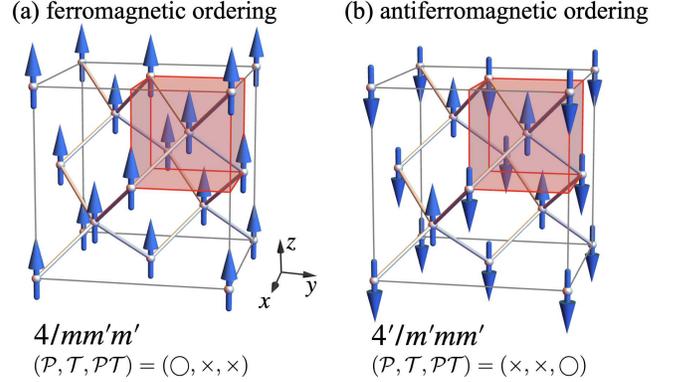}
\caption{(a) Ferromagnetic ordering and (b) antiferromagnetic ordering in the diamond structure.
(a) is characterized by $4/mm'm'$ with the $\mathcal{P}$ symmetry, whereas 
(b) is represented by $4'/m'mm'$ with the $\mathcal{PT}$ symmetry.
 \label{fig:ferro_antiferro}}
\end{figure}

\subsubsection{$\mathcal{P}$-symmetric magnetic structure \label{sec:res_linear_uniform}}
The uniform magnetic structure belonging to the ${\rm T}_{\rm 1g}^-$ representation of $m\bar{3}m1'$ in Fig.~\ref{fig:ferro_antiferro}(a) reduces the symmetry to the $\mathcal{P}$-symmetric $4/mm'm'$.
From Table~\ref{table:active_MP_BWPG_1}, one can find that the active multipoles with rank 0--4 are the even-parity E multipoles $Q_0$ (monopole), $Q_u$ (quadrupole), $Q_{4}, Q_{4u}$ (hexadecapole), even-parity M multipoles $M_z$ (dipole), $M_z^\alpha$ (octupole), and even-parity MT multipole $T_{4z}^\alpha$ (hexadecapole).  
Thus, it is expected to occur the physical responses related to these active multipoles,
such as magnetocaloric response, electric conductivity, and piezomagnetic response from Table~\ref{table:tensor_parity_MP_linear}.

For example, the electric conductivity tensor $J_i=\sum_{j} \sigma_{ij}E_j$ is given by
\begin{align}
\label{eq:con} 
\sigma^{\rm (J)} \leftrightarrow 
\begin{pmatrix}
Q_0-Q_u & 0 &  0 \\
0 & Q_0-Q_u & 0 \\
0 &0 & Q_0+2Q_u \\
\end{pmatrix}, 
\end{align} 
and
\begin{align}
\label{eq:con_anomalous}
\sigma^{\rm (E)} \leftrightarrow 
\begin{pmatrix}
0 & M_z &  0 \\
- M_z & 0 & 0 \\
0 & 0 & 0 \\
\end{pmatrix}, 
\end{align} 
where $\sigma_{ij}=\sigma_{ij}^{\rm (J)}+\sigma_{ij}^{\rm (E)}$.
This means that the system exhibits the anisotropic electric conductivity along the $xy$ and $z$ directions and the anomalous Hall effect in the $xy$ plane.

The other example is the piezomagnetic effect where $M_i = \sum_{jk}\Lambda_{ijk} \tau_{jk}$.
From the active multipoles, the tensor component is given by
\begin{align}
\Lambda^{\rm (E)}
&=
\begin{pmatrix}
0 & 0 & \Lambda_{zxx}^{\rm (E)}
\\
0 & 0 & \Lambda_{zxx}^{\rm (E)}
\\
0 & 0 &\Lambda_{zzz}^{\rm (E)} 
\\
0 & \Lambda_{yyz}^{\rm (E)} 
& 0 \\
\Lambda_{yyz}^{\rm (E)}
 & 0 & 0 \\
 0 & 0 & 0\\
\end{pmatrix}^{\rm T} \\
&\leftrightarrow
\begin{pmatrix}
0 & 0 & \tilde{M}'_z-2  M_{z}^\alpha\\
0 &0 &\tilde{M}'_z  -2M_{z}^\alpha\\
0 & 0 &\tilde{M}_z+4 M_{z}^\alpha \\
0 & -3 M_{z}-2 M_{z}^\alpha & 0 \\
  -3 M_{z}-2 M_{z}^\alpha & 0 & 0 \\
 0 & 0 & 0\\
\end{pmatrix}^{\rm T}, 
\end{align}
by using Eq.~\eqref{eq:rank12}. 
We omit $\Lambda^{\rm (J)}$ by taking $\delta\to 0$. 

\subsubsection{$\mathcal{PT}$-symmetric magnetic structure \label{sec:res_linear_staggered}}
The staggered magnetic structure belonging to the ${\rm T}_{\rm 2u}^-$ representation in Fig.~\ref{fig:ferro_antiferro}(b) has the $\mathcal{PT}$-symmetric $4'/m'mm'$. 
In this case, the odd-parity M multipoles $M_{xy}$ (quadrupole), $M_{4z}^\beta$ (hexadecapole), and odd-parity MT multipole $T_z^\beta$ (octupole) become active in addition to the even-parity E multipole $Q_0$, $Q_u$, $Q_{4}$, and $Q_{4u}$, which are obtained by appropriately replacing the mirror plane of $4'/m'm'm$ in Table~\ref{table:active_MP_BWPG_2}.
The active odd-parity M and MT multipoles induce the cross-correlated responses such as magnetoelectric effect and piezoelectric effect in Table~\ref{table:tensor_parity_MP_linear}.

For example, the magnetoelectric tensor $\alpha_{ij}$ is given by
\begin{align}
\alpha^{\rm (E)}
=\begin{pmatrix}
0 & \alpha^{\rm (E)}_{xy} & 0 \\
\alpha^{\rm (E)}_{xy} & 0 & 0 \\
0 & 0 & 0 \\
\end{pmatrix}
\leftrightarrow
\begin{pmatrix}
0 & M_{xy} & 0 \\
M_{xy}& 0 & 0 \\
0 & 0 & 0 \\
\end{pmatrix}.
\end{align}
It is noted that $\alpha^{\rm (J)}=0$, as no odd-parity E and ET multipoles are active under the magnetic point group $4'/m'm'm$.

Meanwhile, in the (inverse) piezoelectric response, the response tensor $d_{ijk}$ for
$\varepsilon_{ij} = \sum_{k}d_{ijk}E_k$~\cite{watanabe2017magnetic, shiomi2019observation}
is described by $M_{xy}$ and $T_z^\beta$ as
\begin{align}
d^{\rm (J)} &=
\begin{pmatrix}
0 & 0 & d_{xxx}\\
0 & 0 &-d_{xxx}\\
0 & 0 &0 \\
 0 & d_{yzy} & 0 \\
-d_{yzy} & 0 & 0 \\
 0 & 0 & 0 \\
\end{pmatrix}
\notag\\
&\leftrightarrow
\begin{pmatrix}
0 & 0 & -2M_{xy}+2T_{z}^\beta\\
0 & 0 &2M_{xy}-2T_{z}^\beta\\
0 & 0 &0 \\
 0 & -M_{xy}-2 T_{z}^\beta & 0 \\
  M_{xy}+2 T_{z}^\beta & 0 & 0 \\
 0 & 0 & 0 \\
\end{pmatrix}.
\end{align}
Similar to $\alpha_{ij}$, $d^{(\rm E)}_{ijk}=0$ owing to the lack of odd-parity E and ET multipoles.

\subsection{Nonlinear response theory \label{sec:res_nonlinear}}

In a similar way to the linear response tensor $\chi_{i; j}$, 
we consider the second-order nonlinear response tensor $\chi_{i; jk}$ based on the Kubo formula~\cite{kubo_doi:10.1143/JPSJ.12.570}.
\begin{widetext}
The nonlinear complex susceptibility $\chi_{i;jk} (\omega',\omega'')$ satisfies the relation 
\begin{align}
\braket{\hat{B}_{i,\omega}} = \int_{-\infty}^{\infty} \int_{-\infty}^{\infty} \frac{d\omega'}{2\pi}  \frac{d\omega''}{2\pi} 
\chi_{i;jk} (\omega',\omega'') F_{j,\omega'} F_{k,\omega''} \delta(\omega-\omega'-\omega''),
\end{align}
where $\chi_{i;jk}(\omega',\omega'') $ is represented as
\begin{align}
\label{eq:2nd_c_susceptability}
\chi_{i;jk}(\omega',\omega'')  
= \frac{1}{2}\left( \frac{1}{i\hbar} \right)^2  \int_0^\infty  \int_0^\infty  dt'dt''\, {\rm Tr} \left[ 
\hat{B}_i  [\hat{A}_j(-t'), [\hat{A}_k(-t'-t''), \rho_0]] \right]e^{i(\omega'+\omega'')t'-2\delta t'}e^{i\omega'' t'' -\delta t''} 
+ (j, \omega')\leftrightarrow(k, \omega''),
\end{align}
where $\hat{X}(t)$ is the Heisenberg representation of an operator $\hat{X}$.
$\rho_0$ is the density matrix for the nonperturbative state, whose matrix element is represented by the Fermi distribution function $f(\varepsilon_{n})$ as $[\rho_0]_{nm}=f(\varepsilon_{n})\delta_{nm}$.
When we suppose $\hat{A}$ as the well-defined operator in a periodic system, i.e., the matrix element of $\hat{A}$ includes no differential operator with respect to the wave vector, the nonlinear response function in the static limit (${\bm q}\to {\bm 0}$, $\omega \to 0$) is given as follows.
\begin{align}
\chi_{i; jk} & \equiv \chi_{i; jk} (0,0)= \frac{1}{2} \frac{1}{V} \sum_{\bm k} \sum_{lmn} \frac{B_{i {\bm k}}^{nm}\left(A_{j{\bm k}}^{ml} A_{k {\bm k}}^{ln} +A_{k {\bm k}}^{ml} A_{j {\bm k}}^{ln} \right) }{\varepsilon_n({\bm k}) - \varepsilon_m({\bm k})+2i \hbar\delta }
\left\{
\frac{f[\varepsilon_n({\bm k})]-f[\varepsilon_l({\bm k})]}{\varepsilon_n({\bm k})-\varepsilon_l({\bm k}) +i\hbar\delta} - \frac{f[\varepsilon_l({\bm k})]-f[\varepsilon_m({\bm k})]}{\varepsilon_l({\bm k})-\varepsilon_m({\bm k}) +i\hbar\delta} 
\right\}.
\end{align}
The nonlinear response function is also decomposed into the two parts with different time-reversal properties as 
\begin{align}
\chi_{i; jk} =& \chi_{i;jk}^{\rm (Re)} + \chi_{i;jk}^{\rm (Im)},
\end{align}
where
\begin{align}
\chi_{i;jk}^{\rm (Re)} \equiv&  
\frac{1}{V} \sum_{{\bm k}lmn}\frac{ {\rm Re}\left(B_{i {\bm k}}^{nm}A_{j{\bm k}}^{ml} A_{k {\bm k}}^{ln}\right)[\varepsilon_n({\bm k}) - \varepsilon_m({\bm k})]}{[\varepsilon_n({\bm k}) - \varepsilon_m({\bm k})]^2 + (2\hbar\delta)^2}
\left(
\frac{\left\{f[\varepsilon_n({\bm k})]-f[\varepsilon_l({\bm k})]\right\}[\varepsilon_n({\bm k})-\varepsilon_l({\bm k})]}{[\varepsilon_n({\bm k})-\varepsilon_l({\bm k})]^2 +(\hbar\delta)^2}
 - \frac{\left\{f[\varepsilon_l({\bm k})]-f[\varepsilon_m({\bm k})]\right\}[\varepsilon_l({\bm k})-\varepsilon_m({\bm k})]}{[\varepsilon_l({\bm k})-\varepsilon_m({\bm k})]^2+(\hbar\delta)^2} 
\right) \notag\\
&- \frac{2\hbar^2\delta^2}{V} \sum_{{\bm k} lmn} \frac{{\rm Re}\left(B_{i {\bm k}}^{nm}A_{j{\bm k}}^{ml} A_{k {\bm k}}^{ln}\right) }{[\varepsilon_n({\bm k}) - \varepsilon_m({\bm k})]^2 + (2 \hbar\delta)^2}
\left\{
\frac{f[\varepsilon_n({\bm k})]-f[\varepsilon_l({\bm k})]}{[\varepsilon_n({\bm k})-\varepsilon_l({\bm k})]^2 +(\hbar\delta)^2} - \frac{f[\varepsilon_l({\bm k})]-f[\varepsilon_m({\bm k})]}{[\varepsilon_l({\bm k})-\varepsilon_m({\bm k})]^2 +(\hbar\delta)^2} 
\right\}, \\
\chi_{i;jk}^{\rm (Im)} \equiv& 
\frac{\hbar\delta}{V}
\sum_{{\bm k}lmn}  \frac{{\rm Im}\left(B_{i {\bm k}}^{nm} A_{j{\bm k}}^{ml} A_{k {\bm k}}^{ln} \right)[\varepsilon_n({\bm k}) - \varepsilon_m({\bm k})] }{[\varepsilon_n({\bm k}) - \varepsilon_m({\bm k})]^2 + (2 \hbar\delta)^2}
\left\{
\frac{f[\varepsilon_n({\bm k})]-f[\varepsilon_l({\bm k})]}{[\varepsilon_n({\bm k})-\varepsilon_l({\bm k})]^2 +(\hbar\delta)^2} - \frac{f[\varepsilon_l({\bm k})]-f[\varepsilon_m({\bm k})]}{[\varepsilon_l({\bm k})-\varepsilon_m({\bm k})]^2 +(\hbar\delta)^2} 
\right\} \notag\\
&+ \frac{2\hbar\delta}{V} \sum_{{\bm k}lmn} \frac{{\rm Im}\left(B_{i {\bm k}}^{nm}A_{j{\bm k}}^{ml} A_{k {\bm k}}^{ln} \right)}{[\varepsilon_n({\bm k}) - \varepsilon_m({\bm k})]^2 + (2 \hbar\delta)^2}
\left(
\frac{\left\{f[\varepsilon_n({\bm k})]-f[\varepsilon_l({\bm k})]\right\}[\varepsilon_n({\bm k})-\varepsilon_l({\bm k})]}{[\varepsilon_n({\bm k})-\varepsilon_l({\bm k})]^2 +(\hbar\delta)^2}
 - \frac{\left\{f[\varepsilon_l({\bm k})]-f[\varepsilon_m({\bm k})]\right\}[\varepsilon_l({\bm k})-\varepsilon_m({\bm k})]}{[\varepsilon_l({\bm k})-\varepsilon_m({\bm k})]^2+(\hbar\delta)^2} 
\right).
 \end{align}
\end{widetext}
In contrast to the linear response tensor $\chi_{i;j}$, there are complicated intraband and interband processes in both $\chi_{i;jk}^{\rm (Re)}$ and $\chi_{i;jk}^{\rm (Im)}$. 
It is noted that the nonlinear response function of electric field in the length gauge needs rederivation by applying $\hat{A}=-e\hat{r}$ ($\hat{r}$: position operator) in Eq.~\eqref{eq:2nd_c_susceptability}, since the matrix element of $\hat{r}$ in a periodic system includes differential operator of ${\bm k}$~\footnote{
For example, the second-order electric conductivity is obtained by applying $\hat{B}=-e\hat{\varv}$ ($\hat{\varv}$: velocity operator) and $\hat{A}=-e\hat{r}$, where the functional form is consistent to the $\omega \to 0$ limit of the second-order nonlinear optical conductivity~\cite{PhysRevX.11.011001}.
In this case, $\chi_{i;jk}$ has the  following $\delta$ dependences in the clean limit: $\delta^{-2}$ and $\delta^{0}$ 
with the time-inversion property of $\chi_{i;jk}^{\rm (Re)}$ for the Drude and intrinsic terms, respectively, and $\delta^{-1}$ with the time-inversion property of $\chi_{i;jk}^{\rm (Im)}$ for the Berry curvature dipole term~\cite{gao2019semiclassical}.}.

$\chi_{i;jk}^{\rm (Re)}$ and $\chi_{i;jk}^{\rm (Im)}$ show the following relations in the presence of time-reversal symmetry:
\begin{align}
\label{eq:nonlinear_t}
\chi_{i;jk}^{\rm (Re)} = t_{B_i}t_{A_j}t_{A_k}\chi_{i;jk}^{\rm (Re)}, \quad \chi_{i;jk}^{\rm (Im)} = -t_{B_i}t_{A_j}t_{A_k}\chi_{i;jk}^{\rm (Im)}. 
\end{align}
Equation~\eqref{eq:nonlinear_t} indicates that $\chi_{i;jk}^{\rm (Re)}$ and $\chi_{i;jk}^{\rm (Im)}$ are represented by E and ET (M and MT) multipoles and the M and MT (E and ET) multipoles, respectively, for $t_{B_i}t_{A_j}t_{A_k}=+1(-1)$. 
The multipoles relevant with the nonlinear response tensors are shown in Table~\ref{table:tensor_parity_MP_nonlinear}.
In the following, we show a correspondence between the nonlinear responses and multipoles by considering again the ferromagnetic and AFM orderings in the diamond structure in Figs.~\ref{fig:ferro_antiferro}(a) and (b).

\tabcolsep = 5pt
\begin{table*}[htb!]
\centering
\caption{
Correspondence between the second-order nonlinear response functions $\chi^{\rm (Re, Im)}$ and multipoles. 
Nonzero $\chi^{\rm (Re, Im)}$ is shown by the checkmark ($\checkmark$). 
 \label{table:tensor_parity_MP_nonlinear}}
\begin{tabular}{cccccccccccc}
\hline \hline 
& & & & & \multicolumn{5}{l}{$(\mathcal{P}, \mathcal{T}, \mathcal{PT})=$} & examples\\
 rank & & $t_{B_i} t_{A_j} t_{A_k}$ & & multipole & $(\bigcirc, \bigcirc, \bigcirc)$ & $(\times, \bigcirc, \times)$& $(\bigcirc, \times, \times)$  & $(\times, \times, \bigcirc)$ & $(\times, \times, \times)$ 
 & ($B\leftrightarrow F$) \\ \hline 
 
 3 & polar & $+1$ & $\chi^{\rm (Im)}$ & $T_{1m}, M_{2m}, T_{3m}$ & & & & $\checkmark$ &  $\checkmark$ & \\
 & & & $\chi^{\rm (Re)}$ & $Q_{1m}, G_{2m}, Q_{3m}$ &  & $\checkmark$ & & & $\checkmark$ & e.g. ($Q_{1m}\leftrightarrow Q_{0}, Q_{2m}$) \\
 \cline{3-11}
 
 & & $-1$ & $\chi^{\rm (Im)}$ & $Q_{1m}, G_{2m}, Q_{3m}$ &  & $\checkmark$ & & & $\checkmark$ & electric conductivity tensor  \\
 & & & $\chi^{\rm (Re)}$ & $T_{1m}, M_{2m}, T_{3m}$ & & & & $\checkmark$ & $\checkmark$ & ($T_{1m}\leftrightarrow Q_{0}, Q_{2m}$) \\ \cline{2-11}

& axial & $+1$ & $\chi^{\rm (Im)}$ & $M_{1m}, T_{2m}, M_{3m}$ & & & $\checkmark$ & & $\checkmark$ &    Nernst effect tensor\\
& & & $\chi^{\rm (Re)}$ & $G_{1m}, Q_{2m}, G_{3m}$ & $\checkmark$ & $\checkmark$ & $\checkmark$ & $\checkmark$ & $\checkmark$ & ($T_{1m}\leftrightarrow M_{0}, M_{2m}$) \\
\cline{3-11}

& & $-1$ & $\chi^{\rm (Im)}$ & $G_{1m}, Q_{2m}, G_{3m}$ & $\checkmark$ & $\checkmark$ & $\checkmark$ & $\checkmark$ & $\checkmark$ &magnetoelectric tensor \\
& & & $\chi^{\rm (Re)}$ & $M_{1m}, T_{2m}, M_{3m}$ & & & $\checkmark$ & & $\checkmark$ & $(M_{1m} \leftrightarrow Q_0, Q_{2m}$)\\
\hline

4 & polar & $+1$ & $\chi^{\rm (Im)}$ & $T_{0}, M_{1m}, T_{2m},$ & & & $\checkmark$ & & $\checkmark$ & electric striction tensor  \\
& & & & $M_{3m}, T_{4m}$ & & & & & &  $(Q_{0}, Q_{2m} \leftrightarrow Q_0, Q_{2m}$)\\

& & & $\chi^{\rm (Re)}$  & $Q_{0}, G_{1m}, Q_{2m},$ & $\checkmark$ & $\checkmark$ & $\checkmark$ & $\checkmark$ &  $\checkmark$ & \\
& & & & $G_{3m}, Q_{4m}$ \\
\cline{3-11}

& & $-1$ & $\chi^{\rm (Im)}$  &  $Q_{0}, G_{1m}, Q_{2m},$ & $\checkmark$ & $\checkmark$ & $\checkmark$ & $\checkmark$ &  $\checkmark$ &\\ 
& & & & $G_{3m}, Q_{4m}$ & & & & & & e.g. $(Q_{0}, Q_{2m} \leftrightarrow T_0, T_{2m}$) \\

& & & $\chi^{\rm (Re)}$& $T_{0}, M_{1m}, T_{2m},$ & & & $\checkmark$ & & $\checkmark$&  \\ 
& & & & $M_{3m}, T_{4m}$ \\
\cline{2-11}

& axial & $+1$ & $\chi^{\rm (Im)}$ & $M_{0}, T_{1m}, M_{2m},$ & & & & $\checkmark$ & $\checkmark$ & \\
& & & & $T_{3m}, M_{4m}$ & & & & & & e.g. $(Q_{0}, Q_{2m} \leftrightarrow G_0, G_{2m}$) \\

& & & $\chi^{\rm (Re)}$ & $G_{0}, Q_{1m}, G_{2m},$ & & $\checkmark$ & & &  $\checkmark$  &\\
& & & & $Q_{3m}, G_{4m}$ \\
\cline{3-11}
 
& & $-1$ & $\chi^{\rm (Im)}$ & $G_{0}, Q_{1m}, G_{2m},$ & & $\checkmark$ & & &  $\checkmark$ &  \\
& & & & $Q_{3m}, G_{4m}$ & & & & & & e.g. $(Q_{0}, Q_{2m} \leftrightarrow M_0, M_{2m}$) \\
 
& & & $\chi^{\rm (Re)}$ & $M_{0}, T_{1m}, M_{2m},$ & & & & $\checkmark$ & $\checkmark$ & \\
& & & & $T_{3m}, M_{4m}$ \\
 
\hline\hline
\end{tabular}
\end{table*}

\subsubsection{$\mathcal{P}$-symmetric magnetic structure \label{sec:res_nonlinea_uniform}}
In the $\mathcal{P}$-symmetric ferromagnetic structure in Fig.~\ref{fig:ferro_antiferro}(a), the nonlinear responses, such as the Nernst effect and the second-order nonlinear magnetoelectric effect, are expected.
In the case of the nonlinear magnetoelectric effect $M_i =\sum_{jk} \alpha_{ijk}E_jE_k$, 
the tensor $\alpha^{\rm (Re)}_{ijk}$ is induced by the even-parity M and MT multipoles in Table~\ref{table:tensor_parity_MP_nonlinear}. 
Since $M_z$ and $M_z^\alpha$ are active in the $4/mm'm'$ symmetry, the tensor component of $\alpha^{\rm (Re)}_{ijk}$ is represented by
\begin{align}
\alpha^{\rm (Re)}&=
\begin{pmatrix}
0 & 0 & \alpha_{zxx}^{\rm (E)}\\
0 & 0 & \alpha_{zxx}^{\rm (E)}\\
0 & 0 &\alpha_{zzz}^{\rm (E)} \\
0 & \alpha_{yyz}^{\rm (E)} & 0 \\
\alpha_{yyz}^{\rm (E)} & 0 & 0 \\
 0 & 0 & 0\\
\end{pmatrix}^{\rm T} \\
&\leftrightarrow
\begin{pmatrix}
0 & 0 & \tilde{M}'_z-2  M_{z}^\alpha\\
0 &0 &\tilde{M}'_z  -2M_{z}^\alpha\\
0 & 0 &\tilde{M}_z+4 M_{z}^\alpha \\
0 & -3 M_{z}-2 M_{z}^\alpha & 0 \\
  -3 M_{z}-2 M_{z}^\alpha & 0 & 0 \\
 0 & 0 & 0\\
\end{pmatrix}^{\rm T},
\end{align}
where there are three independent matrix elements in $\alpha^{\rm (Re)}$.

\subsubsection{$\mathcal{PT}$-symmetric magnetic structure \label{sec:res_nonlinea_staggered}}
In the AFM ordering with the $4'/m'mm'$ symmetry in Fig.~\ref{fig:ferro_antiferro}(b), 
the second-order nonlinear conductivity, $\sigma_{ijk}$, for $J_i =\sum_{jk} \sigma_{ijk}E_jE_k$ becomes nonzero, which reflects the lack of the spatial inversion symmetry. 
Among the two parts $\sigma_{ijk}^{\rm (Re)}$ and $\sigma_{ijk}^{\rm (Im)}$, $\sigma_{ijk}^{\rm (Re)}$ is induced in the presence of the odd-parity M and MT multipoles.
The finite tensor component of $\sigma_{ijk}^{\rm (Re)}$ is
shown as 
\begin{align}
\sigma^{\rm (Re)}&=
\begin{pmatrix}
0 & 0 & \sigma^{\rm (Re)}_{xxx}\\
0 & 0 &-\sigma^{\rm (Re)}_{xxx}\\
0 & 0 &0 \\
 0 & \sigma^{\rm (Re)}_{yzy} & 0 \\
-\sigma^{\rm (Re)}_{yzy} & 0 & 0 \\
 0 & 0 & 0 \\
\end{pmatrix}^{\rm T}
\notag\\
&\leftrightarrow
\begin{pmatrix}
0 & 0 & -2M_{xy}+2T_{z}^\beta\\
0 & 0 &2M_{xy}-2T_{z}^\beta\\
0 & 0 &0 \\
 0 & -M_{xy}-2 T_{z}^\beta & 0 \\
  M_{xy}+2 T_{z}^\beta & 0 & 0 \\
 0 & 0 & 0 \\
\end{pmatrix}^{\rm T}.
\end{align}
Thus, the nonlinear conductivity, which is known as the nonlinear Drude and/or the intrinsic terms, is expected in the AFM structure in Fig.~\ref{fig:ferro_antiferro}(b)~\cite{gao2019semiclassical}.

\section{Summary \label{sec:summary}}

In summary, we have accomplished the classification of multipole degrees of freedom in accordance with the irreducible (co)representations in 122 magnetic point groups. 
The completed multipole classification enables us to explore further functional multiferroic materials showing the extraordinary cross-correlated phenomena with exotic ordered states more than ordinary
ferromagnetic and AFM orderings.
Our multipole classification has mainly three advantages. 
One is the systematic identification of electronic order parameters to cover unconventional nematic, chiral, excitonic, loop-current, and anisotropic bond ordered states. 
The second is the predictability of overlooked physical phenomena under simple ferromagnetic and AFM orderings which may be hosted by the secondary multipole order parameters. 
The third is the intuitive understanding of the cross-correlated phenomena and nonlinear transports on the basis of the microscopic multipole couplings. 
The present comprehensive study will give an efficient way to explore further exotic physical phenomena induced by higher-rank multipoles, e.g., MT quadrupole and octupole.
Moreover, our result can be applied to investigate the functional magnetic materials in cooperation with
the material database like MAGNDATA~\cite{Gallego:ks5532}, as the multipole analysis is perfectly compatible to group theoretical analysis.

\begin{acknowledgments}
This research was supported by JSPS KAKENHI Grants Numbers JP19K03752, JP19H01834, JP21H01037, and by JST PREST (JPMJPR20L8). 
M.Y. is supported by a JSPS research fellowship and supported by JSPS KAKENHI (Grant No. JP20J12026).
\end{acknowledgments}

\appendix
\section{Multipole notation under cubic and hexagonal point groups \label{sec:ap_real}}

In the main text, we adopt the real expressions of $O_{lm}({\bm r})$ for both atomic multipoles in Eqs.~\eqref{eq:E_def}--\eqref{eq:ET_def} and cluster multipoles in Eqs.~\eqref{eq:Mc_def}--\eqref{eq:MTc_def}, which are given by~\cite{hutchings1964point}
 \begin{align}
 O_{l0}^{\rm (c)}({\bm r}) &\equiv O_{l0}({\bm r}),\\
 \label{eq:Olmc}
 O_{lm}^{\rm (c)}({\bm r}) &\equiv \frac{(-1)^m}{\sqrt{2}}[O_{lm}({\bm r})+O_{lm}^*({\bm r})], \\
 \label{eq:Olms}
 O_{lm}^{\rm (s)}({\bm r}) &\equiv i \frac{(-1)^m}{\sqrt{2}}[O_{lm}({\bm r})-O_{lm}^*({\bm r})].
 \end{align}
A linear combination of $ O_{lm}^{\rm (c)}({\bm r})$ and $ O_{lm}^{\rm (s)}({\bm r})$ gives the expression of multipoles under the magnetic point groups. 
In the case of the cubic and its subgroups, the cubic harmonics are used. 
We show the cubic expressions of $O_{lm}({\bm r})$ up to rank 4 as follows: 
the rank 0 is 
\begin{align}
O_0 = 1,
\end{align}
the rank 1 is 
\begin{align}
(O_x, O_y, O_z)= (x,y,z), 
\end{align}
the rank 2 is 
\begin{align}
O_u &= \frac{1}{2}(3z^2-r^2), \\
O_\varv &= \frac{\sqrt{3}}{2}(x^2-y^2), \\
(O_{yz}, O_{zx}, O_{xy}) &= \sqrt{3}(yz,zx,xy),
\end{align}
the rank 3 is 
\begin{align}
\label{eq:rank3_xyz}
O_{xyz} &= \sqrt{15}xyz, \\
\label{eq:rank3_alpha}
 \left(O_x^\alpha, O_y^\alpha, O_z^\alpha \right)  &= \frac{1}{2}\left( x(5x^2-3r^2), y(5y^2-3r^2), z(5z^2-3r^2) \right),  \\
\label{eq:rank3_beta}
 \left(O_x^\beta, O_y^\beta, O_z^\beta \right)   &= \frac{\sqrt{15}}{2}\left( x(y^2-z^2), y(z^2-x^2), z(x^2-y^2)\right), 
\end{align}
and the rank 4 is 
\begin{align}
\label{eq:rank4_4}
O_4 &= \frac{5\sqrt{21}}{12}\left( x^4+y^4+z^4-\frac{3}{5}r^4 \right), \\
\label{eq:rank4_4u}
O_{4u} &= \frac{7\sqrt{15}}{6}\left[ z^4-\frac{x^4+y^4}{2}-\frac{3}{7}r^2(3z^2-r^2)\right], \\
\label{eq:rank4_4v}
O_{4\varv} &= \frac{7\sqrt{5}}{4}\left[ x^4-y^4-\frac{6}{7}r^2(x^2-y^2) \right], \\
\label{eq:rank4_alpha}
\left(O_{4x}^\alpha, O_{4y}^\alpha, O_{4z}^\alpha \right) &= \frac{\sqrt{35}}{2}\left( yz(y^2-z^2), zx(z^2-x^2), xy(x^2-y^2) \right), \\
\label{eq:rank4_beta}
\left(O_{4x}^\beta, O_{4y}^\beta, O_{4z}^\beta \right) &= \frac{\sqrt{5}}{2} \left( yz(7x^2-r^2), zx(7y^2-r^2), xy(7z^2-r^2) \right), 
\end{align}
where we denote $O_{lm}({\bm r})\to O_{lm}$ for notation simplicity. 
In the case of the hexagonal and trigonal point groups, it is useful to replace 
$O_x^\alpha$, $O_y^\alpha$, $O_x^\beta$, $O_y^\beta$ with rank 3 in Eqs.~\eqref{eq:rank3_alpha} and \eqref{eq:rank3_beta} and $O_{lm}$ with rank 4 in Eqs.~\eqref{eq:rank4_4}--\eqref{eq:rank4_beta} by 
\begin{align}
O_{3a} &= \frac{\sqrt{10}}{4}x(x^2-3y^2), \\
O_{3b} &= \frac{\sqrt{10}}{4}y(3x^2-y^2), \\
(O_{3u}, O_{3\varv}) &=\frac{\sqrt{6}}{4} \left(x(5z^2-r^2),y(5z^2-r^2)\right), 
\end{align}
and 
\begin{align}
O_{40} &= \frac{1}{8}(35z^4-30z^2r^2+3r^4), \\
O_{4a} &=\frac{\sqrt{70}}{4} yz (3x^2-y^2), \\
O_{4b} &=\frac{\sqrt{70}}{4} zx (x^2-3y^2), \\
\left(O_{4u}^\alpha, O_{4\varv}^\alpha \right) &= \frac{\sqrt{10}}{4}\left( zx(7z^2-3r^2), yz(7z^2-3r^2) \right), \\
\left( O_{4u}^{\beta1}, O_{4\varv}^{\beta1} \right) &= \frac{\sqrt{35}}{8} \left(x^4-6x^2y^2+y^4, 4xy(x^2-y^2) \right), \\
\left(O_{4u}^{\beta2}, O_{4\varv}^{\beta2} \right) &= \frac{\sqrt{5}}{4} \left( (x^2-y^2)(7z^2-r^2), 2xy(7z^2-r^2) \right), 
\end{align}
with the use of the tesseral harmonics. 
It is noted that three components $\{O_{xyz}, O_z^\alpha, O_z^\beta\}$ among the rank-3 functions are common to the cubic harmonics in Eqs.~\eqref{eq:rank3_xyz}--\eqref{eq:rank3_beta}.

\section{Black and white point group and its unitary subgroup \label{sec:ap_MPG_subgroup}}

Type-(III) black and white point group ${\bm M}^{\rm(III)}$ is represented by using type-(I) crystallographic
point group ${\bm G}$ and its halving unitary subgroup ${\bm H}$ in Eq.~\eqref{eq:BWPG};
${\bm G}$ is given by removing the prime symbol in ${\bm M}^{\rm (III)}$. 
The correspondence between ${\bm M}^{\rm (III)}$ and ${\bm H}$ is summarized in Table~\ref{table:subgroup}
~\cite{tavger1956magnetic,dimmock1964symmetry,Hamermesh:100343,bradley2009mathematical,dimmock1962irreducible,cracknell1966corepresentations}.

\begin{table*}[htb!]
\centering
\caption{ 
List of 58 black and white point groups ${\bm M}^{\rm (III)}$.
The parentheses represent the corresponding unitary subgroup ${\bm H}$.
\label{table:subgroup}}
\begin{tabular}{ccccccc}
\hline\hline
cubic & tetragonal & orthorhombic & monoclinic & triclinic & hexagonal & trigonal \\ \hline
$m'\bar{3}'m'$ ($432$) &$4/m'm'm'$ ($422$) &  $m'm'm'$ ($222$) & $2'/m'$ ($\bar{1}$) & $\bar{1}'$ ($1$) & $6/m'm'm'$ ($622$) & $\bar{3}m'$ ($\bar{3}$) \rule[0pt]{0pt}{8pt} \\
$m\bar{3}m'$ ($m\bar{3}$) & $4/mm'm'$ ($4/m$) & $m'm'm$ ($2/m$)&  $2/m'$ ($2$) & &$6/mm'm'$ ($6/m$) & $\bar{3}'m'$ ($32$)  \\
$m'\bar{3}'m$ ($\bar{4}3m$) & $4'/m'm'm$ ($\bar{4}2m$) & 
$m'mm$ ($2mm$)
 & $2'/m$ ($m$) & & $6'/m'mm'$ ($\bar{3}m$) & $\bar{3}'m$ ($3m$) \\
$\bar{4}'3m'$ ($23$) & $4'/mm'm$ ($mmm$) & $2'2'2$ ($2$) & $m'$ ($1$) & &$6'/mmm'$ ($\bar{6}m2$) & $32'$ ($3$)\\
$4'32'$ ($23$) & $4/m'mm$ ($4mm$) & $m'm'2$ ($2$) & $2'$ ($1$) & & $6/m'mm$ ($6mm$) & $3m'$ ($3$) \\
$m'\bar{3}'$ ($23$) & $42'2'$ ($4$) &  $m'm2'$ ($m$) & & & $62'2'$ ($6$) & $\bar{3}'$ ($3$) \\
& $4'22'$ ($222$) & & & & $6'22'$ ($32$) & \\
& $\bar{4}2'm'$ ($\bar{4}$) & & & & $\bar{6}m'2'$ ($\bar{6}$)  \\
& $\bar{4}'2m'$ ($222$) & & & & $\bar{6}'m2'$ ($3m$)&\\
& $\bar{4}'2'm$ ($mm2$) & & & & $\bar{6}'m'2$ ($32$) &\\
& $4m'm'$ ($4$) & & & & $6m'm'$ ($6$) &  \\
& $4'm'm$ ($mm2$) & & & & $6'mm'$ ($3m$) & \\
& $4'/m'$ ($\bar{4}$) & & & & $6'/m'$ ($\bar{3}$) & \\
& $4/m'$ ($4$) & & & & $6/m'$ ($6$) & \\
& $4'/m$ ($2/m$) & & & & $6'/m$ ($\bar{6}$) & \\
& $\bar{4}'$ ($2$)& & & & $\bar{6}'$ ($3$) & \\
& $4'$ ($2$) & & & & $6'$ ($3$) & \\
\hline \hline
\end{tabular}
\end{table*}

\section{Multipole classification \label{sec:ap_classification}}

We present the complete tables for the multipole classification in the 32 gray point groups in Sec.~\ref{sec:ap_classification_GPG} and 58 black and white point groups in Sec.~\ref{sec:ap_classification_BWPG}.

\subsection{Gray point groups \label{sec:ap_classification_GPG}}
We show the M and MT multipole classification in all the 32 gray point groups in Table~\ref{table:gray_m3m1} in Sec.~\ref{sec:M_MT_representation} for $m\bar{3}m1'$ and the others in Tables~\ref{table:gray_cubic}--\ref{table:gray_trigonal_2}. 
The primary, secondary, and tertiary axes in Table~\ref{table:op_axis} are used for the point-group operations, unless otherwise mentioned. 
When the primary axis is different from that in Table~\ref{table:op_axis}, secondary and tertiary axes should be transformed by cyclic.

The classifications for cubic systems are shown in Table~\ref{table:gray_cubic},
 tetragonal systems in Tables~\ref{table:gray_tetragonal_1} and \ref{table:gray_tetragonal_2}, orthorhombic systems in Table~\ref{table:gray_ortho}, monoclinic systems in Table~\ref{table:gray_monocli}, triclinic systems in Table~\ref{table:gray_tricli}, hexagonal systems in Tables~\ref{table:gray_hexagonal_1} and \ref{table:gray_hexagonal_2},
 and trigonal systems in Tables~\ref{table:gray_trigonal_1} and \ref{table:gray_trigonal_2}.

\begin{table}[htb!]
\centering
\caption{
Primary, secondary, and tertiary axes with respect to the symmetry operations in the Cartesian coordinates. 
 \label{table:op_axis}}
\begin{tabular}{cccc}
\hline\hline
& primary & secondary & tertiary \\ \hline
cubic & $\left<100\right>$ & $\left<111\right>$ & $\left<110\right>$  \\
tetragonal & $[001]$ & $[100]$ & $[110]$ \\
orthorhombic & $[100]$ & $[010]$ & $[001]$ \\
monoclinic & $[010]$ &  & \\
triclinic &  &  & \\
hexagonal & $[001]$ & $[100]$ & $[010]$ \\
trigonal & $[001]$ & $[010]$ & \\
\hline \hline
\end{tabular}
\end{table}

\begin{table}[h!]
\centering
\caption{IRREPs of magnetic (M) and magnetic toroidal (MT) multipoles ($l\leq 4$) in the cubic gray point groups.
The classification for
$m\bar{3}m1'$ is shown in Table~\ref{table:gray_m3m1}.
\label{table:gray_cubic}}
\begin{tabular}{ccccc}
\multicolumn{5}{l}{$4321'$} \\
\hline\hline
IRREP & MT & M & MPG & P. axis \\
\hline 
A$_{\rm 1}^-$ & $T_0, T_4$ & $M_0, M_4$ & $432$ & $\left<100 \right>$ \\
A$_{\rm 2}^-$ & $T_{xyz}$ & $M_{xyz}$ & $4'32'$ & $\left<100 \right>$ \\
E$^-$ & $T_u, T_{4u}$ & $M_u, M_{4u}$ & $422$ & $[001]$ \\
 & $T_\varv, T_{4\varv}$ & $M_\varv, M_{4\varv}$ & $4'22'$ & $[001]$ \\
T$_{\rm 1}^-$ & $T_x, T_x^\alpha, T_{4x}^\alpha$ & $M_x, M_x^\alpha, M_{4x}^\alpha$ & $42'2'$ & $[100]$ \\
  & $T_y, T_y^\alpha, T_{4y}^\alpha$ & $M_y, M_y^\alpha, M_{4y}^\alpha$ & $42'2'$ & $[010]$ \\
\hline \hline
\end{tabular}
\end{table}

\begin{table}[h!]
\centering
\caption{
IRREPs of magnetic (M) and magnetic toroidal (MT) multipoles in the tetragonal gray point groups with twofold-rotational axes perpendicular to the primary axis or vertical reflection planes. 
\label{table:gray_tetragonal_1}}
\begin{tabular}{ccccc}
\multicolumn{4}{l}{$4/mmm1'$} \\
\hline\hline
IRREP & MT & M & MPG & P. axis \\
\hline 
A$_{\rm 1g}^-$ &  $T_0, T_u, T_4, T_{4u}$ && $4/mmm$ & $[001]$ \\
A$_{\rm 2g}^-$ & $T_{4z}^\alpha$ & $M_z, M_z^\alpha$ & $4/mm'm'$ & $[001]$ \\
B$_{\rm 1g}^-$ & $T_\varv, T_{4\varv}$ & $M_{xyz}$ & $4'/mmm'$ & $[001]$ \\
B$_{\rm 2g}^-$ & $T_{xy}, T_{4z}^\beta$ & $M_{z}^\beta$ & $4'/mm'm$ & $[001]$ \\
E$_{\rm g}^-$ & $T_{yz}, T_{4x}^\alpha, T_{4x}^\beta$ & $M_x, M_x^\alpha, M_x^\beta$ & $mm'm'$ & $[100]$ \\
 & $T_{zx}, T_{4y}^\alpha, T_{4y}^\beta$ & $M_y, M_y^\alpha, M_y^\beta$ & $m'mm'$ & $[100]$ \\
\hline\hline
 \end{tabular}
\end{table}

\begin{table}[h!]
\centering
\caption{IRREPs of magnetic (M) and magnetic toroidal (MT) multipoles in the tetragonal gray point groups without twofold-rotational axes perpendicular to the primary axis or vertical reflection planes. 
\label{table:gray_tetragonal_2}}
\begin{tabular}{ccccc}
\multicolumn{4}{l}{$4/m1'$} \\
\hline\hline
IRREP & MT & M & MPG & P. axis \\
\hline 
A$_{\rm g}^-$ & $T_0, T_u$ & $M_z, M_z^\alpha$ & $4/m$ & $[001]$\\
& $T_4, T_{4u}, T_{4z}^\alpha$ & \\
B$_{\rm g}^-$ & $T_\varv, T_{xy}, T_{4\varv}, T_{4z}^\beta$ & $M_{xyz}, M_z^\beta$ & $4'/m$ & $[001]$\\
E$_{\rm g}$ & $T_{yz}, T_{4x}^\alpha, T_{4x}^\beta$ & $M_x, M_x^\alpha, M_x^\beta$ & $2'/m'$ & [001] \\
 & $T_{zx}, T_{4y}^\alpha, T_{4y}^\beta$ & $M_y, M_y^\alpha, M_y^\beta$ & $2'/m'$ & [001] \\
\hline
A$_{\rm u}^-$ & $T_z, T_z^\alpha$ & $M_0, M_u$ & $4/m'$ & $[001]$ \\
\hline\hline
\end{tabular}
\end{table}

\begin{table}[h!]
\centering
\caption{IRREPs of magnetic (M) and magnetic toroidal (MT) multipoles in orthorhombic gray point groups. 
\label{table:gray_ortho}}
\begin{tabular}{ccccc}
\multicolumn{4}{l}{$mmm1'$} \\
\hline\hline
IRREP & MT & M & MPG & P. axis \\
\hline 
A$_{\rm g}^-$ & $T_0, T_u, T_\varv$ & $M_{xyz}$ & $mmm$ & $[100]$ \\
& $T_{4}, T_{4u}, T_{4\varv}$ & \\
B$_{\rm 1g}^-$ & $T_{xy}, T_{4z}^\alpha, T_{4z}^\beta$ & $M_z, M_z^\alpha, M_z^\beta$ & $m'm'm$ & $[100]$  \\
B$_{\rm 2g}^-$ & $T_{zx}, T_{4y}^\alpha, T_{4y}^\beta$ & $M_y, M_y^\alpha, M_y^\beta$ & $m'mm'$ & $[100]$ \\
B$_{\rm 3g}^-$ & $T_{yz}, T_{4x}^\alpha, T_{4x}^\beta$ & $M_x, M_x^\alpha, M_x^\beta$ & $mm'm'$ & $[100]$ \\
\hline
A$_{\rm u}^-$ & $T_{xyz}$ & $M_0, M_u, M_\varv$ & $m'm'm'$ & $[100]$ \\
\hline\hline
\end{tabular}
\end{table}

\begin{table}[h!]
\centering
\caption{IRREPs of magnetic (M) and magnetic toroidal (MT) multipoles in monoclinic gray point groups. 
\label{table:gray_monocli}}
\begin{tabular}{ccccc}
\multicolumn{4}{l}{$2/m1'$} \\
\hline\hline
IRREP & MT & M & MPG & P. axis \\
\hline 
A$_{\rm g}^-$ & $T_0, T_u, T_\varv, T_{zx}$ & $M_y$ & $2/m$ & $[010]$ \\
& $T_4, T_{4u}, T_{4\varv}$ & $ M_{xyz}, M_y^\alpha, M_y^\beta$ \\
& $T_{4y}^\alpha, T_{4y}^\beta$ & \\
B$_{\rm g}^-$ & $T_{yz}, T_{xy}$ & $M_x, M_z$ & $2'/m'$ & $[010]$ \\
& $T_{4x}^\alpha, T_{4z}^\alpha, T_{4x}^\beta, T_{4z}^\beta$ & $M_x^\alpha, M_z^\alpha, M_x^\beta, M_z^\beta$ & \\
\hline
A$_{\rm u}^-$ & $T_y$ & $M_0, M_u, M_\varv, M_{zx}$ & $2/m'$ & $[010]$ \\
\hline\hline
\end{tabular}
\end{table}

\begin{table}[h!]
\centering
\caption{IRREPs of magnetic (M) and magnetic toroidal (MT) multipoles in triclinic gray point groups. 
\label{table:gray_tricli}}
\begin{tabular}{ccccc}
\multicolumn{4}{l}{$\bar{1}1'$ } \\
\hline\hline
IRREP & MT & M & MPG & P. axis \\
\hline 
A$_{\rm g}^-$ & $T_0, T_u, T_\varv$ & $M_x, M_y, M_z$ & $\bar{1}$ &  \rule[0pt]{0pt}{8pt}\\
& $T_{yz}, T_{zx}, T_{xy}$ & $M_{xyz}, M_x^\alpha, M_y^\alpha, M_z^\alpha$ \\
& $T_4, T_{4u}, T_{4\varv}$ & $M_x^\beta, M_y^\beta, M_z^\beta$ \\
& $T_{4x}^\alpha, T_{4y}^\alpha, T_{4z}^\alpha$ & \\
& $T_{4x}^\beta, T_{4y}^\beta, T_{4z}^\beta$ & \\ 
\hline
A$_{\rm u}^-$ & $T_x, T_y, T_z$ & $M_0, M_u, M_\varv$ & $\bar{1}'$ & \rule[0pt]{0pt}{8pt} \\
& $T_{xyz}, T_x^\alpha, T_y^\alpha, T_z^\alpha$ & $M_{yz}, M_{zx}, M_{xy}$ \\
& $T_x^\beta, T_y^\beta, T_z^\beta$ & $M_4, M_{4u}, M_{4\varv}$ \\
& & $M_{4x}^\alpha, M_{4y}^\alpha, M_{4z}^\alpha$ \\
& & $M_{4x}^\beta, M_{4y}^\beta, M_{4z}^\beta$ \\
\hline\hline
\\
\multicolumn{4}{l}{$11'$ } \\
\hline\hline
IRREP & MT & M & MPG & P. axis \\
\hline 
A$^-$ & $T_0, T_x, T_y, T_z$ & $M_0, M_x, M_y, M_z$ & $1$ &\\
& $T_u, T_\varv$ & $M_u, M_\varv$ \\
& $T_{yz}, T_{zx}, T_{xy}$ & $M_{yz}, M_{zx}, M_{xy}$ \\
& $T_{xyz}, T_x^\alpha, T_y^\alpha, T_z^\alpha$ & $M_{xyz}, M_x^\alpha, M_y^\alpha, M_z^\alpha$ \\
& $T_x^\beta, T_y^\beta, T_z^\beta$ & $M_x^\beta, M_y^\beta, M_z^\beta$ \\
& $T_4, T_{4u}, T_{4\varv}$ & $M_4, M_{4u}, M_{4\varv}$ \\
& $T_{4x}^\alpha, T_{4y}^\alpha, T_{4z}^\alpha$ & $M_{4x}^\alpha, M_{4y}^\alpha, M_{4z}^\alpha$ \\
& $T_{4x}^\beta, T_{4y}^\beta, T_{4z}^\beta$ & $M_{4x}^\beta, M_{4y}^\beta, M_{4z}^\beta$ \\
\hline\hline
\end{tabular}
\end{table}

\tabcolsep = 1pt
\begin{table}[h!]
\centering
\caption{
IRREPs of magnetic (M) and magnetic toroidal (MT) multipoles in the hexagonal gray point groups with twofold-rotational axes perpendicular to the primary axis or vertical reflection planes. 
\label{table:gray_hexagonal_1}}
\begin{tabular}{ccccc}
\multicolumn{4}{l}{$6/mmm1'$} \\
\hline\hline
IRREP & MT & M & MPG & P. axis \\
\hline 
A$_{\rm 1g}^-$ &  $T_0, T_u, T_{40}$ && $6/mmm$ & $[001]$ \\
A$_{\rm 2g}^-$ && $M_z, M_z^\alpha$ & $6/mm'm'$ & $[001]$ \\
B$_{\rm 1g}^-$ & $T_{4a}$ & $M_{3a}$ & $6'/m'mm'$ & $[001]$ \\
B$_{\rm 2g}^-$ & $T_{4b}$ & $M_{3b}$ & $6'/m'm'm$ & $[001]$ \\
E$_{\rm 1g}^-$ & $T_{yz}, T_{4\varv}^\alpha$ & $M_x, M_{3u}$ & $mm'm'$ & $[100]$ \\
 & $T_{zx}, T_{4u}^\alpha$ & $M_y, M_{3\varv}$ & $m'mm'$ & $[100]$ \\
\hline\hline
 \end{tabular}
\end{table}

\begin{table}[h!]
\centering
\caption{
IRREPs of magnetic (M) and magnetic toroidal (MT) multipoles in the hexagonal gray point groups without twofold-rotational axes perpendicular to the primary axis or vertical reflection planes.
\label{table:gray_hexagonal_2}}
\begin{tabular}{ccccc}
\multicolumn{4}{l}{$6/m1'$} \\
\hline\hline
IRREP & MT & M & MPG & P. axis \\
\hline 
A$_{\rm g}^-$ & $T_0, T_u, T_{40}$ & $M_z, M_z^\alpha$ & $6/m$ & $[001]$ \\
B$_{\rm g}^-$ & $T_{4a}, T_{4b}$ & $M_{3a}, M_{3b}$ & $6'/m'$ & $[001]$ \\
E$_{\rm 1g}$ & $T_{yz}, T_{4\varv}^\alpha$ & $M_x, M_{3u}$ & $2'/m'$ & [001] \\
 & $T_{zx}, T_{4u}^\alpha$ & $M_y, M_{3\varv}$ & $2'/m'$ & [001] \\
 E$_{\rm 2g}$ & $T_\varv, T_{4u}^{\beta1}, T_{4u}^{\beta2}$ & $M_{xyz}$ & $2/m$ & $[001]$ \\
 & $T_{xy}, T_{4\varv}^{\beta1}, T_{4\varv}^{\beta2}$ & $M_z^\beta$ & $2/m$ & $[001]$ \\
\hline\hline
\end{tabular}
\end{table}

\begin{table}[h!]
\centering
\caption{
IRREPs of magnetic (M) and magnetic toroidal (MT) multipoles in the trigonal gray point groups with twofold-rotational axes perpendicular to the primary axis or vertical reflection planes. 
\label{table:gray_trigonal_1}}
\begin{tabular}{ccccc}
\multicolumn{4}{l}{$\bar{3}m1'$} \\
\hline\hline
IRREP & MT & M & MPG & P. axis \\
\hline 
A$_{\rm 1g}^-$ &  $T_0, T_u, T_{40}, T_{4b}$ & $M_{3b}$ & $\bar{3}m$ & $[001]$ \rule[0pt]{0pt}{8pt} \\
A$_{\rm 2g}^-$ & $T_{4a}$ & $M_z, M_z^\alpha, M_{3a}$ & $\bar{3}m'$ & $[001]$  \\
E$_{\rm g}^-$ & $T_{yz}, T_{xy}$ & $M_x, M_{3u}, M_z^\beta$ & $2'/m'$ & $[010]$ \\
& $T_{4\varv}^\alpha, T_{4\varv}^{\beta1}, T_{4\varv}^{\beta2}$ & \\
& $T_{zx}, T_\varv$ & $M_y, M_{3\varv}, M_{xyz}$ & $2/m$ & $[010]$ \\
& $T_{4u}^\alpha, T_{4u}^{\beta1}, T_{4u}^{\beta2}$ & \\
\end{tabular}
\end{table}

\begin{table}[h!]
\centering
\caption{
IRREPs of magnetic (M) and magnetic toroidal (MT) multipoles in the trigonal gray point groups without twofold-rotational axes perpendicular to the primary axis or vertical reflection planes. 
\label{table:gray_trigonal_2}}
\begin{tabular}{ccccc}
\multicolumn{4}{l}{$\bar{3}1'$} \\
\hline\hline
IRREP & MT & M & MPG & P. axis \\
\hline 
A$_{\rm g}^-$ &  $T_0, T_u$ & $M_z, M_z^\alpha, M_{3a}, M_{3b}$ & $\bar{3}$ & $[001]$ \\
&  $ T_{40}, T_{4a}, T_{4b}$ &  &   \\
E$_{\rm g}$ & $T_{yz}, T_{xy}$ & $M_x, M_{3u}, M_z^\beta$ & $\bar{1}$ &\\
& $T_{4\varv}^\alpha, T_{4\varv}^{\beta1}, T_{4\varv}^{\beta2}$ & \\
& $T_{zx}, T_\varv$ & $M_y, M_{3\varv}, M_{xyz}$ & $\bar{1}$ & \\
& $T_{4u}^\alpha, T_{4u}^{\beta1}, T_{4u}^{\beta2}$ & \\
\hline
A$_{\rm 1u}^-$ & $T_z, T_z^\alpha, T_{3a}, T_{3b}$ &  $M_0, M_u$  & $\bar{3}'$ & $[001]$  \\
&  & $M_{40}, M_{4a}, M_{4b}$ & \\
E$_{\rm u}$  & $T_x, T_{3u}, T_z^\beta$ & $M_{yz}, M_{xy}$ & $\bar{1}'$ &\\
& & $M_{4\varv}^\alpha, M_{4\varv}^{\beta1}, M_{4\varv}^{\beta2}$ & \\
 & $T_y, T_{3\varv}, T_{xyz}$ & $M_{zx}, M_\varv$ & $\bar{1}'$ &\\
 & & $M_{4u}^\alpha, M_{4u}^{\beta1}, M_{4u}^{\beta2}$ & \\
\hline\hline
\\
\multicolumn{4}{l}{$31'$} \\
\hline\hline
IRREP & MT & M & MPG & P. axis \\
\hline 
A$^-$ & $T_0, T_z, T_u$ &  $M_0, M_z, M_u$ & $3$ & $[001]$ \\
& $T_z^\alpha, T_{3a}, T_{3b}$ & $M_z^\alpha, M_{3a}, M_{3b}$ & \\
& $T_{40}, T_{4a}, T_{4b}$ & $M_{40}, M_{4a}, M_{4b}$ &  \\
E & $T_x, T_{yz}, T_{xy}$ & $M_x, M_{yz}, M_{xy} $ & $1$ & \\
& $T_{3u}, T_z^\beta$ & $M_{3u}, M_z^\beta$ \\
& $T_{4\varv}^\alpha, T_{4\varv}^{\beta1}, T_{4\varv}^{\beta2}$ & $M_{4\varv}^\alpha, M_{4\varv}^{\beta1}, M_{4\varv}^{\beta2}$ & \\
& $T_y, T_{zx}, T_\varv$ & $M_y, M_{zx}, M_\varv$ & $1$ & \\
& $T_{3\varv}, T_{xyz}$ & $M_{3\varv}, M_{xyz}$ \\
& $T_{4u}^\alpha, T_{4u}^{\beta1}, T_{4u}^{\beta2}$ & $M_{4u}^\alpha, M_{4u}^{\beta1}, M_{4u}^{\beta2}$ & \\
\hline\hline
\end{tabular}
\end{table}

\subsection{Black and white point groups \label{sec:ap_classification_BWPG}}

The classification of four types of multipoles, E, ET, MT, and M multipoles, in 58 black and white point groups is shown. 
The classification for cubic systems is shown in Table~\ref{table:BW_cubic}, tetragonal systems in Tables~\ref{table:BW_tetragonal_1} and \ref{table:BW_tetragonal_2}, orthorhombic systems in Table~\ref{table:BW_orthorhombic}, monoclinic and triclinic systems in Table~\ref{table:BW_monocli_tricli}, hexagonal systems in Tables~\ref{table:BW_hexagonal_1} and \ref{table:BW_hexagonal_2}, and trigonal systems in Table~\ref{table:BW_trigonal}.

\tabcolsep = 3pt
\begin{table}[h!]
\centering
\caption{
IRREPs of multipoles in the cubic black and white point groups. 
Corresponding unitary subgroups are also shown.
The sign $\pm$ represents the parity with respect to the anti-unitary operation in the third row, where the direction of the rotational axis or mirror plane are shown in the next row (see Appendix~\ref{sec:ap_MPG_corepresentation}). 
$\{ X_{u}+ iX_{\varv}, X_{u}- iX_{\varv}\}$ and $\{ X_{4u}+ iX_{4\varv}, X_{4u}- iX_{4\varv}\} $ are the basis of E representations in $m\bar{3}m'$, where $X=Q, G, T$, and $M$ for ${\rm E}_{\rm g}^{\rm (1,2)+}, {\rm E}_{\rm u}^{\rm (1,2)+}, {\rm E}_{\rm g}^{\rm (1,2)-}$, and ${\rm E}_{\rm u}^{\rm (1,2)-}$, respectively.
They are also basis of ${\rm E}^{\rm (1,2)+}$($X=Q, G$) and ${\rm E}^{\rm (1,2)-}$($X=T, M$) in $4'32'$, whereas ${\rm E}^{\rm (1,2)+}$($X=Q, M$) and ${\rm E}^{\rm (1,2)-}$($X=G, T$) in $\bar{4}'3m'$.
 \label{table:BW_cubic}}
\vspace{2mm}
\begin{tabular}{cccccccc}\hline \hline
\multicolumn{2}{c}{magnetic point group} &  $m'\bar{3}'m$ & $m\bar{3}m'$ & $m'\bar{3}'m'$ & $4'32'$ &$\bar{4}'3m'$ & $m'\bar{3}'$ \rule[0pt]{0pt}{8pt} \\\cline{3-8}
\multicolumn{2}{c}{unitary subgroup} &   $\bar{4}3m$ & $m\bar{3}$ & $432$ & $23$ & $23$ & $23$ \rule[0pt]{0pt}{8pt} \\
\multicolumn{2}{c}{anti-unitary operation} & $\theta I$ & $\theta C'_{2}$ & $\theta I$ & $\theta C'_{2}$ & $\theta \sigma_{d}$ & $\theta I$ \\ 
\multicolumn{2}{c}{} & & $[110]$ & & $[110]$ & $\perp[110]$ \\\hline
E & ET &   \\ \cline{1-2}
$Q_0,Q_4$ &  & A$_{\rm 1}^+$ & A$_{\rm g}^+$ & A$_1^+$ &  A$^+$ & A$^+$ & A$^+$ \\
 & $G_{xyz}$ & A$_{\rm 2}^+$ & A$_{\rm g}^-$ & A$_2^+$ & A$^-$ & A$^-$ & A$^+$ \\
$Q_u, Q_{4u}$ & & E$^+$ & E$_{\rm g}^{(1, 2)+}$ & E$^+$ & E$^{(1,2)+}$ & E$^{(1,2)+}$ & E \\
$Q_\varv, Q_{4\varv}$ & & &  &  &  &  &   \\
$Q_{4x}^\alpha$ & $G_x, G_x^\alpha$ & T$_1^+$ & T$_{\rm g}^+$ & T$_1^+$&T$^+$ & T$^+$ & T$^+$ \\
$Q_{4y}^\alpha$ & $G_y, G_y^\alpha$ & & & & & & \\
$Q_{4z}^\alpha$ & $G_z, G_z^\alpha$ & & & & & & \\
$Q_{yz}, Q_{4x}^\beta$ & $G_x^\beta$ & T$_2^+$ & T$_{\rm g}^-$ & T$_2^+$ & T$^-$ & T$^-$ & T$^+$ \\
$Q_{zx}, Q_{4y}^\beta$ & $G_y^\beta$ & & & & & & \\
$Q_{xy}, Q_{4z}^\beta$ & $G_z^\beta$ & & & & & & \\
\hline
  & $G_0,G_4$ & A$_2^-$ & A$_{\rm u}^+$ & A$_1^-$ & A$^+$ & A$^-$ & A$^-$ \\
 $Q_{xyz}$ & & A$_1^-$ & A$_{\rm u}^-$ & A$_2^-$ & A$^-$ & A$^+$ & A$^-$ \\
 & $G_u, G_{4u}$ & E$^-$ &  E$_{\rm u}^{(1,2)+}$ & E$^-$ &  E$^{(1,2)+}$ & E$^{(1,2)-}$ & E  \\
 & $G_\varv, G_{4\varv}$ & &  & &  &  &  \\
$Q_x, Q_x^\alpha$ & $G_{4x}^\alpha$ & T$_2^-$ & T$_{\rm u}^+$ & T$_1^-$ & T$^+$ & T$^-$ & T$^-$ \\
$Q_y, Q_y^\alpha$ & $G_{4y}^\alpha$ & & & & & & \\
 $Q_z, Q_z^\alpha$ & $G_{4z}^\alpha$ & & & & & & \\
 $Q_x^\beta$ & $G_{yz}, G_{4x}^\beta$ & T$_1^-$ & T$_{\rm u}^-$ & T$_2^-$ & T$^-$ & T$^+$ & T$^-$ \\
 $Q_y^\beta$ & $G_{zx}, G_{4y}^\beta$ & & & & & & \\
 $Q_z^\beta$ & $G_{xy}, G_{4z}^\beta$ & & & & & & \\ 
 \hline
MT & M & \\\cline{1-2}
$T_0,T_4$ &  & A$_1^-$ & A$_{\rm g}^-$ & A$_1^-$ & A$^-$ & A$^-$ & A$^-$ \\
 & $M_{xyz}$ & A$_2^-$ & A$_{\rm g}^+$ & A$_2^-$ & A$^+$ & A$^+$ & A$^-$ \\
$T_u, T_{4u}$ & &  E$^-$ & E$_{\rm g}^{(1,2)-}$ & E$^-$ & E$^{(1,2)-}$ &  E$^{(1,2)-}$ &  E \\
$T_\varv, T_{4\varv}$ & & &  & &  & & \\
$T_{4x}^\alpha$ & $M_x, M_x^\alpha$ & T$_1^-$ & T$_{\rm g}^-$ & T$_1^-$ &  T$^-$ & T$^-$ & T$^-$ \\
$T_{4y}^\alpha$ & $M_y, M_y^\alpha$ & & & & & & \\
$T_{4z}^\alpha$ & $M_z, M_z^\alpha$ & & & & & & \\
$T_{yz}, T_{4x}^\beta$ & $M_x^\beta$ & T$_2^-$ & T$_{\rm g}^+$ & T$_2^-$ &  T$^+$ & T$^+$ & T$^-$ \\
$T_{zx}, T_{4y}^\beta$ & $M_y^\beta$ & & & & & & \\
$T_{xy}, T_{4z}^\beta$ & $M_z^\beta$ & & & & & & \\
\hline
  & $M_0,M_4$ & A$_2^+$ & A$_{\rm u}^-$ & A$_1^+$ & A$^-$ & A$^+$ & A$^+$ \\
 $T_{xyz}$ &  & A$_1^+$ & A$_{\rm u}^+$ & A$_2^+$ & A$^+$ & A$^-$ & A$^+$ \\
 & $M_u, M_{4u}$ & E$^+$ & E$_{\rm u}^{(1,2)-}$ & E$^+$ & E$^{(1,2)-}$ & E$^{(1,2)+}$ & E \\
 & $M_\varv, M_{4\varv}$ & & & & & & \\
$T_x, T_x^\alpha$ & $M_{4x}^\alpha$ & T$_2^+$ & T$_{\rm u}^-$ & T$_1^+$ & T$^-$ & T$^+$ & T$^+$ \\
$T_y, T_y^\alpha$ & $M_{4y}^\alpha$ & & & & & & \\
 $T_z, T_z^\alpha$ & $M_{4z}^\alpha$ & & & & & & \\
 $T_x^\beta$ & $M_{yz}, M_{4x}^\beta$ & T$_1^+$ & T$_{\rm u}^+$ & T$_2^+$ & T$^+$ & T$^-$ & T$^+$ \\
 $T_y^\beta$ & $M_{zx}, M_{4y}^\beta$ & & & & & & \\
 $T_z^\beta$ & $M_{xy}, M_{4z}^\beta$ & & & & & & \\ 
 \hline\hline
\end{tabular}
\end{table}

\tabcolsep = 3pt
\begin{table*}[h!]
\centering
\caption{
IRREPs of multipoles in tetragonal black and white point groups except for $4'/m'$, $4/m'$, $4'/m$,  $\bar{4}'$, and $4'$ (see also Table~\ref{table:BW_tetragonal_2}).
$\{X_x+iX_y, X_x-iX_y\}$, $\{X_{yz}-iX_{zx}, X_{yz}+iX_{zx}\}$, $\{X_x^\alpha+iX_y^\alpha, X_x^\alpha-iX_y^\alpha\}$, $\{X_x^\beta-iX_y^\beta, X_x^\beta+iX_y^\beta\}$, $\{X_{4x}^\alpha+iX_{4y}^\alpha, X_{4x}^\alpha-iX_{4y}^\alpha\}$, and $\{X_{4x}^\beta-iX_{4y}^\beta, X_{4x}^\beta+iX_{4y}^\beta\}$ ($X=Q,G, T, M$) are the basis of ${\rm E}_{\rm g/u}^{\rm (1,2)\pm}$ in $4/mm'm'$, ${\rm E}^{(1,2)\pm}$ in $42'2'$, $4m'm'$, and $\bar{4}2'm'$.
 \label{table:BW_tetragonal_1}}
\vspace{2mm}
\begin{tabular}{cccccccccccccc}\hline \hline 
\multicolumn{2}{c}{magnetic point group} &  $4/m'm'm'$ & $4/m'mm$ & $4'/m'm'm$ & $4/mm'm'$ & $4'/mmm'$ & $42'2'$ & $4'22'$ & $4m'm'$ & $4'mm'$ & $\bar{4}2'm'$ & $\bar{4}'m2'$ & $\bar{4}'2m'$ \rule[0pt]{0pt}{8pt} \\ \cline{3-14}

\multicolumn{2}{c}{unitary subgroup} & $422$ & $4mm$ & $\bar{4}2m$ & $4/m$ & $mmm$ & $4$ & $222$ & $4$ & $mm2$ & $\bar{4}$ & $mm2$ & $222$ \rule[0pt]{0pt}{8pt} \\ 

\multicolumn{2}{c}{anti-unitary operation} & $\theta I$ & $\theta I$ & $\theta I$ & $\theta C_{2}^\prime$ & $\theta C_{2}^{\prime\prime}$ & $\theta C_{2}^\prime$ &$\theta C_{2}^{\prime\prime}$ & $\theta \sigma_{\varv}$ & $\theta \sigma_{d}$ & $\theta C_{2}^\prime$ & $\theta C_{2}^{\prime\prime}$ & $\theta \sigma_{d}$ \\ 

& & & & & $[100]$ & $[110]$ &  $[100]$ & $[110]$ & $\perp[100]$ & $\perp [110]$ & $[100]$ & $[110]$ & $\perp [110]$ \\\hline
E & ET &   \\ \cline{1-2}
$Q_0,Q_u, $ &  &  A$_{\rm 1}^+$ & A$_{\rm 1}^+$ & A$_{\rm 1}^+$ & A$_{\rm g}^+$  & A$_{\rm g}^+$ &  A$^+$ & A$^+$  & A$^+$ & A$_{1}^+$ & A$^+$ & A$_{1}^+$ & A$^+$ \\
$Q_4, Q_{4u}$ \\
$Q_\varv, Q_{4\varv}$ & $G_{xyz}$ &  B$_{\rm 1}^+$ & B$_{\rm 1}^+$ & B$_{\rm 1}^+$ & B$_{\rm g}^+$ & A$_{\rm g}^-$ & B$^+$ & A$^-$ & B$^+$ & A$_{1}^-$ & B$^+$ & A$_{1}^-$ & A$^-$ \\
$Q_{yz}, Q_{4x}^\alpha, Q_{4x}^\beta$ & $G_x, G_x^\alpha, G_x^\beta$ &  E$^+$ & E$^+$ & E$^+$ & E$_{\rm g}^{(1,2)+}$ & B$_{\rm 2g}$ & E$^{(1,2)+}$ & B$_{2}$ & E$^{(1,2)+}$ & B$_{1}$ & E$^{(1,2)+}$ & B$_{1}$ & B$_{2}$ \\
$Q_{zx}, Q_{4y}^\alpha, Q_{4y}^\beta$ & $G_y, G_y^\alpha, G_y^\beta$ & & & &  & & & & & & & & \\
$Q_{4z}^\alpha$ & $G_z, G_z^\alpha$ &  A$_{\rm 2}^+$ & A$_{\rm 2}^+$ & A$_{\rm 2}^+$ & A$_{\rm g}^-$ & B$_{\rm 1g}^-$ & A$^-$ &  B$_{\rm 1}^-$ & A$^-$ & A$_2^-$ & A$^-$ & A$_2^-$ & B$_{\rm 1}^-$ \\
$Q_{xy}, Q_{4z}^\beta$ & $G_z^\beta$ & B$_{\rm 2}^+$ & B$_{\rm 2}^+$ & B$_{\rm 2}^+$ & B$_{\rm g}^-$ & B$_{\rm 1g}^+$ & B$^-$ & B$_{\rm 1}^+$ & B$^-$ & A$_2^+$ & B$^-$ & A$_2^+$ & B$_{\rm 1}^+$\\
\hline
  & $G_0, G_u,$ &  A$_{\rm 1}^-$ & A$_{\rm 2}^-$ & B$_{\rm 1}^-$ & A$_{\rm u}^+$ & A$_{\rm u}^+$ & A$^+$ &A$^+$ & A$^-$ &  A$_2^-$ & B$^+$ & A$_2^+$ & A$^-$ \\
  & $G_4, G_{4u}$ \\
 $Q_{xyz}$ & $G_\varv, G_{4\varv}$ &  B$_{\rm 1}^-$ & B$_{\rm 2}^-$ & A$_{\rm 1}^-$ & B$_{\rm u}^+$ & A$_{\rm u}^-$ & B$^+$ & A$^-$ & B$^-$ & A$_2^+$ & A$^+$ & A$_2^-$ & A$^+$ \\
$Q_x, Q_x^\alpha, Q_x^\beta$ & $G_{yz}, G_{4x}^\alpha, G_{4x}^\beta$ & E$^-$ & E$^-$ & E$^-$ & E$_{\rm u}^{(1,2)+}$ & B$_{\rm 2u}$& E$^{(1,2)+}$ & B$_{2}$ & E$^{(1,2)-}$& B$_{1}$ & E$^{(2,1)+}$ & B$_{1}$ & B$_{2}$ \\
$Q_y, Q_y^\alpha, Q_y^\beta$ & $G_{zx}, G_{4y}^\alpha, G_{4y}^\beta$ & & & & & & & & & & & & \\
 $Q_z, Q_z^\alpha$ & $G_{4z}^\alpha$ & A$_{\rm 2}^-$ & A$_{\rm 1}^-$ & B$_{\rm 2}^-$ & A$_{\rm u}^-$ & B$_{\rm 1u}^-$ & A$^-$ &  B$_{\rm 1}^-$ & A$^+$ & A$_{1}^+$ & B$^-$ & A$_{1}^-$ &  B$_{\rm 1}^+$\\
 $Q_z^\beta$ & $G_{xy}, G_{4z}^\beta$ &  B$_{\rm 2}^-$ & B$_{\rm 1}^-$ & A$_{\rm 2}^-$ & B$_{\rm u}^-$ & B$_{\rm 1u}^+$ & B$^-$ &  B$_{\rm 1}^+$ & B$^+$ & A$_{1}^-$ & A$^-$ & A$_{1}^+$ & B$_{\rm 1}^-$ \\ 
 \hline
MT & M & \\\cline{1-2}
$T_0, T_u,$  &  & A$_{\rm 1}^-$ & A$_{\rm 1}^-$ &   A$_{\rm 1}^-$ & A$_{\rm g}^-$ & A$_{\rm g}^-$ & A$^-$ & A$^-$ & A$^-$ &  A$_{1}^-$ & A$^-$ & A$_{1}^-$ & A$^-$\\
$T_4, T_{4u}$ & \\
 $T_\varv, T_{4\varv}$ & $M_{xyz}$ & B$_{\rm 1}^-$ &  B$_{\rm 1}^-$ & B$_{\rm 1}^-$ & B$_{\rm g}^-$ & A$_{\rm g}^+$ & B$^-$ & A$^+$ & B$^-$ &  A$_{1}^+$ & B$^-$ & A$_{1}^+$ & A$^+$ \\
$T_{yz}, T_{4x}^\alpha, T_{4x}^\beta$ & $M_x, M_x^\alpha, M_x^\beta$ & E$^-$ & E$^-$ &  E$^-$ &  E$_{\rm g}^{(1,2)-}$ & B$_{\rm 2g}$ & E$^{(1,2)-}$ & B$_{2}$ & E$^{(1,2)-}$ & B$_{1}$ &  E$^{(1,2)-}$ & B$_{1}$ &   B$_{2}$\\
$T_{zx}, T_{4y}^\alpha, T_{4y}^\beta$ & $M_y, M_y^\alpha, M_y^\beta$ & & & & & & & & & & & & \\
$T_{4z}^\alpha$ & $M_z, M_z^\alpha$ & A$_{\rm 2}^-$ & A$_{\rm 2}^-$ & A$_{\rm 2}^-$ & A$_{\rm g}^+$ & B$_{\rm 1g}^+$ & A$^+$ & B$_{\rm 1}^+$ & A$^+$ & A$_2^+$ & A$^+$ & A$_2^+$ & B$_{\rm 1}^+$ \\
$T_{xy}, T_{4z}^\beta$ & $M_z^\beta$ & B$_{\rm 2}^-$ & B$_{\rm 2}^-$ & B$_{\rm 2}^-$ & B$_{\rm g}^+$ & B$_{\rm 1g}^-$ & B$^+$ & B$_1^-$ & B$^+$ & A$_2^-$ & B$^+$ & A$_2^-$ & B$_{\rm 1}^-$ \\
\hline
  & $M_0, M_u,$ & A$_{\rm 1}^+$ & A$_{\rm 2}^+$ & B$_{\rm 1}^+$ & A$_{\rm u}^-$ & A$_{\rm u}^-$ & A$^-$ & A$^-$ & A$^+$ & A$_2^+$ & B$^-$ & A$_2^-$ & A$^+$ \\
 &  $M_4, M_{4u}$ & \\
 $T_{xyz}$ & $M_\varv, M_{4\varv}$ & B$_{\rm 1}^+$ & B$_{\rm 2}^+$ &  A$_{\rm 1}^+$ & B$_{\rm u}^-$ & A$_{\rm u}^+$ & B$^-$ & A$^+$ & B$^+$ & A$_2^-$ & A$^-$ & A$_2^+$ & A$^-$ \\
 $T_x, T_x^\alpha, T_x^\beta$ & $M_{yz}, M_{4x}^\alpha, M_{4x}^\beta$ & E$^+$ & E$^+$ & E$^+$ & E$_{\rm u}^{(1,2)-}$ & B$_{\rm 2u}$ & E$^{(1,2)-}$ & B$_{2}$ & E$^{(1,2)+}$ & B$_{1}$ & E$^{(2,1)-}$ & B$_{1}$ & B$_{2}$ \\
$T_y, T_y^\alpha, T_y^\beta$ & $M_{zx}, M_{4y}^\alpha, M_{4y}^\beta$ & & & &  & & & & & & & & \\
 $T_z, T_z^\alpha$ & $M_{4z}^\alpha$ & A$_{\rm 2}^+$ & A$_{\rm 1}^+$ & B$_{\rm 2}^+$ & A$_{\rm u}^+$ & B$_{\rm 1u}^+$ & A$^+$ & B$_{\rm 1}^+$ & A$^-$ & A$_{1}^-$ & B$^+$ & A$_{1}^+$ & B$_{\rm 1}^-$ \\
 $T_z^\beta$ & $M_{xy}, M_{4z}^\beta$ & B$_{\rm 2}^+$ & B$_{\rm 1}^+$ & A$_{\rm 2}^+$ & B$_{\rm u}^+$ & B$_{\rm 1u}^-$ & B$^+$ & B$_{\rm 1}^-$ & B$^-$ & A$_{1}^+$ & A$^+$ & A$_{1}^-$ & B$_{\rm 1}^+$ \\ 
 \hline\hline
\end{tabular}
\end{table*}

\tabcolsep = 2pt
\begin{table*}[h!]
\centering
\caption{
IRREPs of multipoles in tetragonal black and white point groups for $4'/m'$, $4/m'$, $4'/m$,  $\bar{4}'$, and $4'$.
 \label{table:BW_tetragonal_2}}
\vspace{2mm}
\begin{tabular}{cccccccccccccccccccccc}\hline \hline
\multicolumn{2}{c}{magnetic point group} & $4'/m'$ & $4/m'$ & $4'/m$ & $\bar{4}'$ &  $4'$ \rule[0pt]{0pt}{8pt}  \\\cline{3-7}
\multicolumn{2}{c}{unitary subgroup} & $\bar{4}$ & $4$ & $2/m$ & $2$ &  $2$ \rule[0pt]{0pt}{8pt}  \\

\multicolumn{2}{c}{anti-unitary operation} & $\theta I$ & $\theta I$ & $\theta C_4$ & $\theta IC_4$ & $\theta C_4$  \\
& & & & $[001]$ & $[001]$ & $[001]$ \\\hline
E & ET &   \\ \cline{1-2}
$Q_0,Q_u, Q_4, Q_{4u},Q_{4z}^\alpha$ & $G_z, G_z^\alpha$ &  A$^+$ & A$^+$ & A$_{\rm g}^+$ & A$^+$ & A$^+$ \\
$Q_\varv, Q_{xy}, Q_{4\varv}, Q_{4z}^\beta$ & $G_{xyz}, G_z^\beta$ &  B$^+$ & B$^+$ & A$_{\rm g}^-$ & A$^-$ & A$^-$ \\

$Q_{yz}, Q_{4x}^\alpha, Q_{4x}^\beta$ & $G_x, G_x^\alpha, G_x^\beta$ &  E  & E & B$_{\rm g}$ & B & B \\
$Q_{zx}, Q_{4y}^\alpha, Q_{4y}^\beta$ & $G_y, G_y^\alpha, G_y^\beta$ &  & & & &  \\

\hline
  $Q_z, Q_z^\alpha$  & $G_0, G_u, G_4, G_{4u}, G_{4z}^\alpha$ & B$^-$ & A$^-$ & A$_{\rm u}^+$ & A$^-$ & A$^+$ \\
 $Q_{xyz}, Q_z^\beta$ & $G_\varv, G_{xy}, G_{4\varv}, G_{4z}^\beta$ & A$^-$ &  B$^-$ & A$_{\rm u}^-$ & A$^+$ & A$^-$ \\
 
$Q_x, Q_x^\alpha, Q_x^\beta$ & $G_{yz}, G_{4x}^\alpha, G_{4x}^\beta$ & E & E & B$_{\rm u}$ & B &  B \\
$Q_y, Q_y^\alpha, Q_y^\beta$ & $G_{zx}, G_{4y}^\alpha, G_{4y}^\beta$ &  &  & & &  \\
 \hline
MT & M & \\\cline{1-2}
$T_0, T_u, T_4, T_{4u}, T_{4z}^\alpha$ & $M_z, M_z^\alpha$  & A$^-$ & A$^-$ & A$_{\rm g}^-$ & A$^-$ & A$^-$  \\
$T_\varv, T_{xy}, T_{4\varv}, T_{4z}^\beta$ & $M_{xyz}, M_z^\beta$ & B$^-$ & B$^-$ & A$_{\rm g}^+$ & A$^+$ & A$^+$ \\

$T_{yz}, T_{4x}^\alpha, T_{4x}^\beta$ & $M_x, M_x^\alpha, M_x^\beta$ & E & E & B$_{\rm g}$ & B & B \\
$T_{zx}, T_{4y}^\alpha, T_{4y}^\beta$ & $M_y, M_y^\alpha, M_y^\beta$ &  &  & & &  \\
\hline
 $T_z, T_z^\alpha$ & $M_0,M_u, M_4, M_{4u}, M_{4z}^\alpha$ & B$^+$ & A$^+$ & A$_{\rm u}^-$ & A$^+$ &A$^-$ \\
 $T_{xyz}, T_z^\beta$ & $M_\varv, M_{xy},  M_{4\varv}, M_{4z}^\beta$ & A$^+$ & B$^+$ &  A$_{\rm u}^+$ & A$^-$ &  A$^+$ \\

$T_x, T_x^\alpha, T_x^\beta$ & $M_{yz}, M_{4x}^\alpha, M_{4x}^\beta$ & E & E & B$_{\rm u}$ & B  & B \\
$T_y, T_y^\alpha, T_y^\beta$ & $M_{zx},  M_{4y}^\alpha, M_{4y}^\beta$ &  &  &  &  & & \\
 \hline\hline
\end{tabular}
\end{table*}

\tabcolsep = 3pt
\begin{table*}[h!]
\centering
\caption{
IRREPs of multipoles in orthorhombic black and white point groups.
 \label{table:BW_orthorhombic}}
\vspace{2mm}
\begin{tabular}{cccccccccccccccccccccc}\hline \hline
\multicolumn{2}{c}{magnetic point group} & $m'm'm'$ & $mmm'$ & $m'm'm$ & $2'2'2$ & $m'm'2$ & $m'm2'$ \\\cline{3-8}
\multicolumn{2}{c}{unitary subgroup} &$222$ & $mm2$ & $2/m$ & $2$ & $2$ & $m$  \\

\multicolumn{2}{c}{anti-unitary operation} & $\theta I$ & $\theta I$ & $\theta C_{2x}$ & $\theta C_{2x}$ & $\theta \sigma_{x}$ & $\theta C_{2z}$  \\  \hline
E & ET &   \\ \cline{1-2}
$Q_0, Q_u, Q_\varv, Q_4, Q_{4u}, Q_{4\varv}$ & $G_{xyz}$ 
& A$^+$ & A$_1^+$ & A$_{\rm g}^+$ & A$^+$ & A$^+$ & A$^{\prime+}$  \\

$Q_{yz}, Q_{4x}^\alpha, Q_{4x}^\beta$ & $G_x, G_x^\alpha, G_x^\beta$ 
& B$_3^+$ & B$_2^+$ & B$_{\rm g}^+$ & B$^+$ & B$^+$ & A$^{\prime\prime-}$ \\

$Q_{zx}, Q_{4y}^\alpha, Q_{4y}^\beta$ & $G_y, G_y^\alpha, G_y^\beta$ 
& B$_2^+$ & B$_1^+$ & B$_{\rm g}^-$ & B$^-$ & B$^-$ & A$^{\prime-}$\\

$Q_{xy}, Q_{4z}^\alpha, Q_{4z}^\beta$ & $G_z, G_z^\alpha, G_z^\beta$ 
& B$_1^+$ & A$_2^+$ & A$_{\rm g}^-$ & A$^-$ & A$^-$ & A$^{\prime\prime+}$ \\
\hline
  $Q_{xyz}$ & $G_0, G_u, G_\varv, G_4, G_{4u}, G_{4\varv}$ 
  &  A$^-$ & A$_2^-$ & A$_{\rm u}^+$ & A$^+$ & A$^-$ & A$^{\prime\prime+}$\\
  
$Q_x, Q_x^\alpha, Q_x^\beta$ & $G_{yz}, G_{4x}^\alpha, G_{4x}^\beta$ 
& B$_3^-$ & B$_1^-$ & B$_{\rm u}^+$ & B$^+$ & B$^-$ & A$^{\prime-}$\\

$Q_y, Q_y^\alpha, Q_y^\beta$ & $G_{zx}, G_{4y}^\alpha, G_{4y}^\beta$ 
& B$_2^-$ & B$_2^-$ & B$_{\rm u}^-$ & B$^-$ & B$^+$ & A$^{\prime\prime-}$ \\

 $Q_z, Q_z^\alpha, Q_z^\beta$ & $G_{xy}, G_{4z}^\alpha, G_{4z}^\beta$ 
 & B$_1^-$ & A$_1^-$ & A$_{\rm u}^-$ & A$^-$ & A$^+$ & A$^{\prime+}$ \\

 \hline
MT & M & \\\cline{1-2}
$T_0, T_u, T_\varv, T_4, T_{4u}, T_{4\varv}$ &  $M_{xyz}$ 
& A$^-$ & A$_1^-$ & A$_{\rm g}^-$ & A$^-$ & A$^-$ & A$^{\prime-}$  \\

$T_{yz}, T_{4x}^\alpha, T_{4x}^\beta$ & $M_x, M_x^\alpha, M_x^\beta$ 
&  B$_3^-$ & B$_2^-$ & B$_{\rm g}^-$ & B$^-$ & B$^-$ & A$^{\prime\prime+}$ \\

$T_{zx}, T_{4y}^\alpha, T_{4y}^\beta$ & $M_y, M_y^\alpha, M_y^\beta$ 
&  B$_2^-$ & B$_1^-$ & B$_{\rm g}^+$ & B$^+$ &  B$^+$ & A$^{\prime+}$ \\

 $T_{xy}, T_{4z}^\alpha, T_{4z}^\beta$ & $M_z, M_z^\alpha, M_z^\beta$ 
 & B$_1^-$ & A$_2^-$ & A$_{\rm g}^+$ & A$^+$ & A$^+$ & A$^{\prime\prime-}$ \\
 \hline
 $T_{xyz}$ & $M_0, M_u, M_\varv, M_4, M_{4u}, M_{4\varv}$ 
 & A$^+$ & A$_2^+$ & A$_{\rm u}^-$ & A$^-$ & A$^+$ & A$^{\prime\prime-}$ \\
 
$T_x, T_x^\alpha, T_x^\beta$ & $M_{yz}, M_{4x}^\alpha, M_{4x}^\beta$ 
& B$_3^+$ & B$_1^+$ & B$_{\rm u}^-$ & B$^-$ & B$^+$ & A$^{\prime+}$  \\

$T_y, T_y^\alpha, T_y^\beta$ & $M_{zx}, M_{4y}^\alpha, M_{4y}^\beta$ 
& B$_2^+$ & B$_2^+$ & B$_{\rm u}^+$ & B$^+$ & B$^-$ & A$^{\prime\prime+}$\\

 $T_z, T_z^\alpha, T_z^\beta$ & $M_{xy}, M_{4z}^\alpha, M_{4z}^\beta$ 
 & B$_1^+$ & A$_1^+$ & A$_{\rm u}^+$ & A$^+$ & A$^-$ & A$^{\prime-}$\\
 
  \hline\hline
\end{tabular}
\end{table*}

\tabcolsep = 3pt
\begin{table*}[h!]
\centering
\caption{
IRREPs of multipoles in monoclinic and triclinic black and white point groups.
 \label{table:BW_monocli_tricli}}
\vspace{2mm}
\begin{tabular}{cccccccccccccccccccccc}\hline \hline
\multicolumn{2}{c}{magnetic point group} &  $2'/m$ & $2/m'$ & $2'/m'$ & $2'$ & $m'$ & $\bar{1}'$  \rule[0pt]{0pt}{8pt} \\ \cline{3-8}
\multicolumn{2}{c}{unitary subgroup} & $m$ & $2$ & $\bar{1}$ & $1$ & $1$ & $1$  \rule[0pt]{0pt}{8pt} \\

\multicolumn{2}{c}{anti-unitary operation} & $\theta I$ & $\theta I$ & $\theta C_2$ & $\theta C_2$ & $\theta\sigma$ & $\theta I$ \\\hline
E & ET &   \\ \cline{1-2}
$Q_0, Q_u, Q_\varv, Q_{zx}, Q_4, Q_{4u}, Q_{4\varv}, Q_{4y}^\alpha, Q_{4y}^\beta$ & $G_y, G_{xyz}, G_y^\alpha, G_y^\beta$ 
&  A$^{\prime+}$ & A$^+$ & A$_{\rm g}^+$ &  A$^+$ & A$^+$ &A$^+$ \\

$Q_{yz}, Q_{xy},  Q_{4x}^\alpha, Q_{4z}^\alpha, Q_{4x}^\beta, Q_{4z}^\beta$ & $G_x, G_z, G_x^\alpha, G_z^\alpha, G_x^\beta, G_z^\beta$ 
& A$^{\prime\prime+}$ & B$^+$ & A$_{\rm g}^-$ &  A$^-$ & A$^-$ & A$^+$ \\

\hline
  $Q_y, Q_{xyz}, Q_y^\alpha, Q_y^\beta$ & $G_0, G_u, G_\varv, G_{zx}, G_4, G_{4u}, G_{4\varv}, G_{4y}^\alpha, G_{4y}^\beta$ 
  & A$^{\prime\prime-}$ & A$^-$ & A$_{\rm u}^+$ & A$^+$ & A$^-$ & A$^-$ \\

$Q_x, Q_z,  Q_x^\alpha, Q_z^\alpha, Q_x^\beta, Q_z^\beta$ & $G_{yz}, G_{xy},  G_{4x}^\alpha, G_{4z}^\alpha, G_{4x}^\beta, G_{4z}^\beta$ 
& A$^{\prime-}$ & B$^-$ & A$_{\rm u}^-$ & A$^-$ & A$^+$ & A$^-$ \\

 \hline
MT & M & \\\cline{1-2}
$T_0, T_u, T_\varv, T_{zx}, T_4, T_{4u}, T_{4\varv}, T_{4y}^\alpha, T_{4y}^\beta$ & $M_y, M_{xyz}, M_y^\alpha, M_y^\beta$
& A$^{\prime-}$ & A$^-$ & A$_{\rm g}^-$ & A$^-$ & A$^-$ & A$^-$ \\

$T_{yz}, T_{xy},  T_{4x}^\alpha, T_{4z}^\alpha, T_{4x}^\beta, T_{4z}^\beta$ & $M_x, M_z, M_x^\alpha, M_z^\alpha, M_x^\beta, M_z^\beta$ 
& A$^{\prime\prime-}$ & B$^-$ & A$_{\rm g}^+$ & A$^+$ & A$^+$ & A$^-$ \\

\hline
  $T_y, T_{xyz}, T_y^\alpha, T_y^\beta$ & $M_0, M_u, M_\varv, M_{zx}, M_4, M_{4u}, M_{4\varv}, M_{4y}^\alpha, M_{4y}^\beta$ 
 & A$^{\prime\prime+}$ & A$^+$ & A$_{\rm u}^-$ & A$^-$ & A$^+$ & A$^+$ \\

$T_x, T_z,  T_x^\alpha, T_z^\alpha, T_x^\beta, T_z^\beta$ & $M_{yz}, M_{xy},  M_{4x}^\alpha, M_{4z}^\alpha, M_{4x}^\beta, M_{4z}^\beta$ 
& A$^{\prime+}$ & B$^+$ & A$_{\rm u}^+$ & A$^+$ & A$^-$ & A$^+$ \\

 \hline\hline
\end{tabular}
\end{table*}

\clearpage
\tabcolsep = 3pt
\begin{table*}[h!]
\centering
\caption{
IRREPs of multipoles in hexagonal black and white point groups except for $6'/m$, $6/m'$, $6'/m'$,  $\bar{6}'$, and $6'$ (see also Table~\ref{table:BW_hexagonal_2}).
$\{X_x-iX_y, X_x+iX_y\}$, 
$\{X_{yz}+iX_{zx}, X_{yz}-iX_{zx}\}$, 
$\{X_\varv+iX_{xy}, X_\varv-iX_{xy}\}$, 
$\{X_{3u}-iX_{3\varv}, X_{3u}+iX_{3\varv}\}$, $\{X_{xyz}-iX_{z}^\beta, X_{xyz} +iX_z^\beta\}$, 
$\{X_{4\varv}^\alpha+iX_{4u}^\alpha, X_{4\varv}^\alpha-iX_{4u}^\alpha\}$, $\{X_{4u}^{\beta1}-iX_{4\varv}^{\beta1}, X_{4u}^{\beta1}+iX_{4\varv}^{\beta1}\}$, and $\{X_{4u}^{\beta2}+iX_{4\varv}^{\beta2}, X_{4u}^{\beta2}-iX_{4\varv}^{\beta2}\}$ ($X=Q, G, M, T$) are the basis of E$_{\rm 1g/u}^{(1,2) \pm}$ and  E$_{\rm 2g/u}^{(1,2) \pm}$ representations in $6/mm'm'$, E$_{\rm 1}^{(1,2) \pm}$ and  E$_{\rm 2}^{(1,2) \pm}$ representations in $62'2'$ and $6m'm'$, E$^{\prime(1,2) \pm}$ and  E$^{\prime \prime(1,2) \pm}$ representations in $\bar{6}m'2'$.
 \label{table:BW_hexagonal_1}}
\vspace{2mm}
\begin{tabular}{cccccccccccccccccccccc}\hline \hline
\multicolumn{2}{c}{magnetic point group} & $6/m'm'm'$ & $6/m'mm$ & $6'/mmm'$ & $6/mm'm'$ & $6'/m'mm'$ & $62'2'$ & $6'22'$ & $6m'm'$ & $6'mm'$ & $\bar{6}m'2'$ & $\bar{6}'m2'$ & $\bar{6}'m'2$  \rule[0pt]{0pt}{8pt} \\\cline{3-14}
\multicolumn{2}{c}{unitary subgroup} & $622$ & $6mm$ & $\bar{6}m2$ & $6/m$ & $\bar{3}m$ & $6$ & $32$ & $6$ & $3m$ & $\bar{6}$ & $3m$ & $32$ \rule[0pt]{0pt}{8pt} \\

\multicolumn{2}{c}{anti-unitary operation} & $\theta I$ & $\theta I$ & $\theta I$ & $\theta C_{2x}$ & $\theta C_2$ & $\theta C_{2x}$ & $\theta C_2$ & $\theta \sigma_{x}$ & $\theta C_2$ & $\theta C_{2y}$ & $\theta \sigma_h$ & $\theta \sigma_h$ \\\hline
E & ET &   \\ \cline{1-2}
$Q_0,Q_u, Q_{40}$ & 
& A$_{\rm 1}^+$ & A$_1^+$ & A$_1^{\prime+}$ & A$_{\rm g}^+$ & A$_{\rm 1g}^+$ & A$^+$ & A$_{\rm 1}^+$ & A$^+$ & A$_1^+$ & A$^{\prime+}$ & A$_1^+$ & A$_1^+$  \\

 & $G_z, G_z^\alpha$ 
 & A$_{\rm 2}^+$ & A$_{\rm 2}^+$ & A$_2^{\prime+}$ & A$_{\rm g}^-$ & A$_{\rm 2g}^+$ & A$^-$ & A$_{\rm 2}^+$ & A$^-$ & A$_2^+$ & A$^{\prime-}$ & A$_2^+$ & A$_2^+$ \\
 
$Q_{4a}$ & $G_{3a}$ 
& B$_{\rm 1}^+$ & B$_{\rm 1}^+$ & A$_2^{\prime\prime+}$ & 
B$_{\rm g}^+$ & A$_{\rm 1g}^-$ & B$^+$ & A$_{\rm 1}^-$ & B$^+$ & A$_1^-$ & A$^{\prime\prime-}$ & A$_1^-$  & A$_{2}^-$ \\

$Q_{4b}$ & $G_{3b}$ 
& B$_{\rm 2}^+$ & B$_{\rm 2}^+$ &  A$_1^{\prime\prime+}$ & B$_{\rm g}^-$ & A$_{\rm 2g}^-$ & B$^-$ & A$_{\rm 2}^-$ & B$^-$ & A$_2^-$  & A$^{\prime\prime+}$ & A$_2^-$  & A$_{1}^-$ \\

$Q_{yz}, Q_{4\varv}^\alpha$ & $G_x, G_{3u}$ 
& E$_{\rm 1}^+$ & E$_{\rm 1}^+$ & E$^{\prime\prime+}$ & E$_{\rm 1g}^{(1,2)+}$ & E$_{\rm g}^-$ &  E$_{\rm 1}^{(1,2)+}$ & E$^-$ &  E$_{\rm 1}^{(1,2)+}$ & E$^-$ &  E$^{\prime\prime(1,2)-}$ & E$^-$ & E$^-$\\

$Q_{zx}, Q_{4u}^\alpha
$ & $G_y, G_{3\varv}$ 
& & & &  & & & & & & & &  \\

$Q_\varv, Q_{4u}^{\beta1}, Q_{4u}^{\beta2}$ & $G_{xyz}$
& E$_{\rm 2}^+$ & E$_{\rm 2}^+$ & E$^{\prime+}$ & E$_{\rm 2g}^{(1,2)+}$ & E$_{\rm g}^+$ &  E$_{\rm 2}^{(1,2)+}$ & E$^+$ &  E$_{\rm 2}^{(1,2)+}$ & E$^+$ & E$^{\prime (1,2)+}$ & E$^+$ & E$^+$ \\

$Q_{xy}, Q_{4\varv}^{\beta1}, Q_{4\varv}^{\beta2}$ & $G_{z}^\beta$
& & & &  & & & & & & & & \\
\hline
  & $G_0,G_u, G_{40}$ 
&  A$_{\rm 1}^-$ & A$_{\rm 2}^-$ & A$_1^{\prime\prime-}$ & A$_{\rm u}^+$ & A$_{\rm 1u}^+$ & A$^+$ & A$_{\rm 1}^+$ & A$^-$ & A$_2^+$ & A$^{\prime\prime+}$ & A$_2^-$ & A$_1^-$ \\ 

$Q_z, Q_z^\alpha$ &  
& A$_{\rm 2}^-$ & A$_{\rm 1}^-$ &  A$_2^{\prime\prime-}$ & A$_{\rm u}^-$ & A$_{\rm 2u}^+$ & A$^-$ & A$_{\rm 2}^+$ & A$^+$ & A$_1^+$ & A$^{\prime\prime-}$ & A$_1^-$ & A$_2^-$ \\

 $Q_{3a}$ & $G_{4a}$ 
 & B$_{\rm 1}^-$ & B$_{\rm 2}^-$ &  A$_2^{\prime-}$ & B$_{\rm u}^+$ & A$_{\rm 1u}^-$ & B$^+$ & A$_{\rm 1}^-$ & B$^-$ & A$_2^-$
 & A$^{\prime-}$ & A$_2^+$
& A$_{2}^+$ \\ 
 
$Q_{3b}$ & $G_{4b}$ 
& B$_{\rm 2}^-$ & B$_{\rm 1}^-$ & A$_1^{\prime-}$ & B$_{\rm u}^-$ & A$_{\rm 2u}^-$ & B$^-$ & A$_{\rm 2}^-$ & B$^+$ & A$_1^-$
&  A$^{\prime+}$ &  A$_1^+$
 & A$_{1}^+$ \\

$Q_x, Q_{3u}$ & $G_{yz}, G_{4\varv}^\alpha$
& E$_{\rm 1}^-$ & E$_{\rm 1}^-$ & E$^{\prime-}$ & E$_{\rm 1u}^{(1,2)+}$ & E$_{\rm u}^-$ & E$_{\rm 1}^{(1,2)+}$  & E$^-$ & E$_1^{(1,2)-}$ & E$^-$ & E$^{\prime (1,2)-}$ & E$^+$ & E$^+$  \\

$Q_y, Q_{3\varv}$ & $G_{zx}, G_{4u}^\alpha$ 
& & & & & & & & & & & &  \\
$Q_{xyz}$ & $G_\varv, G_{4u}^{\beta1}, G_{4u}^{\beta2}$ 
 & E$_{\rm 2}^-$ & E$_{\rm 2}^-$ & E$^{\prime\prime-}$ & E$_{\rm 2u}^{(1,2)+}$ & E$_{\rm u}^+$ & E$_{\rm 2}^{(1,2)+}$ & E$^+$ & E$_2^{(1,2)-}$ & E$^+$ & E$^{\prime\prime (1,2)+}$ & E$^-$ & E$^-$  \\
 
$Q_{z}^\beta$ & $G_{xy}, G_{4\varv}^{\beta1}, G_{4\varv}^{\beta2}$ 
& & & &  & & &  & & & & & \\
 \hline
MT & M & \\\cline{1-2}
$T_0,T_u, T_{40}$ &  
& A$_{\rm 1}^-$ & A$_{\rm 1}^-$ & A$_1^{\prime-}$ & A$_{\rm g}^-$ & A$_{\rm 1g}^-$ & A$^-$ & A$_{\rm 1}^-$ & A$^-$ & A$_1^-$ & A$^{\prime-}$ & A$_1^-$ & A$_1^-$ \\

 & $M_z, M_z^\alpha$ 
 & A$_{\rm 2}^-$ & A$_{\rm 2}^-$ & A$_2^{\prime-}$ & A$_{\rm g}^+$ & A$_{\rm 2g}^-$ &  A$^+$ & A$_{\rm 2}^-$ & A$^+$ & A$_2^-$ & A$^{\prime+}$ & A$_2^-$ & A$_2^-$  \\ 
 
$T_{4a}$ & $M_{3a}$ 
& B$_{\rm 1}^-$ & B$_{\rm 1}^-$ &  A$_2^{\prime\prime-}$ & B$_{\rm g}^-$ & A$_{\rm 1g}^+$ & B$^-$ & A$_{\rm 1}^+$ & B$^-$ & A$_1^+$ & A$^{\prime\prime+}$ & A$_1^+$ 
 &  A$_{2}^+$  \\

$T_{4b}$ & $M_{3b}$ 
& B$_{\rm 2}^-$ & B$_{\rm 2}^-$ & A$_1^{\prime\prime-}$ & B$_{\rm g}^+$ & A$_{\rm 2g}^+$ & B$^+$ & A$_{\rm 2}^+$ & B$^+$ & A$_2^+$ &  A$^{\prime\prime-}$ & A$_2^+$ & A$_{1}^+$  \\

$T_{yz}, T_{4\varv}^\alpha$ & $M_x, M_{3u}$ 
& E$_{\rm 1}^-$ & E$_{\rm 1}^-$ & E$^{\prime\prime-}$ & E$_{\rm 1g}^{(1,2)-}$ & E$_{\rm g}^+$ & E$_{\rm 1}^{(1,2)-}$ & E$^+$ & E$_1^{(1,2)-}$ & E$^+$ & E$^{\prime\prime (1,2)+}$ & E$^+$ & E$^+$  \\

$T_{zx}, T_{4u}^\alpha$  & $M_y, M_{3\varv}$ 
& & & & & & & & & & & & \\

$T_\varv, T_{4u}^{\beta1}, T_{4u}^{\beta2}$ & $M_{xyz}$
& E$_{\rm 2}^-$ & E$_{\rm 2}^-$ & E$^{\prime-}$ & E$_{\rm 2g}^{(1,2)-}$ & E$_{\rm g}^-$ &  E$_{\rm 2}^{(1,2)-}$  & E$^-$ & E$_2^{(1,2)-}$ & E$^-$ &  E$^{\prime (1,2)-}$ & E$^-$ & E$^-$ \\

$T_{xy}, T_{4\varv}^{\beta1}, T_{4\varv}^{\beta2}$ & $M_{z}^\beta$
& & & & & & & & & & & & \\

\hline
  & $M_0,M_u, M_{40}$ 
& A$_{\rm 1}^+$ & A$_{\rm 2}^+$ & A$_1^{\prime\prime+}$ & A$_{\rm u}^-$ & A$_{\rm 1u}^-$ & A$^-$ & A$_{\rm 1}^-$ & A$^+$ & A$_2^-$ & A$^{\prime\prime-}$ & A$_2^+$ & A$_1^+$ \\

$T_z, T_z^\alpha$ & 
& A$_{\rm 2}^+$ & A$_{\rm 1}^+$ & A$_2^{\prime\prime+}$ & A$_{\rm u}^+$ & A$_{\rm 2u}^-$ & A$^+$ & A$_{\rm 2}^-$ & A$^-$ & A$_1^-$ & A$^{\prime\prime+}$ & A$_1^+$ & A$_2^+$ \\

 $T_{3a}$ & $M_{4a}$ 
 & B$_{\rm 1}^+$ & B$_{\rm 2}^+$ & A$_2^{\prime+}$ & B$_{\rm u}^-$ & A$_{\rm 1u}^+$ & B$^-$ & A$_{\rm 1}^+$ & B$^+$ & A$_2^+$ & A$^{\prime+}$ & A$_2^-$ & A$_{2}^-$ \\
 
$T_{3b}$ & $M_{4b}$ 
& B$_{\rm 2}^+$ & B$_{\rm 1}^+$ & A$_1^{\prime+}$ & B$_{\rm u}^+$ & A$_{\rm 2u}^+$ & B$^+$ & A$_{\rm 2}^+$ & B$^-$ & A$_1^+$ & A$^{\prime-}$ &  A$_1^-$ & A$_{1}^-$\\

$T_x, T_{3u}$ & $M_{yz}, M_{4\varv}^\alpha$ 
& E$_{\rm 1}^+$ & E$_{\rm 1}^+$ & E$^{\prime+}$ & E$_{\rm 1u}^{(1,2)-}$ & E$_{\rm u}^+$ & E$_{\rm 1}^{(1,2)-}$  & E$^+$ & E$_1^{(1,2)+}$ & E$^+$ & E$^{\prime (1,2)+}$ & E$^-$ & E$^-$  \\

$T_y, T_{3\varv}$ & $M_{zx}, M_{4u}^\alpha$ 
& & & & & & & & & & & &  \\
 $T_{xyz}$ & $M_\varv, M_{4u}^{\beta1}, M_{4u}^{\beta2}$ 
 & E$_{\rm 2}^+$ & E$_{\rm 2}^+$ &  E$^{\prime\prime+}$ & E$_{\rm 2u}^{(1,2)-}$ & E$_{\rm u}^-$ & E$_{\rm 2}^{(1,2)-}$ & E$^-$ & E$_2^{(1,2)+}$ & E$^-$  & E$^{\prime\prime (1,2)-}$ & E$^+$ & E$^+$ \\ 
 
$T_{z}^\beta$ & $M_{xy}, M_{4\varv}^{\beta1}, M_{4\varv}^{\beta2}$ 
& & & & & & & & & & & & \\
 \hline\hline
\end{tabular}
\end{table*}

\clearpage
\tabcolsep = 1pt
\begin{table}[h!]
\centering
\caption{IRREPs of multipoles in hexagonal black and white point groups for $6'/m$, $6/m'$, $6'/m'$,  $\bar{6}'$, and $6'$.
 \label{table:BW_hexagonal_2}}
\vspace{2mm}
\begin{tabular}{cccccccccccccccccccccc}\hline \hline
\multicolumn{2}{c}{magnetic point group} & $6'/m$ & $6/m'$ & $6'/m'$ &  $\bar{6}'$ & $6'$ \\ \cline{3-7}
\multicolumn{2}{c}{unitary subgroup} & $\bar{6}$ & $6$ & $\bar{3}$ & $3$ & $3$ \rule[0pt]{0pt}{8pt}\\

\multicolumn{2}{c}{anti-unitary operation} & $\theta I$ & $\theta I$ & $\theta C_2$ & $\theta \sigma_h$ & $\theta C_2$ \\ 
\hline
E & ET &   \\ \cline{1-2}
$Q_0,Q_u, Q_{40}$ & $G_z, G_z^\alpha$ 
&  A$^{\prime+}$ & A$^+$ & A$_{\rm g}^+$ &  A$^+$ & A$^+$ \\

$Q_{4a}, Q_{4b}$ & $G_{3a}, G_{3b}$ 
& A$^{\prime\prime+}$ & B$^+$ & A$_{\rm g}^-$ &  A$^-$ & A$^-$  \\

$Q_{yz}, Q_{4\varv}^\alpha$ & $G_x, G_{3u}$ 
& E$^{\prime\prime}$ & E$_1$ & E$_{\rm g}$ & E &  E \\

$Q_{zx}, Q_{4u}^\alpha$ & $G_y, G_{3\varv}$ 
& & & & &  \\

$Q_\varv, Q_{4u}^{\beta1}, Q_{4u}^{\beta2}$ & $G_{xyz}$
& E$^{\prime}$ &  E$_2$ & E$_{\rm g}$ & E & E \\

$Q_{xy}, Q_{4\varv}^{\beta1}, Q_{4\varv}^{\beta2}$ & $G_{z}^\beta$
& & & & &  \\

\hline
$Q_z, Q_z^\alpha$   & $G_0,G_u, G_{40}$ 
& A$^{\prime\prime-}$ & A$^-$ & A$_{\rm u}^+$ & A$^-$ &  A$^+$ \\ 

 $Q_{3a}, Q_{3b}$ & $G_{4a}, G_{4b}$ 
 & A$^{\prime-}$ &  B$^-$ & A$_{\rm u}^-$ & A$^+$ & A$^-$ \\

$Q_x, Q_{3u}$ & $G_{yz}, G_{4\varv}^\alpha$ 
& E$^{\prime}$ & E$_1$ & E$_{\rm u}$& E & E  \\

$Q_y, Q_{3\varv}$ & $G_{zx}, G_{4u}^\alpha$ 
& & & & & \\

$Q_{xyz}$  & $G_\varv, G_{4u}^{\beta1}, G_{4u}^{\beta2}$ 
 & E$^{\prime\prime}$ & E$_2$ & E$_{\rm u}$ & E & E  \\
 
$Q_{z}^\beta$ & $G_{xy}, G_{4\varv}^{\beta1}, G_{4\varv}^{\beta2}$ 
& & & & & \\

 \hline
MT & M & \\\cline{1-2}
$T_0,T_u, T_{40}$ &  $M_z, M_z^\alpha$   
& A$^{\prime-}$ & A$^{-}$ & A$_{\rm g}^{-}$ & A$^-$ &  A$^-$  \\ 

$T_{4a}, T_{4b}$ & $M_{3a}, M_{3b}$ 
& A$^{\prime\prime-}$ & B$^{-}$ & A$_{\rm g}^{+}$ & A$^+$ & A$^+$ \\

$T_{yz}, T_{4\varv}^\alpha$ & $M_x, M_{3u}$ 
& E$^{\prime\prime}$ & E$_1$ & E$_{\rm g}$& E & E\\

$T_{zx}, T_{4u}^\alpha$ & $M_y, M_{3\varv}$ 
& & & & & \\

$T_\varv, T_{4u}^{\beta1}, T_{4u}^{\beta2}$ & $M_{xyz}$
& E$^{\prime}$ & E$_2$& E$_{\rm g}$ & E & E\\

$T_{xy}, T_{4\varv}^{\beta1}, T_{4\varv}^{\beta2}$ & $M_{z}^\beta$
& & & & & \\

\hline
$T_z, T_z^\alpha$  & $M_0,M_u, M_{40}$ 
& A$^{\prime\prime+}$ & A$^+$ & A$_{\rm u}^-$ & A$^+$ & A$^-$ \\

 $T_{3a}, T_{3b}$ & $M_{4a}, M_{4b}$ 
 & A$^{\prime+}$ & B$^+$ & A$_{\rm u}^+$ & A$^-$ & A$^+$\\
 
$T_x, T_{3u}$ & $M_{yz}, M_{4\varv}^\alpha$ 
&  E$^{\prime}$ & E$_1$ & E$_{\rm u}$ & E & E\\

$T_y, T_{3\varv}$ & $M_{zx}, M_{4u}^\alpha$ 
& & & & & \\
 
$T_{xyz}$
& $M_\varv, M_{4u}^{\beta1}, M_{4u}^{\beta2}$ 
 & E$^{\prime\prime}$ & E$_2$ & E$_{\rm u}$ & E& E \\ 
 
$T_{z}^\beta$ & $M_{xy}, M_{4\varv}^{\beta1}, M_{4\varv}^{\beta2}$ 
& & & & & \\

 \hline\hline
\end{tabular}
\end{table}

\tabcolsep = 3pt
\begin{table*}[h!]
\centering
\caption{
IRREPs of multipoles in trigonal black and white point groups.
$\{X_y-iX_x, X_y+iX_x\}$, $\{X_{zx}+iX_{yz}, X_{zx}-iX_{yz}\}$, $\{X_\varv-iX_{xy},X_\varv+iX_{xy}\}$, $\{X_{3\varv}-iX_{3u},X_{3\varv}+iX_{3u}\}$, $\{X_{xyz}+iX_{z}^\beta, X_{xyz}-iX_{z}^\beta\}$, $\{X_{4u}^\alpha+iX_{4\varv}^\alpha, X_{4u}^\alpha-iX_{4\varv}^\alpha\}$, $\{X_{4u}^{\beta1}+iX_{4\varv}^{\beta1}, X_{4u}^{\beta1}-iX_{4\varv}^{\beta1}\}$, and $\{X_{4u}^{\beta2}-iX_{4\varv}^{\beta2}, X_{4u}^{\beta2}+iX_{4\varv}^{\beta2}\}$  
 ($X=Q, G, T, M$) are the basis of E$_{\rm g/u}^{(1,2)\pm}$ in $\bar{3}m'$ and E$^{(1,2)\pm}$ in $32'$ and $3m'$.
 \label{table:BW_trigonal}}
\vspace{2mm}
\begin{tabular}{cccccccccccccccccccccc}\hline \hline
\multicolumn{2}{c}{magnetic point group} & $\bar{3}'m'$ & $\bar{3}'m$ & $\bar{3}m'$ & $32'$ & $3m'$ & $\bar{3}'$   \\\cline{3-8}
\multicolumn{2}{c}{unitary subgroup} & $32$ & $3m$ & $\bar{3}$ & $3$ & $3$ & $3$ \\

\multicolumn{2}{c}{anti-unitary operation} & $\theta I$ & $\theta I$ & $\theta C_{2}^\prime$ & $\theta C_{2}^\prime$ & $\theta \sigma_{\varv}$ & $\theta I$ \\ \hline
E & ET &   \\ \cline{1-2}
$Q_0,Q_u, Q_{40}, Q_{4b}$ & $G_{3b}$ 
& A$_{\rm 1}^+$ & A$_{\rm 1}^+$ & A$_{\rm g}^+$ & A$^+$ & A$^+$ & A$^+$  \\

$Q_{4a}$ & $G_z, G_z^\alpha, G_{3a}$ 
& A$_{\rm 2}^+$ &  A$_{\rm 2}^+$ & A$_{\rm g}^-$ & A$^-$ & A$^-$ & A$^+$  \\

$Q_{yz}, Q_{xy}, Q_{4\varv}^\alpha, Q_{4\varv}^{\beta1}, Q_{4\varv}^{\beta2}$ & $G_x, G_{3u}, G_{z}^\beta$ 
& E$^+$ & E$^+$ & E$_{\rm g}^{(1,2)+}$ & E$^{(1,2)+}$ &  E$^{(1,2)+}$ & E \\

$Q_{zx}, Q_\varv, Q_{4u}^\alpha,Q_{4u}^{\beta1}, Q_{4u}^{\beta2}$ & $G_y, G_{3\varv}, G_{xyz}$ 
& & &  &  &  &    \\

\hline
$Q_{3b}$ & $G_0,G_u, G_{40}, G_{4b}$ 
& A$_{\rm 1}^-$ & A$_{\rm 2}^-$ & A$_{\rm u}^+$ & A$^+$ & A$^-$ & A$^-$ \\ 

$Q_z, Q_z^\alpha, Q_{3a}$ & $G_{4a}$ 
 & A$_{\rm 2}^-$ &  A$_{\rm 1}^-$ & A$_{\rm u}^-$ & A$^-$ & A$^+$ & A$^-$  \\ 

$Q_x, Q_{3u}, Q_{z}^\beta$ & $G_{yz}, G_{xy}, G_{4\varv}^\alpha, G_{4\varv}^{\beta1}, G_{4\varv}^{\beta2}$ 
& E$^-$ &E$^-$ & E$_{\rm u}^{(1,2)+}$ & E$^{(1,2)+}$ &  E$^{(1,2)-}$ & E  \\

$Q_y, Q_{3\varv}, Q_{xyz}$ & $G_{zx}, G_\varv, G_{4u}^\alpha, G_{4u}^{\beta1}, G_{4u}^{\beta2}$ 
& & & &  & &  \\

 \hline
MT & M & \\\cline{1-2}
$T_0,T_u, T_{40}, T_{4b}$ & $M_{3b}$   
& A$_{\rm 1}^-$ &  A$_{\rm 1}^-$ & A$_{\rm g}^-$ & A$^-$ & A$^-$ & A$^-$ \\ 

$T_{4a}$ & $M_z, M_z^\alpha, M_{3a}$ 
&  A$_{\rm 2}^-$ &  A$_{\rm 2}^-$ & A$_{\rm g}^+$ & A$^+$ & A$^+$ & A$^-$ \\ 

$T_{yz}, T_{xy}, T_{4\varv}^\alpha, T_{4\varv}^{\beta1}, T_{4\varv}^{\beta2}$ & $M_x, M_{3u}, M_{z}^\beta$ 
& E$^-$ & E$^-$ & E$_{\rm g}^{(1,2)-}$ & E$^{(1,2)-}$ &  E$^{(1,2)-}$ & E   \\

$T_{zx}, T_\varv, T_{4u}^\alpha, T_{4u}^{\beta1}, T_{4u}^{\beta2}$ & $M_y, M_{3\varv}, M_{xyz}$ 
& & & & & &  \\

\hline
 $T_{3b}$  & $M_0,M_u, M_{40}, M_{4b}$ 
& A$_{\rm 1}^+$ & A$_{\rm 2}^+$ & A$_{\rm u}^-$ & A$^-$ & A$^+$ & A$^+$  \\

$T_z, T_z^\alpha, T_{3a}$ & $M_{4a}$ 
& A$_{\rm 2}^+$ & A$_{\rm 1}^+$ & A$_{\rm u}^+$ & A$^+$ & A$^-$ &  A$^+$  \\

$T_x, T_{3u}, T_{z}^\beta$ & $M_{yz}, M_{xy},  M_{4\varv}^\alpha, M_{4\varv}^{\beta1}, M_{4\varv}^{\beta2}$ 
& E$^+$ & E$^+$ & E$_{\rm u}^{(1,2)-}$ & E$^{(1,2)-}$ &  E$^{(1,2)+}$ & E  \\

$T_y, T_{3\varv}, T_{xyz}$ & $M_{zx}, M_\varv,  M_{4u}^\alpha, M_{4u}^{\beta1}, M_{4u}^{\beta2}$ 
& & &  &  &  &   \\
 \hline\hline
\end{tabular}
\end{table*}

\clearpage
\section{Corepresentation of magnetic point group \label{sec:ap_MPG_corepresentation}}

We give the short review of the irreducible corepresentation in magnetic point groups.
For the group with ${\bm G}+\mathcal{A}{\bm G}$ ($\mathcal{A}$ is the anti-unitary operation), the corepresentation for the basis set of the irreducible representation $\Gamma$ of ${\bm G}$, $\bra{\psi}$, and another set obtained as $\bra{\phi}=\mathcal{A}\bra{\psi}$ is given as follows~\cite{wigner1959group,wigner2013collected, dimmock1963representation}
\begin{align}
\mathcal{R} \bra{\psi, \phi} &=
\bra{\psi, \phi}
\begin{pmatrix}
\Delta^\Gamma (\mathcal{R}) & 0 \\
0 & \left[\Delta^{\Gamma}(\mathcal{A}^{-1}\mathcal{R}\mathcal{A})\right]^* \\
\end{pmatrix}
 \ \mbox{for} \ \mathcal{R} \in {\bm G}, \\
\mathcal{B} \bra{\psi, \phi} &= 
\bra{\psi, \phi}
\begin{pmatrix}0 & \Delta^{\Gamma}(\mathcal{B}\mathcal{A}) \\
\left[\Delta^{\Gamma}(\mathcal{A}^{-1}\mathcal{B})\right]^* & 0 \\
\end{pmatrix} \ \mbox{for} \ \mathcal{B} \in \mathcal{A}{\bm G}, 
\end{align}
where $\mathcal{R}$ ($\mathcal{B}$) represents the (anti-)unitary point group operation and $\Delta^{\Gamma}$ is the matrix representation of $\Gamma$.
In the present study, we adopt $\mathcal{A}=\theta$ for the gray point group and $\mathcal{A}=\theta \mathcal{S}$ [$\mathcal{S}$ is the unitary point group operation in (${\bm G}-{\bm H}$) in Eq.~\eqref{eq:BWPG}] for the black and white point group.

By the unitary transformation, we can obtain the irreducible corepresentation $D^{\Gamma}$
~\cite{dimmock1962irreducible, cracknell1966corepresentations, cracknell1967double, bradley1968magnetic, inui1990group, bradley2009mathematical}.
The irreducuble corepresentation can be classified into three types:
\begin{align}
\sum_{\mathcal{B} \in \mathcal{A}{\bm G}} \chi^{\Gamma}(\mathcal{B}^2) = 
\begin{cases}
\,\,+|{\bm G}| \ &: \mbox{case (a)}, \\
\,\,-|{\bm G}| \ &: \mbox{case (b)}, \\
\,\,0 \ &: \mbox{case (c)},  \\
\end{cases}
\end{align}
where $|{\bm G}|$ is the order of ${\bm G}$ and $\chi^{\Gamma}(\mathcal{B}^2)$ is the character with respect to the unitary operation $\mathcal{B}^2$ in $\Gamma$.
For case (a), $D^{\Gamma}$ is expressed as 
\begin{align}
D^{\Gamma}(\mathcal{R}) &= 
\begin{pmatrix}
\Delta^{\Gamma}(\mathcal{R}) & 0\\
0 & \Delta^{\Gamma}(\mathcal{R})\\
\end{pmatrix} \ \mbox{for} \ \mathcal{R} \in {\bm G}, \\
D^{\Gamma}(\mathcal{B}) &= 
\begin{pmatrix}
\Delta^{\Gamma}(\mathcal{B}\mathcal{A}^{-1})N & 0\\
0 & -\Delta^{\Gamma}(\mathcal{B}\mathcal{A}^{-1})N\\
\end{pmatrix} \ \mbox{for} \ \mathcal{B} \in \mathcal{A}{\bm G}, 
\end{align}
where $N$ is the unitary matrix to satisfy the relation $\Delta^\Gamma(\mathcal{R})=N [\Delta^\Gamma(\mathcal{A}^{-1}\mathcal{R}\mathcal{A})]^*N^{-1}$~\cite{bradley2009mathematical}.
Here, the basis set $\bra{\psi,\phi}$ is transformed by the unitary transformation
\begin{align}
V
=\frac{1}{\sqrt{2}}
\begin{pmatrix}
1  & -1\\
 N^{-1} & N^{-1}\\
\end{pmatrix}.
\end{align}
In the present paper, we denote the counterpart of the irreducible corepresentation characterized by $\Delta^{\Gamma}(\mathcal{R})$ for $\mathcal{R}$ and $\pm\Delta^{\Gamma}(\mathcal{B}\mathcal{A}^{-1})N$ for $\mathcal{B}$ as $\Gamma^{\pm}$, e.g., A$_{\rm 1g}^\pm$.

In contrast to case (a), the corepresentation cannot be reduced into even- and odd-parity parts with respect to the anti-unitary operation in cases (b) and (c).
For case (b), $D^{\Gamma}$ is described as 
\begin{align}
D^{\Gamma}(\mathcal{R}) &= 
\begin{pmatrix}
\Delta^{\Gamma}(\mathcal{R}) & 0\\
0 & \Delta^{\Gamma}(\mathcal{R})\\
\end{pmatrix} \ \mbox{for} \ \mathcal{R} \in {\bm G}, \\
D^{\Gamma}(\mathcal{B}) &= 
\begin{pmatrix}
0 & -\Delta^{\Gamma}(\mathcal{B}\mathcal{A}^{-1})N \\
\Delta^{\Gamma}(\mathcal{B}\mathcal{A}^{-1})N & 0 \\
\end{pmatrix} \ \mbox{for} \ \mathcal{B} \in \mathcal{A}{\bm G}, 
\end{align}
where the basis set is transformed by the unitary transformation
\begin{align}
U=\begin{pmatrix}
1  & 0\\
0 & N^{-1}\\
\end{pmatrix}.
\end{align}
Meanwhile, for case (c), $D^{\Gamma}$ is represented as
\begin{align}
D^{\Gamma}(\mathcal{R}) &= 
\begin{pmatrix}
\Delta^{\Gamma}(\mathcal{R}) & 0\\
0 & \left[\Delta^{\Gamma}(\mathcal{A}^{-1}\mathcal{R}\mathcal{A})\right]^*\\
\end{pmatrix} \ \mbox{for} \ \mathcal{R} \in {\bm G}, \\
D^{\Gamma}(\mathcal{B}) &= 
\begin{pmatrix}0 & \Delta^{\Gamma}(\mathcal{B}\mathcal{A}) \\
\left[\Delta^{\Gamma}(\mathcal{A}^{-1}\mathcal{B})\right]^* & 0 \\
\end{pmatrix} \ \mbox{for} \ \mathcal{B} \in \mathcal{A}{\bm G}.
\end{align}
$D^{\Gamma}$ in both cases are not block-diagonal with respect to the anti-unitary operation $\mathcal{B}$.
Thus, we denote the corepresentation in cases (b) and (c) simply as $\Gamma$.

\clearpage
\section{Laue group and magnetic Laue group \label{sec:ap_Laue}}

The correspondence between Laue groups and magnetic point groups is summarized in Table~\ref{table:Laue}, while that between magnetic Laue groups and magnetic point groups is listed in Tables~\ref{table:MLG_GPG} and \ref{table:MLG_CPG}.
Table~\ref{table:MLG_GPG} presents the magnetic Laue group with the $\mathcal{T}$ and/or $\mathcal{PT}$ symmetry, whereas Table~\ref{table:MLG_CPG} presents that without the $\mathcal{PT}$ symmetry.

\begin{table}[h!]
\centering
\caption{Laue group (LG) and the corresponding three types of magnetic point group: gray point group (GPG), crystallographic point group (CPG), and black and white point group (BWPG).
 \label{table:Laue}}
\vspace{2mm}
\begin{tabular}{ccccccccc}\hline \hline
 LG & GPG & CPG & BWPG \\\hline
 
$m\bar{3}m$ & $m\bar{3}m1'$ & $m\bar{3}m$ & $m'\bar{3}'m'$, $m\bar{3}m'$, $m'\bar{3}'m$  \rule[0pt]{0pt}{8pt} \\
 & $4321'$  & $432$ & $4'32'$ &  \\
  & $\bar{4}3m1'$ &  $\bar{4}3m$ & $\bar{4}'3m'$ & \\ \hline
  
$m\bar{3}$ & $m\bar{3}1'$ & $m\bar{3}$ & $\bar{m}'\bar{3}'$  \rule[0pt]{0pt}{8pt} \\
& $231'$ & $23$ & \\\hline

$4/mmm$ & $4/mmm1'$ & $4/mmm$ & $4/m'm'm'$, $4/mm'm'$  \\
& & & $4'/m'm'm$, $4'/mm'm$, $4/m'mm$ &  \\
& $4221'$ & $422$ & $42'2'$, $4'22'$ & \\
& $\bar{4}2m1'$ & $\bar{4}2m$ & $\bar{4}2'm'$, $\bar{4}'2m'$, $\bar{4}'2'm$ & \\
& $4mm1'$ & $4mm$ & $4m'm'$, $4'm'm$ & \\ \hline

$4/m$ & $4/m1'$ & $4/m$ & $4'/m'$, $4/m'$, $4'/m$ \\
& $41'$ & $4$ & $4'$ &  \\
& $\bar{4}1'$ & $\bar{4}$ & $\bar{4}'$ & \\ \hline

$mmm$ & $mmm1'$ & $mmm$ &$m'm'm'$, $m'm'm$, $m'mm$ \\
& $2221'$ & $222$ & $2'2'2$ & \\
& $mm21'$ & $mm2$ & $m'm'2$, $m'm2'$ & \\ \hline

$2/m$ & $2/m1'$ & $2/m$ & $2'/m'$, $2/m'$, $2'/m$ \\
& $21'$ & $2$ & $2'$ & \\ 
& $m1'$ & $m$ & $m'$ & \\ \hline

$\bar{1}$ & $\bar{1}1'$ & $\bar{1}$ & $\bar{1}'$  \rule[0pt]{0pt}{8pt}\\
& $11'$ & $1$ & \\\hline

$6/mmm$ & $6/mmm1'$ & $6/mmm$ & $6/m'm'm'$, $6/mm'm'$   \\
& & & $6'/m'mm'$, $6'/mmm'$, $6/m'mm$ \\
& $6221'$ & $622$ & $62'2'$, $6'22'$ & \\
& $\bar{6}m21'$ & $\bar{6}m2$ & $\bar{6}m'2'$, $\bar{6}'m2'$, $\bar{6}'m'2$ & \\
& $6mm1'$ & $6mm$ & $6m'm'$, $6'mm'$ & \\ \hline

$6/m$ & $6/m1'$ & $6/m$ & $6'/m'$, $6/m'$, $6'/m$ \\
& $61'$ & $6$ & $6'$ & \\
& $\bar{6}1'$ & $\bar{6}$ & $\bar{6}'$ & \\ \hline

$\bar{3}m$ & $\bar{3}m1'$ & $\bar{3}m$ & $\bar{3}m'$, $\bar{3}'m'$, $\bar{3}'m$  \rule[0pt]{0pt}{8pt} \\
& $321'$ & $32$ & $32'$ & \\
& $3m1'$ & $3m$ & $3m'$ & \\ \hline

$\bar{3}$ & $\bar{3}1'$ & $\bar{3}$ & $\bar{3}'$  \rule[0pt]{0pt}{8pt} \\
& $31'$ & $3$ & &  \\ 

\hline\hline
\end{tabular}
\end{table}

\begin{table}[h!]
\centering
\caption{
Magnetic Laue group (MLG) for gray point group (GPG) and the $\mathcal{PT}$-symmetric black and white point group (BWPG).
 \label{table:MLG_GPG}}
\vspace{2mm}
\begin{tabular}{ccccccccc}\hline \hline
 MLG & GPG & BWPG \\\hline
 
$m\bar{3}m1'$ & $m\bar{3}m1'$ & $m'\bar{3}'m', m'\bar{3}'m$ \rule[0pt]{0pt}{8pt} \\ 
& $4321'$ & \\
& $\bar{4}3m1'$ & \\\hline

$m\bar{3}1'$ & $m\bar{3}1'$ & $m'\bar{3}'$  \rule[0pt]{0pt}{8pt}\\
& $231'$ \\\hline

$4/mmm1'$ & $4/mmm1'$ &  $4/m'm'm', 4'/m'm'm, 4/m'mm$\\
& $4221'$ \\
& $\bar{4}2m1'$ \\
& $4mm1'$ \\ \hline

$4/m1'$ & $4/m1'$ & $4'/m', 4/m'$ \\
& $41'$ \\
& $\bar{4}1'$ \\\hline

$mmm1'$ & $mmm1'$ & $m'm'm', m'mm$ \\
& $2221'$ \\
& $mm21'$ \\\hline

$2/m1'$ & $2/m1'$ & $2'/m, 2/m'$ \\
& $21'$ \\
& $m1'$ \\\hline

$\bar{1}1'$ & $\bar{1}1'$ & $\bar{1}'$  \rule[0pt]{0pt}{8pt} \\ 
& $11'$ \\\hline

$6/mmm1'$ & $6/mmm1'$ & $6/m'm'm', 6'/mmm', 6/m'mm$\\
& $6221'$ \\
& $\bar{6}m21'$ \\
& $6mm1'$ \\\hline

 $6/m1'$ & $6/m1'$ & $6'/m, 6/m'$ \\
 & $61'$ & \\
  & $\bar{6}1'$ & \\\hline
 
$\bar{3}m1'$ & $\bar{3}m1'$ & $\bar{3}'m', \bar{3}'m$  \rule[0pt]{0pt}{8pt}\\
& $321'$ \\
& $3m1'$ \\
\hline

$\bar{3}1'$ & $\bar{3}1'$ & $\bar{3}'$  \rule[0pt]{0pt}{8pt}\\
& $31'$ \\

\hline \hline
\end{tabular}
\end{table}

\begin{table}[h!]
\centering
\caption{
Magnetic Laue group (MLG) for crystallographic point group (CPG) and the $\mathcal{PT}$-breaking black and white point group (BWPG).
 \label{table:MLG_CPG}}
\vspace{2mm}
\begin{tabular}{ccccccccc}
\cline{1-2} \noalign{\vspace{2pt}} \cline{1-2} 
\noalign{\vspace{-2pt}}
\cline{4-5} \noalign{\vspace{2pt}} \cline{4-5} 
 MLG & CPG & \hspace{5mm} & MLG & BWPG \\\cline{1-2} \cline{4-5}
$m\bar{3}m$ & $m\bar{3}m$  & &
$m\bar{3}m'$ &  $m\bar{3}m'$
\rule[0pt]{0pt}{8pt} \\
& $432$  & & & $4'32'$ \\
& $\bar{4}3m$ & & & $\bar{4}'3m'$ \\
\cline{1-2} \cline{4-5}

$m\bar{3}$ & $m\bar{3}$ & & 
$4/mm'm'$ &  $4/mm'm'$  \rule[0pt]{0pt}{8pt}\\

&  $23$ & & & $42'2'$ \\
\cline{1-2}

$4/mmm$ & $4/mmm$& &
 & $\bar{4}2'm'$ \\
 
& $422$ & & 
& $4m'm'$ \\
\cline{4-5}

& $\bar{4}2m$ & & 
$4'/mm'm$ & $4'/mm'm$ \\

& $4mm$ &&
& $4'22'$ \\
\cline{1-2}

$4/m$ & $4/m$ & & 
& $\bar{4}'2m', \bar{4}'2'm$\\

& $4$ & & 
 & $4'm'm$ \\
 \cline{4-5}
 
& $\bar{4}$ & &
 $4'/m$ & $4'/m$ \\ \cline{1-2}

$mmm$ & $mmm$ & & 
& $4'$ \\

& $222$ & &
 & $\bar{4}'$ \\
 \cline{4-5}
 
& $mm2$ & & 
$m'm'm$ & $m'm'm$\\
\cline{1-2}

$2/m$ & $2/m$ & & 
& $2'2'2$\\

& $2$ & & 
& $m'm'2, m'm2'$ \\
\cline{4-5}

& $m$ & & 
$2'/m'$ & $2'/m'$ \\
\cline{1-2}

$\bar{1}$ & $\bar{1}$ & &
& $2'$  \rule[0pt]{0pt}{8pt}\\

& $1$ & &
& $m'$\\
\cline{1-2}
\cline{4-5}

$6/mmm$ & $6/mmm$ & & 
$6/mm'm'$ & $6/mm'm'$\\

& $622$ & & 
 & $62'2'$ \\

& $\bar{6}m2$ & & 
& $\bar{6}m'2'$ \\

& $6mm$ & & 
& $6m'm'$ \\
\cline{4-5}

\cline{1-2}
$6/m$ & $6/m$ & & 
$6'/m'mm'$ & $6'/m'mm'$\\

& $6$ & & 
& $6'22'$ \\

& $\bar{6}$ & &
& $\bar{6}'m2', \bar{6}'m'2$ \\
\cline{1-2}

$\bar{3}m$ & $\bar{3}m$ & & 
& $6'mm'$  \rule[0pt]{0pt}{8pt}\\
\cline{4-5}

& $32$ & & 
$6'/m'$ & $6'/m'$\\

& $3m$ & & 
& $6'$ 
\\ 
\cline{1-2}

$\bar{3}$ & $\bar{3}$ & & 
 & $\bar{6}'$  \rule[0pt]{0pt}{8pt}\\
\cline{4-5}

& $3$ & & 
 $\bar{3}m'$ & $\bar{3}m'$  \rule[0pt]{0pt}{8pt}\\
\cline{1-2} \noalign{\vspace{2pt}} \cline{1-2} 
\noalign{\vspace{-2pt}}

& & & & $32'$ \\

& & & & $3m'$ \\ 
\cline{4-5} \noalign{\vspace{2pt}} \cline{4-5} 
\end{tabular}
\end{table}

\clearpage
\section{Derivation of multipoles in response tensors \label{sec:ap_res_mp}}
We present the expressions of the multipoles in the response tensor components in Sec.~\ref{sec:res_symmetry}.
The multipoles in the rank-2 tensors are shown in Secs.~\ref{sec:ap_res_mp_11} and \ref{sec:ap_res_mp_20},  the rank-3 tensors in Secs.~\ref{sec:ap_res_mp_12} and \ref{sec:ap_res_mp_30}, and the rank-4 tensors in Secs.~\ref{sec:ap_res_mp_13} and \ref{sec:ap_res_mp_22}.

\subsection{$\chi^{[1\times1]}$ \label{sec:ap_res_mp_11}}
We decompose the rank-2 tensor $\chi^{[1\times1]}$ into the monopole, dipole, and quadrupole components, which are given as 
\begin{align}
\label{eq:M11}
\chi^{ {\rm M}(1\times1)} &= \frac{1}{3}\sum_{i}\chi_{i;i}^{[1\times1]},\\
\label{eq:D11}
\chi^{{\rm D} (1\times1)}_i &=\frac{1}{2} \sum_{jk} \epsilon_{ijk}\chi_{j;k}^{[1\times1]}, \\
\label{eq:Q11}
\chi^{{\rm Q}(1\times1)}_{ij} &= \frac{1}{2}\left(\chi_{i;j}^{[1\times1]}+\chi_{j;i}^{[1\times1]} \right)= \chi^{{\rm Q}(1\times1)}_{ji}, 
\end{align}
respectively ($i,j=x,y,z$).
$\epsilon_{ijk}$ is the totally antisymmetric tensor (Levi-Civita symbol).
The upper script of $\chi^{{\rm X}(l_{B}\times l_{F})}$ represents the ranks of the response (output), $l_{B}$, and the external field (input), $l_{F}$, in terms of the spherical tensors.
By using Eqs.~\eqref{eq:M11}--\eqref{eq:Q11}, the multipoles in Eq.~\eqref{eq:rank11} are expressed as 
\begin{align}
\label{eq:monopole_11}
X_0 &=\chi^{{\rm M}(1\times1)}, \\
\label{eq:dipole_11}
(Y_x, Y_y, Y_z) &= (\chi^{{\rm D} (1\times1)}_x, \chi^{{\rm D} (1\times1)}_y, \chi^{{\rm D} (1\times1)}_z), \\
X_u &= \displaystyle{\frac{1}{6}\left(3\chi^{{\rm Q}(1\times1)}_{zz}-\sum_{i}\chi^{{\rm Q}(1\times1)}_{ii}\right)}, \notag \\
X_\varv &= \frac{1}{2}\left(\chi^{{\rm Q}(1\times1)}_{xx}-\chi^{{\rm Q}(1\times1)}_{yy}\right), \notag \\
\label{eq:quadrupole_11}
(X_{yz}, X_{zx}, X_{xy}) &= (\chi^{{\rm Q}(1\times1)}_{yz}, \chi^{{\rm Q}(1\times1)}_{zx}, \chi^{{\rm Q}(1\times1)}_{xy}).
\end{align}

\subsection{$\chi^{[0\times2]}$ \label{sec:ap_res_mp_20}}
$\chi^{[0\times2]}$ is decomposed into the monopole and quadrupole in Eq.~\eqref{eq:rank20}, which are represented by
\begin{align}
\label{eq:monopole_20}
X_0 &= \frac{1}{3}\left(\chi^{[0\times2]}_{0;xx}+\chi^{[0\times2]}_{0;yy}+\chi^{[0\times2]}_{0;zz}\right), \\
X_u &= \frac{1}{6}\left(
3\chi^{[0\times2]}_{0;zz}-\sum_i \chi^{[0\times2]}_{0;ii}
\right), \notag \\
X_\varv &= \frac{1}{2}\left(\chi^{[0\times2]}_{0;xx}-\chi^{[0\times2]}_{0;yy}\right),\notag \\
\label{eq:quadrupole_20}
(X_{yz}, X_{zx}, X_{xy}) &= (\chi^{[0\times2]}_{0;yz}, \chi^{[0\times2]}_{0;zx}, \chi^{[0\times2]}_{0;xy}).
\end{align}

\subsection{$\chi^{[1\times2]}$ \label{sec:ap_res_mp_12}}
$\chi^{[1\times2]}$ consists of the dipole, quadrupole, and octupole components, which are represented by using the component $\chi^{[1\times2]}_{i;jk} (=\chi^{[1\times2]}_{i;kj})$ as 
\begin{align}
\label{eq:D12_2}
\chi_i^{{\rm D}(1\times0)}&= \frac{1}{3}\sum_{j}\chi_{i;jj}^{[1\times2]}, \\
\label{eq:D12}
\chi_i^{{\rm D}(1\times 2)}&= \sum_{j}\left(\frac{1}{3}\chi_{i;jj}^{[1\times2]} -\chi_{j;ij}^{[1\times2]} \right), \\
\label{eq:Q12}
\chi_{ij}^{{\rm Q}(1\times2)} &= \frac{1}{2}\sum_{kl} \left(\epsilon_{ikl}\chi_{k;lj}^{[1\times2]} +\epsilon_{jkl}\chi_{k;li}^{[1\times2]}\right)=\chi_{ji}^{{\rm Q}(1\times2)} , \\
\label{eq:O12}
\chi_{ijk}^{{\rm O}(1\times2)} &= \frac{1}{3} \left(\chi_{i;jk}^{[1\times2]}+\chi_{j;ki}^{[1\times2]}+\chi_{k;ij}^{[1\times2]}\right)=\chi_{jki}^{{\rm O}(1\times2)}=\chi_{jik}^{{\rm O}(1\times2)}.
\end{align}
It is noted that there are two dipole components in $\chi^{[1\times2]}$, as the symmetric tensor field $F^{[2]}=(F_{xx}, F_{yy}, F_{zz}, F_{yz}, F_{zx}, F_{xy})$ is decomposed into the components with $l_{F}=0$ and with $l_{F}=2$. 
The $l_{F}=0$ component in $F^{[2]}$, 
leads to 
$\chi_i^{{\rm D}(1\times0)}$ in Eq.~\eqref{eq:D12_2} and the $l_{F}=2$ component in $F^{[2]}$ leads to 
$\chi_i^{{\rm D}(1\times 2)}$, $\chi_{ij}^{{\rm Q}(1\times2)}$, and $\chi_{ijk}^{{\rm O}(1\times2)}$ in Eqs.~\eqref{eq:D12}--\eqref{eq:O12}.

The corresponding multipoles in Eq.~\eqref{eq:rank12} are expressed by $\chi_i^{{\rm D}(1\times0)}$, $\chi_i^{{\rm D}(1\times 2)}$, $\chi_{ij}^{{\rm Q}(1\times2)}$, and $\chi_{ijk}^{{\rm O}(1\times2)}$ in Eqs.~\eqref{eq:D12_2}--\eqref{eq:O12} as 
\begin{align}
\label{eq:dipole_12}
(Y_x, Y_y, Y_z) &= \frac{1}{10}(\chi^{{\rm D}(1\times2)}_x, \chi^{{\rm D}(1\times2)}_y, \chi^{{\rm D}(1\times2)}_z), \\
(Y'_x, Y'_y, Y'_z) &= (\chi^{{\rm D}(1\times0)}_x, \chi^{{\rm D}(1\times0)}_y, \chi^{{\rm D}(1\times0)}_z), \\
X_u &= \frac{1}{6}\left(3\chi^{{\rm Q}(1\times2)}_{zz}-\sum_{i} \chi^{{\rm Q}(1\times2)}_{ii}\right), \notag \\
X_\varv &= \frac{1}{6}\left(\chi^{{\rm Q}(1\times2)}_{xx}-\chi^{{\rm Q}(1\times2)}_{yy}\right), \notag\\
\label{eq:quadrupole_12}
(X_{yz}, X_{zx}, X_{xy}) &= \frac{1}{3}(\chi^{{\rm Q}(1\times2)}_{yz}, \chi^{{\rm Q}(1\times2)}_{zx}, \chi^{{\rm Q}(1\times2)}_{xy}), \\
Y_{xyz} &= \displaystyle{\chi^{{\rm O}(1\times2)}_{xyz}}, \notag\\
Y_x^\alpha &= \frac{1}{20}\left(5\chi^{{\rm O}(1\times2)}_{xxx}-3\sum_{i}\chi^{{\rm O}(1\times2)}_{xii}\right), \notag\\
Y_y^\alpha &= \frac{1}{20}\left(5\chi^{{\rm O}(1\times2)}_{yyy}-3\sum_{i}\chi^{{\rm O}(1\times2)}_{yii}\right), \notag\\
Y_z^\alpha &= \frac{1}{20}\left(5\chi^{{\rm O}(1\times2)}_{zzz}-3\sum_{i}\chi^{{\rm O}(1\times2)}_{zii}\right), \notag\\
Y_x^\beta &= \frac{1}{4}\left(\chi^{{\rm O}(1\times2)}_{xyy}-\chi^{{\rm O}(1\times2)}_{zzx}\right), \notag\\
Y_y^\beta &= \frac{1}{4}\left(\chi^{{\rm O}(1\times2)}_{yzz}-\chi^{{\rm O}(1\times2)}_{xxy}\right), \notag\\
\label{eq:octupole_12}
Y_z^\beta &= \frac{1}{4}\left(\chi^{{\rm O}(1\times2)}_{zxx}-\chi^{{\rm O}(1\times2)}_{yyz}\right). 
\end{align}
We set 
\begin{align}
\label{eq:dipole_12_tilde}
(\tilde{X}_x, \tilde{X}'_x) \equiv (X_{x}^\prime-4X_{x}, X_{x}^\prime+2X_{x}),  \mbox{(cyclic)},
\end{align}
for notational simplicity.

\subsection{$\chi^{[0\times3]}$  \label{sec:ap_res_mp_30}}
The multipoles in Eq.~\eqref{eq:rank30} are represented by using $\chi^{[0\times3]}_{0;ijk}$, which is totally symmetric for the permutation of $i$, $j$, and $k$, as follows:
\begin{align}
\label{eq:dipole_30}
(X_x, X_y, X_z) &= \frac{1}{5}\left(\sum_{i}\chi^{[0\times3]}_{0;xii}, \sum_{i}\chi^{[0\times3]}_{0;yii}, \sum_{i}\chi^{[0\times3]}_{0;zii}\right), \\
X_{xyz} &= \chi^{[0\times3]}_{0;xyz}, \notag\\
X_x^\alpha &= \frac{1}{10}\left(5\chi^{[0\times3]}_{0;xxx}-3\sum_{i} \chi^{[0\times3]}_{0;xii}\right), \notag\\
X_y^\alpha &= \frac{1}{10}\left(5\chi^{[0\times3]}_{0;yyy}-3\sum_{i} \chi^{[0\times3]}_{0;yii}\right), \notag\\
X_z^\alpha &= \frac{1}{10}\left(5\chi^{[0\times3]}_{0;zzz}-3\sum_{i} \chi^{[0\times3]}_{0;zii}\right), \notag\\
X_x^\beta &= \frac{1}{2}\left(\chi^{[0\times3]}_{0;xyy}-\chi^{[0\times3]}_{0;zzx}\right), \notag\\
X_y^\beta &= \frac{1}{2}\left(\chi^{[0\times3]}_{0;yzz}-\chi^{[0\times3]}_{0;xxy}\right), \notag\\
\label{eq:octupole_30}
X_z^\beta &= \frac{1}{2}\left(\chi^{[0\times3]}_{0;zxx}-\chi^{[0\times3]}_{0;yyz}\right).
\end{align}

\begin{widetext}
\subsection{$\chi^{[1\times3]}$ \label{sec:ap_res_mp_13}}
The monopole, dipole, quadrupole, octupole, and hexadecapole components of $\chi^{[1\times3]}$ are represented by using  $\chi^{[1\times3]}_{i;jkl}$, which is totally symmetric by the permutation of $j$, $k$, and $l$, as
\begin{align}
\label{eq:M13}
\chi^{{\rm M}(1\times1)} &= \frac{1}{3}\sum_{ij}\chi_{i;ijj}^{[1\times3]}, \\ 
\label{eq:D13}
\chi^{{\rm D}(1\times1)}_i &=  \frac{1}{2}\sum_{jkl} \epsilon_{ijk}\chi_{j;kll}^{[1\times3]}, \\
\label{eq:Q13_2}
\chi^{{\rm Q}(1\times1)}_{ij} &= \frac{1}{6} \sum_{k}\left(\chi_{i;jkk}^{[1\times3]}+\chi_{j;ikk}^{[1\times3]} \right)=\chi^{{\rm Q}(1\times1)}_{ji},\\
\label{eq:Q13}
\chi^{{\rm Q}(1\times 3)}_{ij} &= \frac{1}{2}\sum_{k}\left[\left(\chi_{k;ijk}^{[1\times3]}+\chi_{k;jik}^{[1\times3]} \right)-\frac{2}{5}\left(\chi_{i;jkk}^{[1\times3]}+\chi_{j;ikk}^{[1\times3]}\right)\right] =\chi^{{\rm Q}(1\times 3)}_{ji}, \\
\label{eq:O13}
\chi^{{\rm O}(1\times3)}_{ijk} &=
\frac{1}{6}\sum_{lm}\left(\epsilon_{klm}\chi_{l;ijm}^{[1\times3]} + \epsilon_{ilm}\chi_{l;jkm}^{[1\times3]}+\epsilon_{jlm}\chi_{l;kim}^{[1\times3]} \right) =\chi^{{\rm O}(1\times3)}_{jki}=\chi^{{\rm O}(1\times3)}_{jik}, \\
\label{eq:H13}
\chi^{{\rm H}(1\times3)}_{ijkl} &=\frac{1}{4}\left(\chi_{i;jkl}^{[1\times3]}+\chi_{j;kli}^{[1\times3]}+\chi_{k;lij}^{[1\times3]}+\chi_{l;ijk}^{[1\times3]}\right) =\chi^{{\rm H}(1\times3)}_{jkli}
= \chi^{{\rm H}(1\times3)}_{jikl}.
\end{align} 
The field $F^{[3]}=(F_{xxx}, F_{yyy}, F_{zzz}, F_{yyz}, F_{zzx}, F_{xxy}, F_{yzz}, F_{zxx}, F_{xyy}, F_{xyz})$ is decomposed into the $l_{F}=1$ and $3$ components.
The $l_{F}=1$ field in $F^{[3]}$ leads to the monopole, dipole, and quadrupole components, 
$\chi^{{\rm M}(1\times1)}$,  $\chi^{{\rm D}(1\times1)}$, and $\chi^{{\rm Q}(1\times1)}$, whereas the $l_{F}=3$ field in $F^{[3]}$ results in the quadrupole, octupole, and hexadecapole components, $\chi^{{\rm Q}(1\times 3)}$, $\chi^{{\rm O}(1\times3)}$, and $\chi^{{\rm H}(1\times3)}$.

By using Eqs.~\eqref{eq:M13}--\eqref{eq:H13}, the multipoles in Eq.~\eqref{eq:rank13} are shown as 
\begin{align}
\label{eq:monopole_13}
X_0 &= \frac{1}{5}\chi^{{\rm M}(1\times1)}, \\
\label{eq:dipole_13}
Y_x &= \frac{1}{5} \chi^{{\rm D}(1\times1)}_x, \ 
Y_y = \frac{1}{5} \chi^{{\rm D}(1\times1)}_y, \ 
Y_z = \frac{1}{5} \chi^{{\rm D}(1\times1)}_z, \\ 
\label{eq:quadrupole_13}
X_u& = \frac{1}{42}\left(3\chi^{{\rm Q}(1\times3)}_{zz}-\sum_{i}\chi^{{\rm Q}(1\times3)}_{ii}\right), \
X_\varv = \frac{1}{42}\left(\chi^{{\rm Q}(1\times3)}_{xx}-\chi^{{\rm Q}(1\times3)}_{yy}\right), \
X_{yz} =\frac{1}{21}\chi^{{\rm Q}(1\times3)}_{yz}, \ 
X_{zx} =\frac{1}{21}\chi^{{\rm Q}(1\times3)}_{zx}, \ 
X_{xy} =\frac{1}{21}\chi^{{\rm Q}(1\times3)}_{xy}, \\
\label{eq:monopole_13_prime}
X'_u &= \frac{1}{10}\left(3\chi^{{\rm Q}(1\times1)}_{zz}-\sum_{i}\chi^{{\rm Q}(1\times1)}_{ii}\right),\
X'_\varv = \frac{3}{10}\left(\chi^{{\rm Q}(1\times1)}_{xx}-\chi^{{\rm Q}(1\times1)}_{yy}\right), \
X'_{yz} = \frac{3}{5}\chi^{{\rm Q}(1\times1)}_{yz}, \ 
X'_{zx} = \frac{3}{5}\chi^{{\rm Q}(1\times1)}_{zx}, \ 
X'_{xy} = \frac{3}{5}\chi^{{\rm Q}(1\times1)}_{xy}, \\
Y_{xyz} &= \chi^{{\rm O}(1\times3)}_{xyz}, \notag\\
Y_x^\alpha &= \frac{1}{20}\left(5\chi^{{\rm O}(1\times3)}_{xxx}-3\sum_{i}\chi^{{\rm O}(1\times3)}_{xii}\right), \
Y_y^\alpha = \frac{1}{20}\left(5\chi^{{\rm O}(1\times3)}_{yyy}-3\sum_{i}\chi^{{\rm O}(1\times3)}_{yii}\right), \
Y_z^\alpha = \frac{1}{20}\left(5\chi^{{\rm O}(1\times3)}_{zzz}-3\sum_{i}\chi^{{\rm O}(1\times3)}_{zii}\right), \notag\\
\label{eq:octupole_13}
Y_x^\beta &= \frac{1}{4}\left(\chi^{{\rm O}(1\times3)}_{xyy}-\chi^{{\rm O}(1\times3)}_{zzx}\right), \
Y_y^\beta = \frac{1}{4}\left(\chi^{{\rm O}(1\times3)}_{yzz}-\chi^{{\rm O}(1\times3)}_{xxy}\right), \
Y_z^\beta = \frac{1}{4}\left(\chi^{{\rm O}(1\times3)}_{zxx}-\chi^{{\rm O}(1\times3)}_{yyz}\right), \\
X_{4} &= \frac{1}{15}\left[\sum_{i} \chi^{{\rm H}(1\times3)}_{iiii}-3\left(\chi^{{\rm H}(1\times3)}_{yyzz}+\chi^{{\rm H}(1\times3)}_{zzxx}+\chi^{{\rm H}(1\times3)}_{xxyy} \right) \right], \notag\\
X_{4u} &= \frac{1}{42}\left[3\chi^{{\rm H}(1\times3)}_{zzzz}-\sum_{i}\chi^{{\rm H}(1\times3)}_{iiii} +6\left(2\chi^{{\rm H}(1\times3)}_{xxyy} -\chi^{{\rm H}(1\times3)}_{yyzz}-\chi^{{\rm H}(1\times3)}_{zzxx}\right) \right], \
X_{4\varv} = \frac{1}{14}\left[\chi^{{\rm H}(1\times3)}_{xxxx}-\chi^{{\rm H}(1\times3)}_{yyyy} +6 \left( \chi^{{\rm H}(1\times3)}_{yyzz} - \chi^{{\rm H}(1\times3)}_{zzxx} \right) \right], \notag\\
X_{4x}^\alpha &= \frac{1}{2}\left(\chi^{{\rm H}(1\times3)}_{yyyz}-\chi^{{\rm H}(1\times3)}_{yzzz}\right), \
X_{4y}^\alpha = \frac{1}{2}\left(\chi^{{\rm H}(1\times3)}_{zzzx}-\chi^{{\rm H}(1\times3)}_{zxxx}\right), \
X_{4z}^\alpha = \frac{1}{2}\left(\chi^{{\rm H}(1\times3)}_{xxxy}-\chi^{{\rm H}(1\times3)}_{xyyy}\right), \notag\\
\label{eq:hexadecapole_13}
X_{4x}^\beta &= \frac{1}{14}\left(7\chi^{{\rm H}(1\times3)}_{xxyz}-\sum_{i}\chi^{{\rm H}(1\times3)}_{iiyz}\right), \
X_{4y}^\beta = \frac{1}{14}\left(7\chi^{{\rm H}(1\times3)}_{yyzx}-\sum_{i}\chi^{{\rm H}(1\times3)}_{iizx}\right), \
X_{4z}^\beta = \frac{1}{14}\left(7\chi^{{\rm H}(1\times3)}_{zzxy}-\sum_{i}\chi^{{\rm H}(1\times3)}_{iixy}\right).
\end{align}
For notational simplicity, we take
\begin{align}
(\tilde{X}_u,\tilde{X}'_u) &\equiv (X_u+X'_u, 4X_u-X'_u), \ 
\tilde{X}''_u\equiv \tilde{X}_u-\tilde{X}'_u, \  
(\tilde{X}_\varv, \tilde{X}'_\varv) \equiv (3 {X}_{\varv} + {X}'_{\varv}, 2 {X}_{\varv}-{X}'_{\varv}), \ 
\tilde{X}''_\varv \equiv \tilde{X}_\varv +2\tilde{X}'_\varv, \notag\\
\label{eq:quadrupole_13_tilde}
(\tilde{X}_{yz}, \tilde{X}'_{yz}) &\equiv (2 {X}_{yz}- {X}'_{yz}, 8 {X}_{yz}+ {X}'_{yz}),  \mbox{(cyclic)}.
\end{align}

\subsection{$\chi^{[2\times2]}$ \label{sec:ap_res_mp_22}}
For the rank-4 tensor $\chi^{[2\times2]}$
\begin{align}
\chi^{[2\times 2]}=
\begin{pmatrix}
\chi_{ll} & \chi_{lt}\\
\chi_{tl} & \chi_{tt}\\
\end{pmatrix},
\end{align}
\begin{align}
\chi_{ll}=
\begin{pmatrix}
\chi_{xx;xx}^{[2\times2]} & \chi_{xx;yy}^{[2\times2]} & \chi_{xx;zz}^{[2\times2]}\\
\chi_{yy;xx}^{[2\times2]} & \chi_{yy;yy}^{[2\times2]} & \chi_{yy;zz}^{[2\times2]}\\
\chi_{zz;xx}^{[2\times2]} & \chi_{zz;yy}^{[2\times2]} & \chi_{zz;zz}^{[2\times2]}\\
\end{pmatrix}, \ 
\chi_{lt}=
\begin{pmatrix}
\chi_{xx;yz}^{[2\times2]} & \chi_{xx;zx}^{[2\times2]} & \chi_{xx;xy}^{[2\times2]} \\
\chi_{yy;yz}^{[2\times2]} & \chi_{yy;zx}^{[2\times2]} & \chi_{yy;xy}^{[2\times2]} \\
\chi_{zz;yz}^{[2\times2]} & \chi_{zz;zx}^{[2\times2]} & \chi_{zz;xy}^{[2\times2]} \\
\end{pmatrix}, \ 
\chi_{tl}=
\begin{pmatrix}
\chi_{yz;xx}^{[2\times2]} & \chi_{yz;yy}^{[2\times2]} & \chi_{yz;zz}^{[2\times2]} \\
\chi_{zx;xx}^{[2\times2]} & \chi_{zx;yy}^{[2\times2]} & \chi_{zx;zz}^{[2\times2]} \\
\chi_{xy;xx}^{[2\times2]} & \chi_{xy;yy}^{[2\times2]} & \chi_{xy;zz}^{[2\times2]} \\
\end{pmatrix}, \ 
\chi_{tt}=
\begin{pmatrix}
\chi_{yz;yz}^{[2\times2]} & \chi_{yz;zx}^{[2\times2]} & \chi_{yz;xy}^{[2\times2]} \\
\chi_{zx;yz}^{[2\times2]} & \chi_{zx;zx}^{[2\times2]} & \chi_{zx;xy}^{[2\times2]} \\
\chi_{xy;yz}^{[2\times2]} & \chi_{xy;zx}^{[2\times2]} & \chi_{xy;xy}^{[2\times2]} \\
\end{pmatrix},
\end{align}
the monopole, dipole, quadrupole, octupole, and hexadecapole components are expressed by using the tensor component $\chi_{ij;kl}^{[2\times2]} (=\chi_{ji;kl}^{[2\times2]}=\chi_{ij;lk}^{[2\times2]}$) as
\begin{align}
\label{eq:M22_2}
\chi^{{\rm M}(0\times0)} &= \frac{1}{3} \sum_{ij} \chi_{ii;jj}^{[2\times 2]}, \\
\label{eq:Q22_2}
\chi_{ij}^{{\rm Q}(0\times2, \pm)}&= \frac{1}{6}\sum_{k}\left(\chi_{kk;ij}^{[2\times2]}\pm\chi_{ij;kk}^{[2\times2]}\right), \\
\label{eq:M22}
\chi^{{\rm M}(2\times2)} &= \frac{1}{3} \sum_{ij}\left(\chi_{ij;ji}^{[2\times 2]}-\frac{1}{3}\chi_{ii;jj}^{[2\times2]}\right), \\
\label{eq:D22}
\chi_{i}^{{\rm D}(2\times2)} &= \frac{1}{2}\sum_{jkl}\epsilon_{ijk}\chi_{lj;kl}^{[2\times 2]}, \\
\label{eq:Q22}
\chi_{ij}^{{\rm Q}(2\times2)}&= \frac{1}{2}\sum_{k}\left[\left(\chi_{ik;kj}^{[2\times 2]}+\chi_{jk;ki}^{[2\times 2]}\right)
-\frac{2}{3}\left(\chi_{ij;kk}^{[2\times2]}+\chi_{kk;ij}^{[2\times2]}\right)\right]
=\chi_{ji}^{{\rm Q}(2\times2)} , \\
\label{eq:O22}
\chi_{ijk}^{{\rm O}(2\times2)}
&=\frac{1}{6}\sum_{lm}
\left(
\epsilon_{ilm}\chi_{jl;mk}^{[2\times 2]}+ \epsilon_{jlm}\chi_{kl;mi}^{[2\times 2]}+\epsilon_{klm}\chi_{il;mj}^{[2\times 2]}
+\epsilon_{ilm}\chi_{kl;mj}^{[2\times 2]}+\epsilon_{jlm}\chi_{il;mk}^{[2\times 2]}+\epsilon_{klm}\chi_{jl;mi}^{[2\times 2]}
 \right)
=\chi_{jki}^{{\rm O}(2\times2)}=\chi_{jik}^{{\rm O}(2\times2)}, \\
\label{eq:H22}
\chi_{ijkl}^{{\rm H}(2\times2)} &=\frac{1}{6} \left(\chi_{ij;kl}^{[2\times 2]}+\chi_{ik;jl}^{[2\times 2]}+\chi_{il;kj}^{[2\times 2]}+\chi_{kj;il}^{[2\times 2]} +\chi_{lj;ki}^{[2\times 2]}+\chi_{kl;ij}^{[2\times 2]}\right) 
=\chi_{jkli}^{{\rm H}(2\times2)}
=\chi_{jikl}^{{\rm H}(2\times2)}.
\end{align}
Since both $B^{[2]}=(B_{xx}, B_{yy}, B_{zz}, B_{yz}, B_{zx}, B_{xy})$ and $F^{[2]}=(F_{xx}, F_{yy}, F_{zz}, F_{yz}, F_{zx}, F_{xy})$ contain
$l_{B}, l_{F}=0$ and $2$ components, there are two types of monopole components $\chi^{{\rm M}(0\times0)}$ and $\chi^{{\rm M}(2\times2)}$ and three types of quadrupole component $\chi_{ij}^{{\rm Q}(0\times2, \pm)}$ and $\chi_{ij}^{{\rm Q}(2\times2)}$.
By using Eqs.~\eqref{eq:M22_2}--\eqref{eq:H22}, the multipoles in Eqs.~\eqref{eq:rank22ll}--\eqref{eq:rank22tt} are represented by
\begin{align}
\label{eq:monopole_22}
X_0 &= \frac{1}{10}\chi^{{\rm M}(2\times2)}, \
X'_0 = \frac{1}{3}\chi^{{\rm M}(0\times0)}, \\
\label{eq:dipole_22}
Y_x &= \frac{1}{5} \chi^{{\rm D}(2\times2)}_x, \
Y_y = \frac{1}{5} \chi^{{\rm D}(2\times2)}_y, \
Y_z = \frac{1}{5} \chi^{{\rm D}(2\times2)}_z, \\
\label{eq:quadrupole_22}
X_u &= \frac{1}{42}\left(3\chi^{{\rm Q}(2\times2)}_{zz} - \sum_i \chi^{{\rm Q}(2\times2)}_{ii}
\right), \
X_\varv = \frac{1}{14}\left(\chi^{{\rm Q}(2\times2)}_{xx}-\chi^{{\rm Q}(2\times2)}_{yy}\right), \
X_{yz} = \frac{1}{7} \chi^{{\rm Q}(2\times2)}_{yz}, \ 
X_{zx} = \frac{1}{7} \chi^{{\rm Q}(2\times2)}_{zx}, \ 
X_{xy} = \frac{1}{7} \chi^{{\rm Q}(2\times2)}_{xy}, \\
\label{eq:quadrupole_22_prime}
X^{(\pm)}_u &= \frac{1}{6}\left(3 \chi^{{\rm Q}(0\times2,\pm)}_{zz} - \sum_i \chi^{{\rm Q}(0\times2,\pm)}_{ii}
\right), \
X^{(\pm)}_\varv = \frac{1}{2}\left(\chi^{{\rm Q}(0\times2,\pm)}_{xx}-\chi^{{\rm Q}(0\times2,\pm)}_{yy}\right), \
X^{(\pm)}_{yz} = \chi^{{\rm Q}(0\times2,\pm)}_{yz}, \ 
X^{(\pm)}_{zx} = \chi^{{\rm Q}(0\times2,\pm)}_{zx}, \ 
X^{(\pm)}_{xy} = \chi^{{\rm Q}(0\times2,\pm)}_{xy}, \\
Y_{xyz} &= \chi^{{\rm O}(2\times2)}_{xyz}, \notag \\
Y_x^\alpha &= \frac{1}{20}\left(
5\chi^{{\rm O}(2\times2)}_{xxx}-3\sum_i \chi^{{\rm O}(2\times2)}_{xii}
\right), \
Y_y^\alpha = \frac{1}{20}\left(5\chi^{{\rm O}(2\times2)}_{yyy}-3\sum_i \chi^{{\rm O}(2\times2)}_{yii}
\right), \
Y_z^\alpha = \frac{1}{20}\left(5\chi^{{\rm O}(2\times2)}_{zzz}-3\sum_i \chi^{{\rm O}(2\times2)}_{zii}
\right), \notag\\
\label{eq:octupole_22}
Y_x^\beta &= \frac{1}{4}\left(\chi^{{\rm O}(2\times2)}_{xyy}-\chi^{{\rm O}(2\times2)}_{zzx}\right), \
Y_y^\beta = \frac{1}{4}\left(\chi^{{\rm O}(2\times2)}_{yzz}-\chi^{{\rm O}(2\times2)}_{xxy}\right), \
Y_z^\beta = \frac{1}{4}\left(\chi^{{\rm O}(2\times2)}_{zxx}-\chi^{{\rm O}(2\times2)}_{yyz}\right), \\
X_{4} &= \frac{1}{6}\left(\sum_{i}\chi^{{\rm H}(2\times2)}_{iiii}-\frac{3}{5}\sum_{ij}\chi^{{\rm H}(2\times2)}_{iijj} \right),
\
X_{4u} = \frac{1}{6}\left[ \left(3\chi^{{\rm H}(2\times2)}_{zzzz}-\sum_{i}\chi^{{\rm H}(2\times2)}_{iiii}\right)-
\frac{6}{7}\sum_{i}\left(2\chi^{{\rm H}(2\times2)}_{iizz} -\chi^{{\rm H}(2\times2)}_{iixx}-\chi^{{\rm H}(2\times2)}_{iiyy}\right) \right],
 \notag\\
X_{4\varv} &= \frac{1}{2}\left[\chi^{{\rm H}(2\times2)}_{xxxx}-\chi^{{\rm H}(2\times2)}_{yyyy} -
\frac{6}{7} \sum_{i}\left( \chi^{{\rm H}(2\times2)}_{iixx} - \chi^{{\rm H}(2\times2)}_{iiyy} \right) \right],
\notag\\
X_{4x}^\alpha &= \frac{1}{2}\left(\chi^{{\rm H}(2\times2)}_{yyyz}-\chi^{{\rm H}(2\times2)}_{yzzz}\right), \
X_{4y}^\alpha = \frac{1}{2}\left(\chi^{{\rm H}(2\times2)}_{zzzx}-\chi^{{\rm H}(2\times2)}_{zxxx}\right), \
X_{4z}^\alpha = \frac{1}{2}\left(\chi^{{\rm H}(2\times2)}_{xxxy}-\chi^{{\rm H}(2\times2)}_{xyyy}\right), \notag\\
\label{eq:hexadecapole_22}
X_{4x}^\beta &= \frac{1}{2}\left(6\chi^{{\rm H}(2\times2)}_{xxyz}-
\frac{1}{7}\sum_{i}\chi^{{\rm H}(2\times2)}_{yzii}\right),
\
X_{4y}^\beta = \frac{1}{2}\left(6\chi^{{\rm H}(2\times2)}_{yyzx}-
\frac{1}{7}\sum_{i}\chi^{{\rm H}(2\times2)}_{zxii}\right),
\
X_{4z}^\beta = \frac{1}{2}\left(6\chi^{{\rm H}(2\times2)}_{zzxy}-\frac{1}{7}\sum_{i}\chi^{{\rm H}(2\times2)}_{xyii}\right).
\end{align}
We use the notation 
\begin{align}
(\tilde{X}_0, \tilde{X}_0^{\prime}) &\equiv (4X_0+X'_0, -2X_0+X'_0), \notag\\ 
(\tilde{X}_u, \tilde{X}_u^{(\pm)}) &\equiv (-4 {X}_{u}-2 {X}_{u}^{(+)},  -4 {X}_{u}+{X}_{u}^{(+)} \pm 3 {X}_{u}^{(-)}), \ \tilde{X}_u^{\prime}\equiv-\tilde{X}_u^{(+)}-\tilde{X}_u^{(-)}, \notag\\
(\tilde{X}_\varv, \tilde{X}_\varv^{(\pm)}) &\equiv (4 {X}_{\varv}+2 {X}^{(+)}_{\varv}, -4 {X}_{\varv}+{X}^{(+)}_{\varv} \pm {X}^{(-)}_{\varv}),  \notag\\
\label{eq:monopole_quadrupole_22_tilde}
(\tilde{X}_{yz}^{(\pm)}, \tilde{X}_{yz}^{\prime(\pm)}) &\equiv (-4X_{yz}+X^{(+)}_{yz}\pm X^{(-)}_{yz}, 2X_{yz}+X^{(+)}_{yz}\pm X^{(-)}_{yz}), \mbox{(cyclic)},
\end{align}
 for simplicity.
 \end{widetext}

\section{Tensor expression in hexagonal/trigonal system \label{sec:ap_hexagonal}}

In the hexagonal and trigonal systems, multipoles are expressed by the tesseral harmonics.
Since the tesseral harmonics have the different functional form from the cubic harmonics for the rank $l \geq 3$, we show the response tensors in the hexagonal and trigonal systems with rank $l\geq 3$ multipoles. 
The corresponding rank-3 and -4 tensors are represented by
\begin{align}
\chi^{[0\times3]}
=
\begin{pmatrix}
 3 {X}_{x}+{X}_{3a}-3 {X}_{3u} \\
 3 {X}_{y}-{X}_{3b}-3 {X}_{3\varv} \\
 3 {X}_{z}+2 {X}_{z}^\alpha \\
 {X}_{z}-{X}_{z}^\alpha-{X}_{z}^\beta \\ 
  {X}_{x}+4 {X}_{3u} \\
   {X}_{y}+{X}_{3b}-{X}_{3\varv} \\ 
 {X}_{y}+4 {X}_{3\varv} \\
 {X}_{z}-{X}_{z}^\alpha+{X}_{z}^\beta \\
  {X}_{x}-{X}_{3a}-{X}_{3u} \\ 
 {X}_{xyz} \\
\end{pmatrix}^{\rm T},
\end{align}
\begin{widetext}
\begin{align}
\chi^{[1\times2]}
=
\begin{pmatrix}
 \tilde{X}_{x}+X_{3a}-3X_{3u}
 & \tilde{X}'_{y}
 -2Y_{zx}+X_{3b} - X_{3\varv}
 & \tilde{X}'_{z}
 +2Y_{xy}-2  X_{z}^\alpha+2X_{z}^\beta\\
\tilde{X}'_{x}
+2 Y_{yz} -X_{3a} - X_{3u}
 &\tilde{X}_{y}-X_{3b} - 3 X_{3\varv}
 &\tilde{X}'_{z}
 -2Y_{xy}-2  X_{z}^\alpha-2X_{z}^\beta\\
 \tilde{X}'_{x}
 -2Y_{yz}+4X_{3u}
 & \tilde{X}'_{y}
 +2Y_{zx}+4 X_{3\varv}
 & \tilde{X}_{z}+4 X_{z}^\alpha \\
Y_{u}
+Y_{\varv}+ X_{xyz}
 & -3 X_{z}
 +Y_{xy}-2 X_{z}^\alpha-2 X_{z}^\beta
 & -3 X_{y}
 -Y_{zx}+4 X_{3\varv} \\
  -3 X_{z}
  -Y_{xy}-2 X_{z}^\alpha+2 X_{z}^\beta
 & -Y_{u}
 +Y_{\varv}+X_{xyz}
 & -3 X_{x}
 +Y_{yz}+4 X_{3u} \\
 -3 X_{y}
 +Y_{zx}+X_{3b} - X_{3\varv}
 & -3 X_{x}
 -Y_{yz}-X_{3a} - X_{3u}
  & -2 Y_{\varv} +X_{xyz}\\
\end{pmatrix}^{\rm T},
\end{align}
\begin{align}
&
\chi^{[1\times3]}
=
\begin{pmatrix}
 3 ({X}_{0}-\tilde{X}_u+\tilde{X}_\varv)+3 {X}_{40}+{X}_{4u}^{\beta1}-{X}_{4u}^{\beta2} 
 & 3 (-{Y}_{z}-\tilde{X}_{xy}
 +{Y}_{z}^\alpha 
 -{Y}_{z}^\beta)+{X}_{4\varv}^{\beta1}-{X}_{4\varv}^{\beta2} 
 & 3(
 {Y}_{y}-\tilde{X}_{zx} 
 +{Y}_{3b}
 -{Y}_{3\varv})+{X}_{4b}-3 {X}_{4u}^\alpha \\
 3(
 {Y}_{z}-\tilde{X}_{xy}
 - {Y}_{z}^\alpha
 -{Y}_{z}^\beta) -{X}_{4\varv}^{\beta1}-{X}_{4\varv}^{\beta2} 
 &3( {X}_{0} - \tilde{X}_u-\tilde{X}_\varv)+3 {X}_{40}+{X}_{4u}^{\beta1}+{X}_{4u}^{\beta2} 
 & 3 (-{Y}_{x}-\tilde{X}_{yz}
 +{Y}_{3a}
 +{Y}_{3u})-{X}_{4a}-3 {X}_{4\varv}^\alpha \\
 -3( {Y}_{y}
 +\tilde{X}_{zx}+4 {Y}_{3\varv})+4 {X}_{4u}^\alpha 
 & 3(
 {Y}_{x}-\tilde{X}_{yz}
 +4 {Y}_{3u})+4 {X}_{4\varv}^\alpha 
 & 3( {X}_{0}+2\tilde{X}_u) +8 {X}_{40} \\
-{Y}_{y}-\tilde{X}_{zx}
+{Y}_{3b}
+11 {Y}_{3\varv}-{X}_{4b}-{X}_{4u}^\alpha 
 & 
 {Y}_{x}+\tilde{X}'_{yz}
 -{Y}_{3a}
 -{Y}_{3u}-{X}_{4a}-3 {X}_{4\varv}^\alpha 
 &{X}_{0}+\tilde{X}''_u-5 {X}_{\varv}
 -{Y}_{xyz}-4 {X}_{40}-{X}_{4u}^{\beta2} \\
 {X}_{0}+\tilde{X}'_u-\tilde{X}'_\varv
 -{Y}_{xyz}-4 {X}_{40}+{X}_{4u}^{\beta2} 
 &-{Y}_{z}-\tilde{X}_{xy}
 -4 {Y}_{z}^\alpha
 +2 {Y}_{z}^\beta+2 {X}_{4\varv}^{\beta2} 
 & 
 {Y}_{y}+\tilde{X}'_{zx}
 +4 {Y}_{3\varv}+4 {X}_{4u}^\alpha \\
{Y}_{z}+\tilde{X}'_{xy}
-{Y}_{z}^\alpha
+{Y}_{z}^\beta+{X}_{4\varv}^{\beta1}-{X}_{4\varv}^{\beta2} 
 &{X}_{0}-\tilde{X}_u+\tilde{X}''_\varv
 -{Y}_{xyz}+{X}_{40}-{X}_{4u}^{\beta1} 
 &-{Y}_{x}-\tilde{X}_{yz}
 -3 {Y}_{3a}
 +{Y}_{3u}+{X}_{4a}-{X}_{4\varv}^\alpha \\ 
 {Y}_{z}-\tilde{X}_{xy}
 +4 {Y}_{z}^\alpha
 +2 {Y}_{z}^\beta+2 {X}_{4\varv}^{\beta2} 
 & {X}_{0}+\tilde{X}'_u+\tilde{X}'_\varv
 +{Y}_{xyz}-4 {X}_{40}-{X}_{4u}^{\beta2} 
 &-{Y}_{x}+\tilde{X}'_{yz} 
 -4 {Y}_{3u}+4 {X}_{4\varv}^\alpha \\
-{Y}_{y}+\tilde{X}'_{zx}
-{Y}_{3b}
+{Y}_{3\varv}+{X}_{4b}-3 {X}_{4u}^\alpha 
 & 
 {Y}_{x}-\tilde{X}_{yz}
 +{Y}_{3a}
 -11 {Y}_{3u}+{X}_{4a}-{X}_{4\varv}^\alpha 
 & {X}_{0}+\tilde{X}''_u+5 {X}_{\varv}
 +{Y}_{xyz}-4 {X}_{40}+{X}_{4u}^{\beta2} \\
{X}_{0} -\tilde{X}_u-\tilde{X}''_\varv
+{Y}_{xyz}+{X}_{40}-{X}_{4u}^{\beta1} 
 &-{Y}_{z}+\tilde{X}'_{xy}
 +{Y}_{z}^\alpha
 +{Y}_{z}^\beta-{X}_{4\varv}^{\beta1}-{X}_{4\varv}^{\beta2} 
 & 
 {Y}_{y}-\tilde{X}_{zx}
 -3 {Y}_{3b}
 -{Y}_{3\varv}-{X}_{4b}-{X}_{4u}^\alpha \\
 5 {X}_{yz}
 +{Y}_{3a}
 +5 {Y}_{3u}+{X}_{4a}-{X}_{4\varv}^\alpha 
 & 5 {X}_{zx}
 +{Y}_{3b}
 -5 {Y}_{3\varv}-{X}_{4b}-{X}_{4u}^\alpha 
 & 5 {X}_{xy}
 -2 {Y}_{z}^\beta+2 {X}_{4\varv}^{\beta2} \\
\end{pmatrix}^{\rm T},
\end{align}
%
\begin{align}
\chi^{[2\times 2]}=
\begin{pmatrix}
\chi_{ll} & \chi_{lt}\\
\chi_{tl} & \chi_{tt}\\
\end{pmatrix}, 
\end{align}
\begin{align}
\chi_{ll}&=
\begin{pmatrix}
\tilde{X}_0+\tilde{X}_u+\tilde{X}_{\varv}+3 {X}_{40}+{X}_{4u}^{\beta1}- {X}_{4u}^{\beta2} 
 & \tilde{X}'_0+\tilde{X}'_u-2 \tilde{X}_{\varv}^{(-)}+{Y}_{xyz}+{X}_{40}-{X}_{4u}^{\beta1} 
 & \tilde{X}'_0+\tilde{X}_u^{(+)}+\tilde{X}_\varv^{(-)}-{Y}_{xyz}-4 {X}_{40}+ {X}_{4u}^{\beta2} 
 \\
\tilde{X}'_0+\tilde{X}'_u+2 \tilde{X}_{\varv}^{(-)}-{Y}_{xyz}+{X}_{40}-{X}_{4u}^{\beta1} 
 & \tilde{X}_0+\tilde{X}_u-\tilde{X}_{\varv}+3 {X}_{40}+{X}_{4u}^{\beta1}+ {X}_{4u}^{\beta2} 
 & \tilde{X}'_0+\tilde{X}_u^{(+)}-\tilde{X}_\varv^{(-)}+{Y}_{xyz}-4 {X}_{40}- {X}_{4u}^{\beta2} 
\\
\tilde{X}'_0+\tilde{X}_u^{(-)}+\tilde{X}_{\varv}^{(+)}+{Y}_{xyz}-4 {X}_{40}+ {X}_{4u}^{\beta2} 
 & \tilde{X}'_0+\tilde{X}_u^{(-)}-\tilde{X}_{\varv}^{(+)}-{Y}_{xyz}-4 {X}_{40}- {X}_{4u}^{\beta2} 
 & \tilde{X}_0-2\tilde{X}_u+8 {X}_{40} 
 \\
\end{pmatrix}, \\
\chi_{lt}&=
\begin{pmatrix}
\tilde{X}_{yz}^{(+)}+{Y}_{3a}+5 {Y}_{3u}+{X}_{4a}-{X}_{4\varv}^{\alpha} 
 & -2 {Y}_{y}+\tilde{X}_{zx}^{\prime(+)}-{Y}_{3b}+{Y}_{3\varv}+{X}_{4b}-3 {X}_{4u}^{\alpha} 
 & 2 {Y}_{z}+\tilde{X}_{xy}^{\prime(+)}-{Y}_{z}^\alpha+ {Y}_{z}^\beta+{X}_{4\varv}^{\beta1}-{X}_{4\varv}^{\beta2} \\
 2 {Y}_{x}+\tilde{X}_{yz}^{\prime(+)}-{Y}_{3a}-{Y}_{3u}-{X}_{4a}-3 {X}_{4\varv}^{\alpha} 
 & \tilde{X}_{zx}^{(+)}+{Y}_{3b}-5 {Y}_{3\varv}-{X}_{4b}-{X}_{4u}^{\alpha} 
 & -2 {Y}_{z}+\tilde{X}_{xy}^{\prime(+)}+ {Y}_{z}^\alpha+ {Y}_{z}^\beta-{X}_{4\varv}^{\beta1}-{X}_{4\varv}^{\beta2} \\
 -2 {Y}_{x}+\tilde{X}_{yz}^{\prime(+)}-4 {Y}_{3u}+4 {X}_{4\varv}^{\alpha} 
 & 2 {Y}_{y}+\tilde{X}_{zx}^{\prime(+)}+ 4 {Y}_{3\varv}+4 {X}_{4u}^{\alpha} 
 & \tilde{X}_{xy}^{(+)} -2 {Y}_{z}^\beta+2 {X}_{4\varv}^{\beta2} \\
\end{pmatrix}, \\
\chi_{tl}&=
\begin{pmatrix}
\tilde{X}_{yz}^{(-)}-{Y}_{3a}-5 {Y}_{3u}+{X}_{4a}-{X}_{4\varv}^{\alpha} 
 &-2 {Y}_{x} +\tilde{X}_{yz}^{\prime(-)}+{Y}_{3a}+{Y}_{3u}-{X}_{4a}-3 {X}_{4\varv}^{\alpha} 
 & 2 {Y}_{x} +\tilde{X}_{yz}^{\prime(-)}+4 {Y}_{3u}+4 {X}_{4\varv}^{\alpha} 
 \\
 2 {Y}_{y}+\tilde{X}_{zx}^{\prime(-)}+{Y}_{3b}-{Y}_{3\varv}+{X}_{4b}-3 {X}_{4u}^{\alpha} 
 & \tilde{X}_{zx}^{(-)} -{Y}_{3b}+5 {Y}_{3\varv}-{X}_{4b}-{X}_{4u}^{\alpha} 
 & -2 {Y}_{y}+ \tilde{X}_{zx}^{\prime(-)}-4 {Y}_{3\varv}+4 {X}_{4u}^{\alpha} 
\\
 -2 {Y}_{z}+\tilde{X}_{xy}^{\prime(-)}+
  {Y}_{z}^\alpha- 
  {Y}_{z}^\beta+{X}_{4\varv}^{\beta1}-{X}_{4\varv}^{\beta2} 
 & 2 {Y}_{z}+\tilde{X}_{xy}^{\prime(-)}- {Y}_{z}^\alpha- {Y}_{z}^\beta-{X}_{4\varv}^{\beta1}-{X}_{4\varv}^{\beta2} 
 & \tilde{X}_{xy}^{(-)}+2 {Y}_{z}^\beta+2 {X}_{4\varv}^{\beta2} 
  \\
\end{pmatrix},\\
\chi_{tt}&=
\begin{pmatrix}
 3 {X}_{0}+3 {X}_{u}-3 {X}_{\varv}-4 {X}_{40}- {X}_{4u}^{\beta2}
  & -{Y}_{z}+3 {X}_{xy}-2 {Y}_{z}^\alpha+2 {X}_{4\varv}^{\beta2} 
  & {Y}_{y}+3 {X}_{zx}-{Y}_{3b}-3 {Y}_{3\varv}-{X}_{4b}-{X}_{4u}^{\alpha} \\
  {Y}_{z}+3 {X}_{xy}+2 {Y}_{z}^\alpha+2 {X}_{4v}^{\beta2} 
 & 3 {X}_{0}+3 {X}_{u}+3 {X}_{\varv}-4 {X}_{40}+ {X}_{4u}^{\beta2} 
 & -{Y}_{x}+3 {X}_{yz}-{Y}_{3a}+3 {Y}_{3u}+{X}_{4a}-{X}_{4\varv}^{\alpha} \\
  -{Y}_{y}+3 {X}_{zx}+{Y}_{3b}+3 {Y}_{3\varv}-{X}_{4b}-{X}_{4u}^{\alpha}
 & {Y}_{x}+3 {X}_{yz}+{Y}_{3a}-3 {Y}_{3u}+{X}_{4a}-{X}_{4\varv}^{\alpha} 
 & 3 {X}_{0}-6 {X}_{u}+{X}_{40}-{X}_{4u}^{\beta1} \\
\end{pmatrix},
\end{align}
where the following relations with respect to the octupoles and hecadecapoles are used as 
\begin{align}
Y_{3a} = \frac{1}{4}(5Y_x^\alpha-3Y_x^\beta),\
Y_{3b} =  -\frac{1}{4}(5Y_y^\alpha+3Y_y^\beta),\
Y_{3u} = -\frac{1}{4}(Y_x^\alpha+Y_x^\beta), \
Y_{3\varv} = -\frac{1}{4}(Y_y^\alpha-Y_y^\beta),  \notag \\
X_{40} = \frac{1}{4}(X_4+X_{4u}), \
X_{4u}^\alpha = \frac{1}{4}(X_{4y}^\alpha-X_{4y}^\beta), \
X_{4\varv}^\alpha = -\frac{1}{4}(X_{4x}^\alpha+X_{4x}^\beta), \
X_{4u}^{\beta1} =\frac{1}{4}(5X_4-7X_{4u}), \
X_{4\varv}^{\beta1} = X_{4z}^\alpha,  \notag \\
X_{4u}^{\beta2} =  -X_{4\varv}, \
X_{4\varv}^{\beta2} = X_{4z}^\beta, \
X_{4a} =-\frac{1}{4}(X_{4x}^\alpha-7X_{4x}^\beta), \
X_{4b} = -\frac{1}{4}(X_{4y}^\alpha+7X_{4y}^\beta),
\label{eq:cubic_hexa_1}
\end{align}
in $\chi^{[0\times3]}$, $\chi^{[1\times3]}$ and $\chi^{[2\times2]}$, while 
\begin{align}
X_{3a} = \frac{1}{2}(5X_x^\alpha-3X_x^\beta),\
X_{3b} =  -\frac{1}{2}(5X_y^\alpha+3X_y^\beta),\
X_{3u} = -\frac{1}{2}(X_x^\alpha+X_x^\beta), \
X_{3\varv} = -\frac{1}{2}(X_y^\alpha-X_y^\beta),
\label{eq:cubic_hexa_2}
\end{align}
\end{widetext}
in  $\chi^{[1\times2]}$.

\clearpage
\bibliographystyle{apsrev4-2}
\bibliography{ref}
\end{document}